\documentclass[aps,prx,reprint,twocolumn,floatfix,superscriptaddress,amsmath,amssymb,amsfonts,nofootinbib]{revtex4-2}

\usepackage{url}
\usepackage{bm}
\usepackage{graphicx}
\usepackage{amsmath}
\usepackage{amstext}
\usepackage{amssymb}
\usepackage{amsfonts}
\usepackage{amsbsy}
\usepackage{verbatim}
\usepackage{color}
\usepackage[colorlinks=true, urlcolor=blue, linkcolor=blue, citecolor=blue, pdftex]{hyperref}
\usepackage{floatrow}
\usepackage{float}
\usepackage{tikz}
\usepackage{gensymb}
\usepackage{textcomp}
\usepackage{enumitem}
\usepackage[version=3]{mhchem}
\usepackage{afterpage}
\usepackage{pbox}
\usepackage{dsfont}
\usepackage{siunitx}
\usepackage{stackengine,wasysym,scalerel}
\usepackage{latexsym}
\usepackage{cancel}
\usepackage{comment}
\usepackage{array, multirow, bigdelim, makecell, booktabs}
\usepackage[misc]{ifsym}
\usepackage[capitalise]{cleveref}
\usepackage{sidecap}
\usepackage{xcolor}
\usepackage{lipsum}

\usepackage{color}
\usepackage[colorlinks=true, urlcolor=blue, linkcolor=blue, citecolor=blue, pdftex]{hyperref}

\usepackage[normalem]{ulem}
\newcommand{\pdagger}{\phantom{\dagger}}

\definecolor{darkblue}{HTML}{004D6B}
\definecolor{darkred}{HTML}{8c1515}
\definecolor{darkgreen}{HTML}{006400}
\usepackage{hyperref}
\hypersetup{
	colorlinks=true,
	urlcolor=blue,
	citecolor=blue,
	linkcolor=blue,
	breaklinks
}

\graphicspath{{figures/}}
\allowdisplaybreaks

\begin{document}
\title{Semi-Dirac spin liquids and frustrated quantum magnetism on the trellis lattice}
\author{Sourin Chatterjee}
\affiliation{Department of Physics and Quantum Centre of Excellence for Diamond and Emergent Materials (QuCenDiEM), Indian Institute of Technology Madras, Chennai 600036, India}
\author{Atanu Maity}
\affiliation{Institut f\"ur Theoretische Physik und Astrophysik and W\"urzburg-Dresden Cluster of Excellence ct.qmat, Julius-Maximilians-Universit\"at W\"urzburg, Am Hubland, Campus S\"ud, W\"urzburg 97074, Germany}
\affiliation{Department of Physics and Quantum Centre of Excellence for Diamond and Emergent Materials (QuCenDiEM), Indian Institute of Technology Madras, Chennai 600036, India}
\author{Janik Potten}
\affiliation{Institut f\"ur Theoretische Physik und Astrophysik and W\"urzburg-Dresden Cluster of Excellence ct.qmat, Julius-Maximilians-Universit\"at W\"urzburg, Am Hubland, Campus S\"ud, W\"urzburg 97074, Germany}
\author{Tobias M\"uller}
\affiliation{Institut f\"ur Theoretische Physik und Astrophysik and W\"urzburg-Dresden Cluster of Excellence ct.qmat, Julius-Maximilians-Universit\"at W\"urzburg, Am Hubland, Campus S\"ud, W\"urzburg 97074, Germany}
\author{Andreas Feuerpfeil}
\affiliation{Institut f\"ur Theoretische Physik und Astrophysik and W\"urzburg-Dresden Cluster of Excellence ct.qmat, Julius-Maximilians-Universit\"at W\"urzburg, Am Hubland, Campus S\"ud, W\"urzburg 97074, Germany}
\author{Ronny Thomale}
\affiliation{Institut f\"ur Theoretische Physik und Astrophysik and W\"urzburg-Dresden Cluster of Excellence ct.qmat, Julius-Maximilians-Universit\"at W\"urzburg, Am Hubland, Campus S\"ud, W\"urzburg 97074, Germany}
\affiliation{Department of Physics and Quantum Centre of Excellence for Diamond and Emergent Materials (QuCenDiEM), Indian Institute of Technology Madras, Chennai 600036, India}
\author{Karlo Penc}
\affiliation{Institute for Solid State Physics and Optics, HUN-REN Wigner Research Centre for Physics, P.O. Box 49, H-1525 Budapest,  Hungary}
\affiliation{Department of Physics and Quantum Centre of Excellence for Diamond and Emergent Materials (QuCenDiEM), Indian Institute of Technology Madras, Chennai 600036, India}
\author{Harald O. Jeschke}
\affiliation{Research Institute for Interdisciplinary Science, Okayama University, Okayama 700-8530, Japan}
\affiliation{Department of Physics and Quantum Centre of Excellence for Diamond and Emergent Materials (QuCenDiEM), Indian Institute of Technology Madras, Chennai 600036, India}
\author{Rhine Samajdar}
\email{rhine\_samajdar@princeton.edu}
\affiliation{Department of Physics, Princeton University, Princeton, New Jersey 08544, USA}
\affiliation{Princeton Center for Theoretical Science, Princeton University, Princeton, New Jersey 08544, USA}
\author{Yasir Iqbal}
\email{yiqbal@physics.iitm.ac.in}
\affiliation{Department of Physics and Quantum Centre of Excellence for Diamond and Emergent Materials (QuCenDiEM), Indian Institute of Technology Madras, Chennai 600036, India}

\begin{abstract}
Geometrical frustration in quantum magnets provides a fertile setting for unconventional phases of matter, including quantum spin liquids (QSLs). The trellis lattice, with its complex site arrangements and edge-sharing triangular motifs, presents a promising platform for such physics. In this work, we undertake a comprehensive classification of all fully symmetric QSLs on the trellis lattice using the projective symmetry group approach within the Abrikosov-fermion representation. We find 7 U(1) and 25 $\mathbb{Z}_{2}$ short-ranged \textit{Ans\"atze} and analyze the phase diagram in the mean-field parameter space, uncovering both gapped and Dirac QSLs as well as a semi-Dirac spin liquid that emerges at the level of projective symmetry group classification and mean-field band structure, in which the spinon dispersion is linear along one momentum direction but quadratic along the orthogonal one. We demonstrate that such dispersions can occur only at high-symmetry points in the Brillouin zone with $C_{2v}$ little groups and analyze their characteristic correlation signatures. Moreover, by optimizing over all symmetry-allowed mean-field states, we map out a phase diagram---featuring six distinct phases---of the nearest-neighbor Heisenberg Hamiltonian on the trellis lattice. Among these, we find four quasi-one-dimensional QSL phases, one dimer phase, and one Dirac QSL phase. Going beyond mean field, we also assess equal-time and dynamical spin structure factors of these phases using density-matrix renormalization group and Keldysh pseudofermion functional renormalization group calculations and compare qualitative momentum-space features of these spectra with those obtained at the mean-field level. Finally, we identify four cuprate and vanadate compounds as promising experimental realizations and provide spectroscopic predictions, based on first-principles Hamiltonians, as a guide for future neutron-scattering studies on these materials.
\end{abstract}

\date{\today}

\maketitle

\section{Introduction}

The investigation of magnetic properties of quantum materials has long been a central focus in condensed matter physics, particularly the intricate relationship between lattice topology and quantum orders~\cite{auerbach2012interacting, sachdev2023quantum}. A key aspect of this interplay arises when specific geometric arrangements of magnetic moments give rise to \textit{competing} interactions---an effect known as geometric frustration. Lattices containing triangular motifs serve as paradigmatic examples of such frustration, often exhibiting complex magnetic behavior such as spin-spiral states, non-coplanar orders, and, in certain cases, extensive ground-state degeneracy that suppresses conventional long-range magnetic order~\cite{Gang2007,yasir2013,yasir2016,yasir2014,yasir2015,Hu2019,Hu2015,Leon2015,He2017}.

Generically, quantum fluctuations at low temperature tend to destabilize classical magnetic order, pushing the system into a quantum paramagnetic regime. When such fluctuations are sufficiently strong---as alluded to above---they can lead to exotic phases of matter known as quantum spin liquids (QSLs), characterized by the absence of any spontaneous symmetry breaking even at zero temperature~\cite{balents2010,savary2016,kivelson2023}. In recent years, the study of frustrated magnets has garnered significant attention, fueled by both theoretical and experimental evidence for QSL behavior in various two- and three-dimensional lattice geometries, including kagome, triangular, and pyrochlore structures~\cite{arh2022,gao2019,lu2022,gaulin2014,Changlani2019,Smith2022,sibille2020}.

Two-dimensional (2D) Heisenberg antiferromagnets on Archimedean lattices provide a rich platform for exploring the influence of lattice geometry and quantum fluctuations on magnetic phases~\cite{wannier1950,Vannimenus1977,toulouse1980,Farnll2014}.  Of the eleven Archimedean lattices, four---namely the square~\cite{Oitmaa-1978,Trivedi-1989,Sandvik-1997}, honeycomb~\cite{Castrohoneycomb,Albuquerque_PRB.84.024406_honeycomb}, square-octagon, and square-hexagon-dodecagon lattices---are unfrustrated and, as expected, support long-range magnetic order for the Heisenberg model with only nearest-neighbor interactions. The remaining seven incorporate geometrical frustration due to their corner- or edge-sharing triangular motifs. Among these, considering again nearest-neighbor antiferromagnetic couplings of equal strength, the Shastry-Sutherland~\cite{Shastry-1981} and triangular~\cite{Singh-1992,Bernu-1992,Capriotti-1999} lattices maintain long-range order despite strong frustration, whereas the kagome and star lattices host magnetically disordered ground states. The remaining three lattices---maple leaf, trellis, and bounce---lie in an intermediate regime. In this list of candidates, the maple-leaf lattice, in particular, has been intensely studied in recent years~\cite{Misguich1999,Richter2002,Farnell2011,beck2024}, revealing a variety of exotic phases, including an exact dimerized ground state~\cite{Ghosh2022} and possible quantum spin liquid behavior~\cite{lasse2023,ghosh2024,gembe2024,sonnenschein2024}. 

In contrast, the trellis lattice has remained largely unexplored~\cite{Richter2004}, despite its complex site arrangements and edge-sharing triangular connectivity that can host unconventional magnetic structures not typically found in simpler geometries. 
An early study employing various numerical and analytical techniques revealed a range of diverse magnetic and nonmagnetic phases on the trellis lattice~\cite{karlo1997}. It identified a nondegenerate valence-bond solid state in the dimer limit and a subextensively degenerate (gapless) Majumdar-Ghosh-like phase in the regime of weakly coupled zigzag chains, both of which have a finite singlet-triplet gap. Between these two extremes, gapless magnetically ordered phases emerge. Specifically, antiferromagnetic order appears near the honeycomb-lattice limit, while a helical state with incommensurate spin correlations develops along the chains when all three magnetic interactions compete. Extensions of this framework to include ferromagnetic couplings have also been explored~\cite{Dalosto2000,Maiti2019_2,Rabuffo}. 
The aforementioned gapped rung-dimer phase of the trellis lattice has been extensively investigated using perturbative methods and quantum Monte Carlo (QMC) simulations~\cite{Gopalan, Johnston, Miyahara_1998, Riera2001}. In particular, the dynamical structure factor (DSF)---interpreted in terms of triplon excitations---was computed in Ref.~\onlinecite{Schmidt2007}.  Additionally, several works have examined coupled zigzag chains, which form the trellis lattice backbone, using methods such as the coupled-cluster approach~\cite{Zinke5} and the density-matrix renormalization group (DMRG) combined with linearized spin-wave analysis~\cite{Maiti2019}. 
Notably, the trellis lattice can be regarded as a special case of a spatially anisotropic triangular lattice~\cite{Sakakida2017}. When third-nearest-neighbor couplings are set to zero on certain triangles of this anisotropic lattice, the system reduces to a trellis lattice antiferromagnet that supports a stripe like phase composed of alternately stacked one-dimensional (1D) spiral spin chains. These investigations already highlight the rich variety of ground states and excitation spectra that arise from the interplay between frustration, dimensional crossover, and quantum fluctuations in trellis lattice antiferromagnets. Even more interestingly, at the isotropic point, the ground state could be a quantum paramagnet~\cite{Farnll2014}, and possibly, a quantum spin liquid. 

On the experimental front, the trellis lattice garnered significant attention following the synthesis of the layered cuprate compound \ce{SrCu2O3}~\cite{Azuma1994}, in which magnetic Cu$^{2+}$ ions form spin-$1/2$ Heisenberg ladders weakly coupled via frustrated interladder exchanges. Strong antiferromagnetic interactions favor singlet formation on the rungs of the ladders, resulting in a spin gap in the excitation spectrum~\cite{Gopalan}. Under appropriate doping or applied pressure, related cuprate ladder systems, such as Sr$_{14-x}$Ca$_x$Cu$_{24}$O$_{41}$, also exhibit superconductivity, suggesting an intrinsic tendency toward pair formation in these quasi-one-dimensional systems~\cite{Kontani1998}. 
Subsequent studies extended trellis-lattice physics to layered vanadate compounds such as \ce{CaV2O5}~\cite{Iwase1996}, \ce{MgV2O5}~\cite{Millet}, and Pb$_{0.55}$Cd$_{0.45}$V$_2$O$_5$~\cite{Tsirlin2007}. In these materials, V$^{4+}$ ions form spin-$1/2$ planes with a trellis lattice geometry. Unlike in the cuprates, the interladder couplings in vanadates can be appreciable, as revealed by \emph{ab initio} calculations~\cite{Korotin1999,Korotin2000}, magnetic susceptibility measurements~\cite{Johnston}, and Raman spectroscopy~\cite{Lemmens,Konstantinovi}.
More recently, quasi-one-dimensional organic magnets such as $\alpha$-2-Cl-4-F-V~\cite{Kono} and $\beta$-2,3,5-Cl$_3$-V~\cite{Yamaguchi} have been synthesized. These molecular crystals, built from verdazyl radicals carrying spin-$1/2$, offer a platform for investigating quantum magnetism in low dimensions.

Despite these advances, our understanding of \textit{nonmagnetic} phases on the trellis lattice remains limited. An important open question is whether the combined effects of geometrical frustration, quantum fluctuations, and lattice topology can stabilize exotic phases on this geometry, in particular, a quantum spin liquid at and around the isotropic point. Motivated by these considerations, our present work is devoted to a comprehensive investigation of the properties of various potential QSL ground states on the trellis lattice. Owing to their paramagnetic nature, QSLs cannot be described using the conventional Landau theory of spontaneous symmetry breaking. Instead, their classification involves internal symmetry structures governed by emergent SU(2) gauge degrees of freedom coupled to matter fields~\cite{misguich2010}. The projective symmetry group (PSG) framework~\cite{Wen2002,Wenbook2007,Yang2012} provides a powerful approach for the systematic identification and classification of distinct QSL phases by exploiting these emergent gauge structures.
This formalism has been successfully employed for many 2D Archimedean lattices, including the square~\cite{Wen2002, Wang2016}, triangular~\cite{Ashvin_2006, Lu2016, Lu2018}, honeycomb~\cite{Wang2010,Ying2011}, kagome~\cite{lu2011, Bieri2015,Bieri2016}, 
star~\cite{choy2009,yang2010},
square-octagon~\cite{Atanu2023}, 
Shastry-Sutherland~\cite{liu2024,maity2024_3}, maple-leaf~\cite{sonnenschein2024}, and bounce~\cite{maity2024_1,maity2024_2} lattices.

\subsection{Outline of the manuscript}

This paper is organized as follows. In Sec.~\ref{sec:PSG}, we provide a brief introduction to the fermionic representation of spin-$1/2$ operators and outline the fundamentals of the PSG construction. Section~\ref{sec:symmetry} presents a detailed analysis of the full symmetry group of the trellis lattice, along with its projective extensions, defined up to elements of the $\mathrm{U}(1)$ and $\mathbb{Z}_2$ gauge groups.
Focusing on the three symmetry-inequivalent nearest-neighbor bonds, we classify all distinct mean-field states corresponding to different QSLs that arise from the projective extensions of the full symmetry group. These states, along with their associated gauge structures and flux patterns, are summarized in Sec.~\ref{sec:ansatze}. In Sec.~\ref{sec:dispersion}, we investigate the phases that emerge as the hopping parameters are varied within the mean-field description.
We also construct a phase diagram in Sec.~\ref{sec:PD} based on a spin-$1/2$ antiferromagnetic Heisenberg model, with exchange couplings $J_v$, $J_z$, and $J_h$ defined on the three symmetry-inequivalent nearest-neighbor bonds of the trellis lattice. The spectral properties of the various QSL states are then analyzed in Sec.~\ref{sec:SF}, where we compute both the dynamical and equal-time structure factors (EQSFs) from our mean-field \textit{Ans\"atze} as well as using state-of-the-art Keldysh pseudofermion functional renormalization group (pf-FRG) and  DMRG calculations. Thereafter, Sec.~\ref{sec:mat} provides an in-depth discussion on candidate compounds that realize the trellis lattice, using density functional theory (DFT) and Keldysh pf-FRG, thus outlining a pathway
for the observation of this physics in future experiments on quantum materials.
Finally, we conclude with a brief discussion of our results and their implications in Sec.~\ref{sec:end}.

\subsection{Summary of main results}
As mentioned above, in this work, we implement the PSG construction within the Abrikosov-fermion (parton) representation of spin-$1/2$ operators to classify all distinct fully symmetric QSL states at the mean-field level that can be realized on the trellis lattice. Subsequently, we analyze the corresponding fermionic band structures and map out the associated phase diagram. Besides fully gapped and Dirac QSLs, we discover an intriguing \textit{semi}-Dirac spin liquid state~\cite{Shao-2024,Dietl2008,Singh2009,Montambaux2009,Delplace2010,Moessner2013,Saha2016,Roy2018,Uryszek2019,Kotov2021,Yuan2016_1,Uryszek2020,Jimin2015,Kim2017,Rudenko2015,Bellec2013,Rechtsman2013,Real2020}, for which the spinon dispersion is observed to be linear along one momentum direction but quadratic in the direction orthogonal to it.

For itinerant electrons, semi-Dirac dispersions have recently attracted considerable attention, as they arise from the merging of multiple Dirac points into a single one~\cite{Dietl2008}. Semi-Dirac fermions naturally display strongly anisotropic characteristics: Along one direction the dispersion is quadratic, reflecting Galilean invariance, while along the orthogonal direction it remains linear, reflecting Lorentz invariance. A direct consequence of this anisotropy is the unconventional Landau-level scaling of  $B^{2/3}$ with a magnetic field $B$, which has recently been reported in magneto-optical spectroscopy of ZrSiS and microscopically linked to nodal-line crossings in the material~\cite{Shao-2024}. Thermal or charge transport, being inherently direction sensitive, offers an additional potential probe of semi-Dirac points. However, such measurements require a spectrally isolated semi-Dirac point to achieve a satisfactory signal-to-noise ratio. Despite their charge neutrality, many features of itinerant electrons can carry over to elementary spinon excitations in spin liquids, including their response to external magnetic fields~\cite{PhysRevB.107.195155}. In particular, thermal conductivity may serve as a probe of transport signatures from low-lying spinon excitations, although such measurements remain experimentally challenging~\cite{PhysRevLett.117.267202,PhysRevX.9.041051,doi:10.1126/science.1188200}.

Here, we demonstrate that semi-Dirac spinon dispersions can occur only at high-symmetry points in the Brillouin zone where the little group is $C_{2v}$, i.e., characterized by a twofold rotation symmetry and two orthogonal reflection symmetries. In contrast, momentum points in the Brillouin zone with larger little groups cannot host semi-Dirac dispersions. Scaling analysis further reveals that the equal-time spin-spin correlations in real space decay algebraically as $\sim 1/r^3$ and $\sim 1/r^6$ along the two orthogonal directions. Finally, we present both dynamical and equal-time spin structure factors, at the mean-field level, for semi-Dirac spin liquids (s-DSLs).

Within a self-consistent mean-field framework, we map out the quantum phase diagram of the most general nearest-neighbor Heisenberg antiferromagnetic Hamiltonian. This procedure entails optimizing over all symmetry-allowed U(1) and $\mathbb{Z}_{2}$ mean-field \textit{Ans\"atze} in order to identify the energetically favored solutions. Our analysis reveals six distinct phases, characterized by correlation patterns indicative of (I) zigzag-chain, (II) zigzag-ladder, (III) rung-chain, (IV) rung-ladder, (V) ladder-dimer, and (VI) honeycomb structures. For representative points within these phases, we further assess the role of gauge fluctuations beyond mean-field: at zero temperature using DMRG, and at finite temperature via  Keldysh pf-FRG. These calculations reveal significant redistribution of spectral weight in particular within the ladder-dimer and honeycomb phases.

Furthermore, we identify four candidate materials---two cuprates and two vanadates---that can be modeled to good accuracy as two-dimensional systems with dominant Heisenberg interactions along the ladder ($J_v$), rung ($J_h$), and zigzag ($J_z$) bonds. Using DFT combined with an energy-mapping approach, we extract the effective model parameters for CaCu$_2$O$_3$, SrCu$_2$O$_3$, MgV$_2$O$_5$, and CaV$_2$O$_5$. In these compounds, the Cu$^{2+}$ and V$^{4+}$ magnetic ions host $S=1/2$ moments forming the underlying magnetic lattice, which takes the form of a flat trellis lattice in SrCu$_2$O$_3$ and a buckled trellis lattice in the other three compounds.  
The hierarchy of couplings places CaCu$_2$O$_3$ in close proximity to a one-dimensional limit, while SrCu$_2$O$_3$ and CaV$_2$O$_5$ reduce to coupled two-leg ladder systems, with the latter exhibiting substantially stronger rung couplings. In contrast, MgV$_2$O$_5$ displays comparable rung and chain couplings. These effective Hamiltonians are then analyzed using Keldysh pf-FRG, enabling the computation of dynamical spin structure factors at finite temperature. For CaCu$_2$O$_3$, we find excellent agreement between our calculated spectral functions and previously reported neutron scattering data, thereby validating our effective Hamiltonian. For SrCu$_2$O$_3$ and CaV$_2$O$_5$, the dynamical and equal-time structure factors closely resemble those expected for the rung-ladder and ladder-dimer phases, respectively, while MgV$_2$O$_5$ shows clear signatures of magnetic ordering. Our predictions for the dynamical structure factors of SrCu$_2$O$_3$ and CaV$_2$O$_5$ provide concrete benchmarks for future inelastic neutron scattering experiments.

\section{Theoretical framework} 
\label{sec:PSG}

In this section, we outline the generic theoretical framework employed to investigate quantum spin liquid phases. We begin with the Heisenberg Hamiltonian,
\begin{equation}\label{Hamiltonian}
    \hat{\mathcal{H}}= \sum_{\langle i j \rangle} J_{i j} \hat{\mathbf{S}}_i\cdot\hat{\mathbf{S}}_j,
\end{equation}
where $\hat{\mathbf{S}}_i$ denotes the spin operator at site $i$ and $J_{ij}$ represents the exchange coupling between sites $i$ and $j$.

To capture the fractionalized nature of excitations characteristic of QSLs, we introduce charge-neutral, spin-$1/2$ quasiparticles known as spinons or partons. These quasiparticles can obey either bosonic or fermionic statistics. In this work, we adopt the fermionic representation, which conveniently allows for a unified treatment of both gapped and gapless QSLs. In the Abrikosov fermion  formalism, the spin operators are expressed as~\cite{Baskaran1988,Affleck1965}
\begin{equation}\label{spinon}
    \hat{S}_i^{\alpha} = \frac{1}{2} \sum_{\sigma \sigma'} \hat{f}^{\dagger}_{i \sigma} \tau^{\alpha}_{\sigma \sigma'} \hat{f}^{\phantom{\dagger}}_{i \sigma'},
\end{equation}
where $\alpha = x, y, z$, $\sigma,\sigma' \in \{\uparrow,\downarrow\}$, and $\tau^\alpha$ are the usual Pauli matrices.

While this representation captures the essence of fractionalization, it enlarges the local Hilbert space. The physical spin Hilbert space at each site has dimension 2, corresponding to spin-up and spin-down states. However, in the fermionic representation, the local Hilbert space includes four states: two singly occupied (physical), and one each of doubly occupied and empty (unphysical). To project back to the physical Hilbert space, we impose the following local constraints:
\begin{align}\label{constraint}
    \sum_{\sigma} \hat{f}^{\dagger}_{i\sigma} \hat{f}^{\phantom{\dagger}}_{i\sigma} = 1, \quad
    \sum_{\sigma,\sigma'} \hat{f}^{\phantom{\dagger}}_{i\sigma} \hat{f}^{\phantom{\dagger}}_{i\sigma'} \varepsilon^{\phantom{\dagger}}_{\sigma\sigma'} = 0,
\end{align}
where $\varepsilon_{\sigma\sigma'}$ is the rank-2 antisymmetric tensor.

The fermionic parton construction naturally gives rise to an emergent SU(2) gauge symmetry. This becomes explicit by introducing the SU(2) doublet field
\begin{equation}\label{eq:psi}
    \hat{\psi}^{\pdagger}_i =
    \begin{bmatrix}
        \hat{f}^{\phantom{\dagger}}_{i \uparrow} & \phantom{-}\hat{f}^\dagger_{i \downarrow} \\
        \hat{f}^{\phantom{\dagger}}_{i \downarrow} & -\hat{f}^\dagger_{i \uparrow}
    \end{bmatrix},
\end{equation}
in terms of which the spin operators can be compactly written as
\begin{equation}
    \hat{S}_{i}^{\alpha} = \frac{1}{2} \mathrm{Tr} \left[\hat{\psi}^{\dagger}_i \tau^{\alpha} \hat{\psi}^{\phantom{\dagger}}_i\right].
\end{equation}
This representation reveals the invariance of spin operators under local SU(2) gauge transformations $\hat{\psi}_i \rightarrow \hat{\psi}_i W_i$, with $W_i \in \mathrm{SU}(2)$.

Substituting Eq.~\eqref{spinon} into the Heisenberg Hamiltonian \eqref{Hamiltonian} results in a quartic fermionic interaction term. To obtain a solvable theory, we perform a mean-field decoupling in the spin-singlet channel, which is appropriate for describing spin-rotation-invariant states. The relevant mean-field parameters are the singlet hopping, $\chi_{ij}$, and singlet pairing, $\Delta_{ij}$, fields, defined as
\begin{align}
    \chi^{\phantom{\dagger}}_{ij} &= \left\langle \sum_{\sigma} \hat{f}^{\dagger}_{i \sigma} \hat{f}^{\phantom{\dagger}}_{j \sigma} \right\rangle, \\
    \Delta_{ij}^\dagger &= -\left\langle \sum_{\sigma,\sigma'} \hat{f}^{\phantom{\dagger}}_{i \sigma} \hat{f}^{\phantom{\dagger}}_{j \sigma'} \varepsilon^{\phantom{\dagger}}_{\sigma \sigma'} \right\rangle. \label{eq: pairing}
\end{align}

The resulting mean-field Hamiltonian takes the form
\begin{align}
    \hat{H}^{\phantom{\dagger}}_{\mathrm{MF}} &= \frac{3}{8} \sum_{i,j} J^{\pdagger}_{ij} \left[ \frac{1}{2} \mathrm{Tr}(u^\dagger_{ij} u^{\phantom{\dagger}}_{ij}) 
    - \mathrm{Tr} \left(\hat{\psi}^{\pdagger}_i u^{\phantom{\dagger}}_{ij} \hat{\psi}^{\dagger}_j + \mathrm{h.c.} \right) \right] \notag \\
    &+ \sum_{i,\mu} a^{\pdagger}_\mu(i) \, \mathrm{Tr} \left[ \hat{\psi}^{\pdagger}_i \tau^{\mu} \hat{\psi}^{\dagger}_i \right], \label{eq:quadraticH}
\end{align}
where $u_{ij}$ is the SU(2) matrix-valued link field associated with the bond $\langle ij \rangle$, given by
\begin{equation}\label{eq:ansatz}
    u^{\pdagger}_{ij} =
    \begin{bmatrix}
        \chi^{\dagger}_{ij} & \Delta^{\phantom{\dagger}}_{ij} \\
        \Delta^{\dagger}_{ij} & -\chi^{\pdagger}_{ij}
    \end{bmatrix}.
\end{equation}
The last term in Eq.~\eqref{eq:quadraticH} imposes the single-occupancy constraint~\eqref{constraint} at the mean-field level using Lagrange multipliers $a_\mu(i)$ for $\mu = 1,2,3$. Together, the combination $\{u_{ij}, a_\mu(i)\}$ defines a mean-field \textit{Ansatz} characterizing a quantum paramagnetic state. 

Now, it is straightforward to verify that the mean-field Hamiltonian in Eq.~\eqref{eq:quadraticH} is invariant under local SU(2) gauge transformations of the form
\begin{align}
    \hat{\psi}^{\pdagger}_i \rightarrow \hat{\psi}^{\pdagger}_i W^{\pdagger}_i, \quad 
    u^{\pdagger}_{ij} \rightarrow W_i^\dagger u^{\pdagger}_{ij} W^{\pdagger}_j, \quad 
    a^{\pdagger}_\alpha \tau^\alpha \rightarrow a^{\pdagger}_\alpha W_i^\dagger \tau^\alpha W^{\pdagger}_i,
\end{align}
where $W_i \in \mathrm{SU}(2)$ is a local gauge transformation matrix.

The physical implication of this gauge redundancy is that a given mean-field \textit{Ansatz} $\{u^{}_{ij}, a^{}_\alpha\}$ and its gauge-transformed counterpart $\{u'_{ij}, a'_\alpha\}$ describe the same physical spin liquid state. In other words, different \textit{Ans\"atze} related by local SU(2) gauge transformations are physically equivalent, whereas those not connected by such transformations correspond to distinct quantum paramagnetic phases. This redundancy can be exploited to classify and characterize QSL phases using projective symmetry considerations, as we elaborate on next.

An important consequence of the SU(2) gauge structure is that global and lattice symmetries may act on the fermionic operators in a \emph{projective} manner. To illustrate this, consider a symmetry operation $\mathcal{O}$ from the lattice space group. Its naive action on the mean-field variables is to map $u_{ij} \rightarrow u_{\mathcal{O}(i)\mathcal{O}(j)}$. If it happens that $u_{ij} \neq u_{\mathcal{O}(i)\mathcal{O}(j)}$, one might incorrectly conclude that the \textit{Ansatz} breaks the symmetry $\mathcal{O}$. However, the symmetry can still be preserved projectively if there exists a local SU(2) gauge transformation $W_{\mathcal{O}}(i)$ associated with $\mathcal{O}$ such that
\begin{equation}\label{gauge_uij}
    W_{\mathcal{O}}^\dagger(\mathcal{O}(i))\, u^{\pdagger}_{\mathcal{O}(i)\mathcal{O}(j)}\, W^{\pdagger}_{\mathcal{O}}(\mathcal{O}(j)) = u^{\pdagger}_{ij}.
\end{equation}
In this case, the combined operation of the lattice symmetry $\mathcal{O}$ and the corresponding gauge transformation $W_{\mathcal{O}}$ leaves the \textit{Ansatz} invariant. The full set of such symmetry-gauge combinations $\{\mathcal{O}, W_{\mathcal{O}}\}$ defines the PSG. The PSG provides a classification scheme for QSLs, playing a role analogous to that of symmetry groups in the Landau paradigm of symmetry-breaking phases.

Within the PSG framework, one can also define an analog of the identity operation. Specifically, we define the invariant gauge group (IGG) as the subgroup of SU(2) gauge transformations $\mathcal{G}_i$ that leave the mean-field \textit{Ansatz} invariant:
\begin{equation}
    \mathcal{G}_i^\dagger u^{\pdagger}_{ij} \mathcal{G}^{\pdagger}_j = u^{\pdagger}_{ij}, \qquad \mathcal{G}^{\pdagger}_i \in \mathrm{SU}(2).
\end{equation}
The IGG thus characterizes the residual gauge freedom of the \textit{Ansatz} and corresponds to the projective realization of the identity element in the symmetry group.

Although the IGG describes, in general, a \emph{local} symmetry operation, one can always choose a gauge such that the IGG only involves \emph{global} elements of either $\mathrm{SU}(2)$ or its subgroups, $\mathrm{U}(1)$, or $\mathbb{Z}_2$. For such a gauge choice, the \textit{Ansatz} is said to be in its canonical form, and its IGG structure becomes manifest. To make this explicit, we parametrize the SU(2) link variable $u_{ij}$ in terms of four real numbers $\lambda^\mu_{ij}$ as
\begin{equation}
    u^{\pdagger}_{ij} = i \lambda^0_{ij} \tau^0 + \sum_{\mu=1}^{3} \lambda^\mu_{ij} \tau^\mu.
\end{equation}
Based on this parametrization, different IGG structures emerge:
\begin{enumerate}[label=(\alph*)]
    \item If $\lambda^\mu_{ij} = 0$ for $\mu=1,2,3$, then the IGG is global $\mathrm{SU}(2)$.
    \item If $\lambda^1_{ij} = \lambda^2_{ij} = 0$ and $\lambda^3_{ij} \neq 0$, then the IGG is global $\mathrm{U}(1)$.
    \item If $\lambda^\mu_{ij} \neq 0\, \forall \, \mu$, then the IGG is $\mathbb{Z}_2$.
\end{enumerate}
Since the IGG is a gauge-invariant property of the \textit{Ansatz}, it is conventional to label the \textit{Ans\"atze} according to their IGG structure [e.g., $\mathrm{U}(1)$ QSL, $\mathbb{Z}_2$ QSL, etc.], thereby identifying the type of emergent gauge theory associated with each QSL phase.

\begin{figure}[t]
\includegraphics[width=1.0\linewidth]{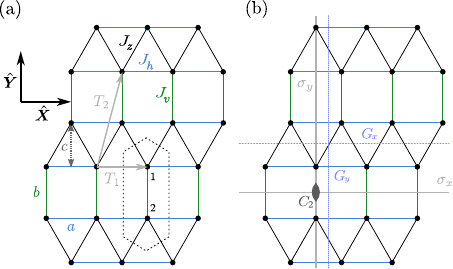}
\caption{(a) Trellis lattice with exchange couplings $J_v$ (green), $J_z$ (black), and $J_h$ (blue). The unit cell contains two sites labeled ``1'' and ``2.'' The lattice parameters are denoted by $a$, $b$, and $c$. (b) Illustration of the lattice space-group symmetries. $C_2$ represents a twofold rotation about an axis perpendicular to the lattice plane. $\sigma_x$ and $\sigma_y$ denote reflections about the horizontal and vertical solid lines, respectively. $G_x$ is a glide symmetry consisting of a reflection about the horizontal dashed line followed by a translation $\frac{a}{2}\hat{\boldsymbol{X}} = \frac{1}{2}\boldsymbol{T}_1$. Similarly, $G_y$ combines a reflection along the vertical dashed line with a translation $(b+c)\hat{\boldsymbol{Y}} = -\frac{1}{2}\boldsymbol{T}_1 + \boldsymbol{T}_2$.}
\label{fig:lattice}
\end{figure}

\section{Symmetries and projective realizations}\label{sec:symmetry}
\subsection{Lattice and time-reversal symmetries}
\label{sec:lattice_time}

On the trellis lattice, any site can be described by the position vector $\boldsymbol{r} = x\,\boldsymbol{T}_1 + y\,\boldsymbol{T}_2 + \bm{\epsilon}_u$, where we choose the lattice vectors in the Cartesian basis as $\boldsymbol{T}_1 = a\hat{\boldsymbol{X}}$ and $\boldsymbol{T}_2 = (a/2)\hat{\boldsymbol{X}} + (b + c)\hat{\boldsymbol{Y}}$, with $a$, $b$, and $c$ denoting the bond lengths as illustrated in Fig.~\ref{fig:lattice}. The vectors $\bm{\epsilon}_u$ represent the positions of the sublattices within the unit cell. The unit cell consists of two sites located at $\bm{\epsilon}_1 = (b/2) \hat{\boldsymbol{Y}}$ and $\bm{\epsilon}_2 = -(b/2) \hat{\boldsymbol{Y}}$.

The symmetry group of the trellis lattice corresponds to the nonsymmorphic wallpaper group \textit{cmm}, which can be generated by two translations ($T_1$ and $T_2$), a twofold rotation ($C_2$), two mirror reflections ($\sigma_x$ and $\sigma_y$), and two glide reflections ($G_x$ and $G_y$). A pictorial representation of these symmetry elements is provided in Fig.~\ref{fig:lattice}(b). The action of each symmetry operation on a lattice site labeled by $(x, y, u)$---where $(x, y)$ denotes the unit cell coordinate and $u$ specifies the sublattice index---is given by
\begin{equation}
\left.\begin{aligned}\label{eq:generators}
& T^{}_1(x, y, u) \rightarrow (x+1, y, u), \\
& T^{}_2(x, y, u) \rightarrow (x, y+1, u), \\
& C^{}_2(x, y, u) \rightarrow (-x, -y, \bar{u}), \\
& \sigma^{}_x(x, y, u) \rightarrow (x+y, -y, \bar{u}), \\
& \sigma^{}_y(x, y, u) \rightarrow (-x-y, y, u), \\
& G^{}_x(x, y, u) \rightarrow (x+y, -y+1, \bar{u}), \\
& G^{}_y(x, y, u) \rightarrow (-x-y, y+1, u),
\end{aligned}\right.
\end{equation}
with $\bar{u} \equiv 3 - u$.

However, one can verify that several of these symmetry operations can be expressed in terms of a minimal generating set consisting of $T_2$, $C_2$, and $\sigma_x$ as 
\begin{equation}\label{eq:symmetries}
\left.\begin{aligned}
& T^{}_1 = G_x^2, \,\, G^{}_x = T^{}_2 \sigma^{}_x,\,\, G^{}_y = T_2^2 C^{}_2 G^{}_x,  \,\, \sigma^{}_y = C^{}_2 \sigma^{}_x.
\end{aligned}\right.
\end{equation}
Therefore, the minimal set of generators required for constructing symmetric \textit{Ans\"atze} invariant under the entire space group comprises $T_2$, $C_2$, and $\sigma_x$. Nonetheless, we include $T_1$ in our analysis for convenience.

In addition to these lattice symmetries, we also incorporate time-reversal symmetry $\mathcal{T}$, given our focus on classifying fully symmetric (i.e., nonchiral) spin liquid \textit{Ans\"atze}. The time-reversal operation does not alter the spatial position of the lattice sites and hence, commutes with all space-group elements. However, its action on a mean-field \textit{Ansatz} is nontrivial~\cite{Wen2002,Bieri2016} and is given by
\begin{equation}
    \mathcal{T}(u^{\pdagger}_{ij}, a^{\pdagger}_\mu) = - (u^{\pdagger}_{ij}, a^{\pdagger}_\mu).
\end{equation}
Accordingly, the projective symmetry condition in Eq.~\eqref{gauge_uij} implies that the time-reversal symmetry imposes
\begin{equation}
    G_{\mathcal{T}}^\dagger(i) u^{\pdagger}_{ij} G^{\pdagger}_{\mathcal{T}}(j) = -u^{\pdagger}_{ij}, \quad \text{for } \mathcal{O} = \mathcal{T}.
\end{equation}
Taken together with Eq.~\eqref{eq:generators}, this completely specifies the full symmetry group relevant for the trellis lattice.

Last, we note that there exist specific combinations of symmetry operations under which any given lattice site remains invariant. These combinations effectively act as the identity and are summarized below:
\begin{align}
T^{}_1 T^{}_2 T^{-1}_1 T_2^{-1}&=  \mathcal{I},  \label{eq:translations} \\
T^{}_1 C_2^{-1} T^{}_1 C^{}_2 &= \mathcal{I}, \label{eq:C_2 T_1} \\
T^{}_2 C_2^{-1} T^{}_2 C^{}_2 &= \mathcal{I}, \label{eq:C_2 T_2} \\
T_1^{-1} \sigma_x^{-1} T^{}_1 \sigma^{}_x &= \mathcal{I}, \label{eq:T_1} \\
T_1^{-1} T^{}_2 \sigma_x^{-1} T_2 \sigma^{}_x &= \mathcal{I}, \label{eq: T_2} \\
C^{}_2 \sigma^{}_x C^{}_2 \sigma^{}_x &= \mathcal{I}, \label{eq:C_2 sigma_x} \\
C_2^2 &= \mathcal{I}, \label{eq:C22a} \\
\sigma_x^2 &= \mathcal{I}, \label{eq:sigma_x2} \\
T_1^{-1} T^{}_2 \sigma^{}_x T^{}_1 \sigma_x^{-1} T_2^{-1} &= \mathcal{I}, \label{eq:sigma_T1_T2} \\
C^{}_2 T^{}_1 T_2^{-1} \sigma^{}_x C^{}_2 T^{}_2 \sigma^{}_x &= \mathcal{I}, \label{eq:C_2 sigma_x2} \\
\mathcal{T}^2 &= \mathcal{I}, \label{eq:cyclic:TR} \\
\mathcal{T} \mathcal{O} \mathcal{T}^{-1} \mathcal{O}^{-1} &= \mathcal{I}, \label{eq:TRO}
\end{align}
where $\mathcal{O} \in \{T_1, T_2, \sigma_x, C_2\}$.

\begin{table*}
\caption{PSG solutions for the trellis lattice with invariant gauge group U(1). For rows 1--4 and 9--12, $\rho_{\mathcal{T},u} = \{0, \pi\}$; for rows 5--8, $\rho_{\mathcal{T},u} = 0$; and for rows 13--16, $\rho_{\mathcal{T},u} = \{0, n_{\mathcal{T}}\pi\}$. In total, 448 gauge-inequivalent PSG solutions exist for the trellis lattice with IGG U(1).}
\begin{ruledtabular}
\begin{tabular}{ccccccccccccccc}
PSG No. & $w^{}_{T_2}$ & $w^{}_{C_2}$ & $w^{}_{\sigma_x}$ & $w^{}_{\mathcal{T}}$ & $\theta$ & $\theta^{}_{C_2 T_1}$ & $\theta^{}_{C_2 T_2}$ & $\theta^{}_{\sigma_x T_1}$ & $\theta^{}_{\sigma_x T_2}$ & $\theta^{}_{\mathcal{T} T_2}$ & $\rho^{}_{C_2,u}$ & $\rho^{}_{\sigma_x,u}$ & No. of PSGs \\
\hline
1 & 0 & 0 & 0 & 0 & $n \pi$ & 0 & 0 & $n \pi$ & $n^{\pdagger}_{\sigma_x T_2}+n\pi/2$ & $n \pi$ & 0 & $\{0,n^{\pdagger}_{\sigma_x} \pi\}$ & $2^3$ \\
2 & 0 & 0 & 1 & 0 & $\theta$ & 0 & 0 & 0 & $n^{\pdagger}_{\sigma_x T_2}\pi - \theta/2$ & $n \pi$ & 0 & $\{0,n^{\pdagger}_{\sigma_x} \pi\}$ & $2^3$ \\
3 & 0 & 1 & 0 & 0 & $n \pi$ & $n \pi$ & $n^{\pdagger}_{C_2 T_2} \pi$ & $n \pi$ & 0 & $n \pi$ & $\{0,n^{\pdagger}_{C_2} \pi\}$ & $\{0,\theta^{\pdagger}_{\sigma_x}\}$ & $2^3$ \\
4 & 0 & 1 & 1 & 0 & $n\pi$ & $n \pi$ & $n^{\pdagger}_{C_2 T_2} \pi$ & $n \pi$ & 0 & $n \pi$ & $\{0,n^{\pdagger}_{C_2}  \pi\}$ & $\{0,n^{\pdagger}_{\sigma_x} \pi \}$ & $2^4$ \\
\hline
5 & 0 & 0 & 0 & 1 & $n \pi$ & $n\pi$ & $n^{\pdagger}_{T T_2} \pi$ & $n \pi$ & $n^{\pdagger}_{T T_2} \pi +n^{\pdagger}_{\sigma_x T_2} \pi$ & 0 & $\{0,n^{\pdagger}_{C_2} \pi\}$ & $\{0,n^{\pdagger}_{\sigma_x} \pi\}$ & $2^5$ \\
6 & 0 & 0 & 1 & 1 & $n \pi$ & $n\pi$ & $n^{\pdagger}_{T T_2} \pi$ & $n\pi$ & $n^{\pdagger}_{\sigma_x T_2} \pi$ & 0 & $\{0,n^{\pdagger}_{C_2} \pi\}$ & $\{0,n^{\pdagger}_{\sigma_x} \pi\}$ & $2^5$ \\
7 & 0 & 1 & 0 & 1 & $n \pi$ & $n \pi$ & $n^{\pdagger}_{C_2 T_2} \pi$ & $n \pi$ & 0 & 0 & $\{0,n^{\pdagger}_{C_2} \pi\}$ & $\{0,\theta^{\pdagger}_{\sigma_x} \}$ & $2^3$ \\
8 & 0 & 1 & 1 & 1 & $n \pi$ & $n \pi$ & $n^{\pdagger}_{C_2 T_2} \pi$ & $n \pi$ & 0 & 0 & $\{0,n^{\pdagger}_{C_2} \pi\}$ & $\{0,n^{\pdagger}_{\sigma_x} \pi \}$ & $2^4$ \\
\hline
\hline
9  & 1 & 0 & 0 & 0 & $n\pi$ & 0 & $n^{\pdagger}_{C_2 T_2}\pi$ & 0 & $n^{\pdagger}_{\sigma_x T_2}\pi$ & $n^{\pdagger}_{\mathcal{T} T_2}\pi$ & $\{0,\theta^{\pdagger}_{C_2}\}$ & $\{0,\theta^{\pdagger}_{\sigma_x}\}$ & $2^4$ \\
10 & 1 & 0 & 1 & 0 & $n\pi$ & 0 & $n^{\pdagger}_{C_2 T_2}\pi$ & 0 & 0 & $n^{\pdagger}_{\mathcal{T} T_2}\pi$ & $\{0,n^{\pdagger}_{C_2} \pi\}$ & $\{0,n^{\pdagger}_{\sigma_x}\pi\}$ & $2^5$ \\
11 & 1 & 1 & 0 & 0 & $n\pi$ & 0 & 0 & 0 & $n^{\pdagger}_{\sigma_x T_2}\pi$ & $n^{\pdagger}_{\mathcal{T} T_2}\pi$ & $\{0,n^{\pdagger}_{C_2} \pi\}$ & $\{0,n^{\pdagger}_{\sigma_x}\pi\}$ & $2^5$ \\
12 & 1 & 1 & 1 & 0 & $n\pi$ & 0 & 0 & 0 & $n^{\pdagger}_{\sigma_x T_2}\pi$ & $n^{\pdagger}_{\mathcal{T} T_2}\pi$ & $\{0,n^{\pdagger}_{C_2} \pi\}$ & $\{0,n^{\pdagger}_{\sigma_x}\pi\}$ & $2^5$ \\
\hline
13 & 1 & 0 & 0 & 1 & $n\pi$ & 0 & $n^{\pdagger}_{C_2 T_2}\pi$ & 0 & $n^{\pdagger}_{\sigma_x T_2}\pi$ & $\theta^{\pdagger}_{\mathcal{T} T_2}\pi$ & $\{0,\theta^{\pdagger}_{C_2}\}$ & $\{0,\theta^{\pdagger}_{\sigma_x}\}$ & $2^3 \times 2$ \\
14 & 1 & 0 & 1 & 1 & $n\pi$ & 0 & $n^{\pdagger}_{C_2 T_2}\pi$ & 0 & 0 & $n^{\pdagger}_{\mathcal{T} T_2}\pi$ & $\{0,n^{\pdagger}_{C_2} \pi\}$ & $\{0,n^{\pdagger}_{\sigma_x}\pi\}$ & $2^5 \times 2$ \\
15 & 1 & 1 & 0 & 1 & $n\pi$ & 0 & 0 & 0 & $n^{\pdagger}_{\sigma_x T_2}\pi$ & $n^{\pdagger}_{\mathcal{T} T_2}\pi$ & $\{0,n^{\pdagger}_{C_2}\pi\}$ & $\{0,n^{\pdagger}_{\sigma_x}\pi\}$ & $2^5 \times 2$ \\
16 & 1 & 1 & 1 & 1 & $n\pi$ & 0 & 0 & 0 & $n^{\pdagger}_{\sigma_x T_2}\pi$ & $n^{\pdagger}_{\mathcal{T} T_2}\pi$ & $\{0,n^{\pdagger}_{C_2}\pi\}$ & $\{0,n^{\pdagger}_{\sigma_x}\pi\}$ & $2^5 \times 2$ \\
\end{tabular}
\end{ruledtabular}
\label{tab:U1_PSG_combined}
\end{table*}

\subsection{Projective symmetry groups}
\label{sec:psg_sol}

To implement the projective realization of the symmetry elements $\mathcal{O} \in \{T_1, T_2, \sigma_x, C_2, \mathcal{T}\}$, each symmetry operation must be associated with a corresponding SU(2) gauge transformation. These projective gauge transformations must satisfy the same algebraic relations as the original symmetry group generators. Consequently, the symmetry relations in Eqs.~\eqref{eq:translations}–\eqref{eq:TRO} impose a set of constraints on the operators $W_{\mathcal{O}}$. The allowed PSG solutions are then obtained by solving the resulting algebraic conditions, as outlined in Appendix~\ref{sec:genric_gauge_con}. Different choices of the IGG lead to distinct classes of PSG solutions, which we describe below.

\subsubsection{$\mathrm{U(1)}$ solutions}
\label{sec:u1_sol}

Given that the trellis lattice is nonbipartite, SU(2) \textit{Ans\"atze} with nonvanishing mean-field parameters on all nearest-neighbor bonds are not possible. We therefore begin with the IGG set to U(1), such that the identity element is defined up to a U(1) phase. Realizing such U(1) \textit{Ans\"atze} requires including only hopping terms in the mean-field Hamiltonian introduced above.

The generic form of the PSG elements in this case is given by
\begin{equation}
    W^{\pdagger}_{\mathcal{O}} = \mathcal{F}^{\pdagger}_z(\phi^{\pdagger}_{\mathcal{O}}(x, y, u)) (i \tau^x)^{w^{\pdagger}_{\mathcal{O}}},
\end{equation}
where $\mathcal{F}_z(\xi)$\,$\equiv$\,$\exp({i \xi\, \tau^z})$ and $w_{\mathcal{O}}$\,$\in$\,$\{0, 1\}$. This structure ensures that the \textit{Ans\"atze} are in their canonical form. Importantly, the symmetry relation in Eq.~\eqref{eq: T_2}, where $T_1$ appears only once, immediately excludes the possibility $w_{T_1}$\,$=$\,$ 1$. Therefore, only the two cases $(w_{T_1}, w_{T_2}) = (0, 0)$ and $(0, 1)$ are allowed for arbitrary values of $w_{\sigma_x}$, $w_{C_2}$, and $w_{\mathcal{T}}$.

The algebraic PSG solutions are then given by the following expressions (see Appendix~\ref{app:u1_PSG_derivation} for details):

\noindent
\paragraph{Case I: $w^{\pdagger}_{T_1} = 0$, $w^{\pdagger}_{T_2} = 0$}
\begin{align}
\phi^{\pdagger}_{T_1}(x, y, u) &= y\,\theta, \label{eq:U1_wT1=0,U1_wT2=0} \\
\phi^{\pdagger}_{T_2}(x, y, u) &= 0, \\
\phi^{\pdagger}_{C_2}(x, y, u) &= -(-1)^{w^{\pdagger}_{C_2}}(x\,\theta^{\pdagger}_{C_2 T_1} + y\,\theta^{\pdagger}_{C_2 T_2}) + \rho^{\pdagger}_{C_2,u}, \\
\phi^{\pdagger}_{\sigma_x}(x, y, u) &= -(-1)^{w^{\pdagger}_{\sigma_x}}(x\,\theta^{\pdagger}_{\sigma_x T_1} + y\,\theta^{\pdagger}_{\sigma_x T_2}) \notag \\
&\quad + (-1)^{w^{\pdagger}_{\sigma_x}} \frac{1}{2} y(y - 1) \theta + \rho^{\pdagger}_{\sigma_x,u}, \\
\phi^{\pdagger}_{\mathcal{T}}(x, y, u) &= y\,\theta^{\pdagger}_{\mathcal{T} T_2} + \rho^{\pdagger}_{\mathcal{T},u}. \label{eq:U1_wT1=0,U1_wT2=02}
\end{align}

\noindent
\paragraph{Case II: $w^{\pdagger}_{T_1} = 0$, $w^{\pdagger}_{T_2} = 1$}
\begin{align}
\phi^{\pdagger}_{T_1}(x, y, u) &= (-1)^y \theta, \\
\phi^{\pdagger}_{T_2}(x, y, u) &= 0, \\
\phi^{\pdagger}_{C_2}(x, y, u) &= (-1)^{w^{\pdagger}_{C_2}}(\zeta^{\pdagger}_y\,\theta^{\pdagger}_{C_2 T_2}) + \rho^{\pdagger}_{C_2,u}, \\
\phi^{\pdagger}_{\sigma_x}(x, y, u) &= (-1)^{w^{\pdagger}_{\sigma_x}}(\zeta^{\pdagger}_y\,\theta^{\pdagger}_{\sigma_x T_2} - y\,\theta) + \rho^{\pdagger}_{\sigma_x,u}, \\
\phi^{\pdagger}_{\mathcal{T}}(x, y, u) &= \zeta^{\pdagger}_y\,\theta^{\pdagger}_{\mathcal{T} T_2} + \rho^{\pdagger}_{\mathcal{T},u},
\end{align}
where $\zeta^{\pdagger}_y = \frac{1}{2}(1 + (-1)^y)$.

All allowed U(1) phase parameters $\theta$ and $\rho$ for each choice of $\{w_{T_2}, w_{\sigma_x}, w_{C_2}, w_{\mathcal{T}}\}$ are enumerated in Table~\ref{tab:U1_PSG_combined}.
Thus, any U(1) PSG solution can be uniquely identified by three binary integers $\{w^{\pdagger}_{C_2}, w^{\pdagger}_{\sigma_x}, w^{\pdagger}_{\mathcal{T}}\}$ and nine U(1) phases:
$
\left\{\theta, \theta^{\pdagger}_{C_2 T_1}, \theta^{\pdagger}_{C_2 T_2}, \theta^{\pdagger}_{\sigma_x T_1}, \theta^{\pdagger}_{\sigma_x T_2}, \theta^{\pdagger}_{\mathcal{T} T_2}, \rho^{\pdagger}_{C_2,u}, \rho^{\pdagger}_{\sigma_x,u}, \rho^{\pdagger}_{\mathcal{T},u}\right\},
$
each taking values in $[0, 2\pi)$. However, these parameters are not all independent due to gauge redundancy. The gauge-inequivalent combinations of $\theta$ and $\rho$ for various choices of $\{w^{\pdagger}_{C_2}, w^{\pdagger}_{\sigma_x}, w^{\pdagger}_{\mathcal{T}}\}$ are summarized in Table~\ref{tab:U1_PSG_combined}, enlisting a total of 448 distinct U(1) PSGs. Hence, the full symmetry group of the trellis lattice admits 448 inequivalent U(1) projective symmetry group realizations.

\begin{table}
    \caption{$\mathcal{W}_{\mathcal{O},u}$ matrices, which, together with $\eta_{T_1}$, $\eta_{C_2T_2}$, $\eta_{\sigma_x}$, and $\eta_{\mathcal{T}x}$, yield a total of $256$ gauge-inequivalent PSG solutions.}
	\begin{ruledtabular}
		\begin{tabular}{cccc} PSG no.&$\mathcal{W}_{C_2,u}$&$\mathcal{W}_{\mathcal{T},u}$ & No. of PSGs\\
			\hline
1&$\{\tau^0,\eta^{\pdagger}_{C_2}\tau^0\}$& $\{\tau^0, -\tau^0\}$& $2^4\times 2$\\
2&$\{\tau^0,\eta^{\pdagger}_{C_2}\tau^0\}$& $\{\dot\iota \tau^y,\eta^{\pdagger}_{\mathcal{T}} i \tau^y\}$& $2^4\times 2^2$\\
3&$\{\dot\iota \tau^z,\eta^{\pdagger}_{C_2} \dot\iota \tau^z\}$& $\{\tau^0,-\tau^0\}$& $2^4\times 2$\\
4&$\{\dot\iota \tau^z,\eta^{\pdagger}_{C_2} \dot\iota \tau^z\}$& $\{\dot\iota \tau^z,\eta^{\pdagger}_{\mathcal{T}} \dot\iota \tau^z\}$& $2^4\times 2^2$\\
5&$\{\dot\iota \tau^z,\eta^{\pdagger}_{C_2} \dot\iota \tau^z\}$& $\{\dot\iota \tau^y,\eta^{\pdagger}_{\mathcal{T}} \dot\iota \tau^y\}$& $2^4\times 2^2$
		\end{tabular}
	\end{ruledtabular}
	\label{table:z2_psg1}
\end{table}

\subsubsection{$\mathbb{Z}_2$ solutions}
\label{sec:z2_sol}

Breaking down the IGG from U(1) to $\mathbb{Z}_2$ necessitates the inclusion of both hopping and pairing amplitudes in the link fields, thereby requiring the most general form of mean-field \textit{Ans\"atze}. Within this framework, projective realizations of the symmetry group are defined up to $\mathbb{Z}_2$ gauge transformations, i.e., global signs $\pm 1$. Solving the associated algebraic constraints (see Appendix~\ref{app:z2_PSG_derivation}), we obtain a total of 256 gauge-inequivalent $\mathbb{Z}_2$ PSG solutions. These can be summarized as follows:
\begin{align}
    W^{\pdagger}_{T_1}(x, y, u) &= \eta_{T_1}^{\,y} \tau^0, \\
    W^{\pdagger}_{T_2}(x, y, u) &= \tau^0, \\
    W^{\pdagger}_{C_2}(x, y, u) &= \eta_{C_2 T_2}^{\,y} \eta_{T_1}^{\,x} \mathcal{W}^{\pdagger}_{C_2,u}, \\
    W^{\pdagger}_{\sigma_x}(x, y, u) &= \eta_{T_1}^{\,-y(y-1)/2 + x} \eta^{u}_{\sigma_x} \tau^0, \\
    W^{\pdagger}_{\mathcal{T}}(x, y, u) &= \eta^{\,y}_{\mathcal{T}y}  \mathcal{W}^{\pdagger}_{\mathcal{T},u},
\end{align}
where all the parameters $\eta_{\ldots}$ take values $\pm 1$. The sublattice-dependent PSG elements $ \mathcal{W}_{C_2,u}$ and $ \mathcal{W}_{\mathcal{T},u} \in$ SU(2) are naturally constrained by the gauge-enriched algebraic relations imposed by the symmetry group. All gauge-inequivalent choices for these sublattice-dependent elements are provided in Table~\ref{table:z2_psg1}. From the enumeration therein, we see that when the IGG is set to $\mathbb{Z}_2$, the full symmetry group admits 256 distinct projective realizations.

\begin{figure}[t]
\includegraphics[width=0.9\linewidth]{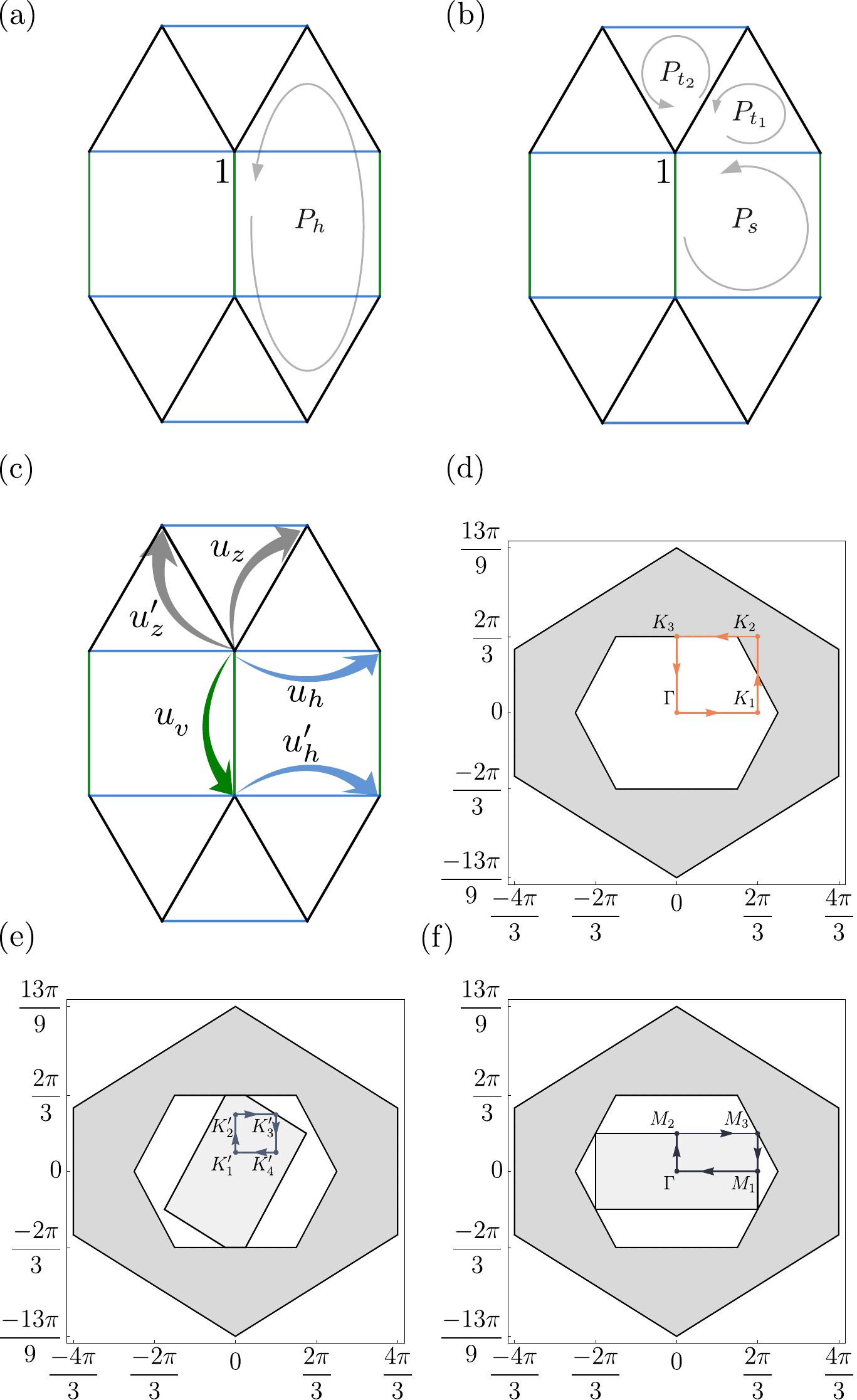}
\caption{(a), (b) Graphical illustration of the SU(2) flux operators on the trellis lattice with the base site labeled ``1.'' Loop operators $P_h$, $P_s$, and $P_t$ correspond to hexagonal, square, and triangular plaquettes, respectively. (c) Definitions of reference bonds within the unit cell at $(x,y)=(0,0)$. (d)--(f) The white (gray) hexagons denote the first (extended) Brillouin zones for lattice parameters $b=1$, $c=1/2$, and $a=3/2$, with the extended zone obtained by a scaling factor of 3. Panels (d)--(f) indicate the paths along which dispersions are plotted for the U(1) \textit{Ans\"atze}: (d) U1 and U2, (e) U3 and U4, where the light gray region shows the reduced Brillouin zone for doubling along $\boldsymbol{T}_1$, and (f) U5 and U6, with the reduced Brillouin zone for doubling along $\boldsymbol{T}_2$. $\boldsymbol{T}_1$ and $\boldsymbol{T}_2$ are defined in Fig.~\ref{fig:lattice}.}
\label{fig:flux}
\end{figure}

\section{Short-ranged mean-field {Ans\"atze}}
\label{sec:ansatze}

Given the PSG solutions obtained in Secs.~\ref{sec:u1_sol} and~\ref{sec:z2_sol}, we now turn to the construction of mean-field \textit{Ans\"atze} for QSL states by imposing symmetry constraints on the link fields. The detailed derivation of symmetry-allowed forms for the link variables is presented in Appendix~\ref{app:u1_ansatz}. While our algebraic analysis yielded 448 U(1) and 256 $\mathbb{Z}_2$ PSG classes, the number of distinct realizable short-ranged \textit{Ans\"atze} on the trellis lattice is significantly fewer. This reduction arises from restricting the link fields to nonzero values on only the three symmetry-inequivalent nearest-neighbor bonds---$J_h$, $J_v$, and $J_z$.

Within this short-range framework, we find a total of $6+1$ U(1) \textit{Ans\"atze}, where ``$+1$'' corresponds to an infinite class of gauge-inequivalent states stemming from a continuous U(1) degree of freedom associated with the PSG class given in the second row of Table~\ref{tab:U1_PSG_combined}. Additionally, we identify 25 distinct $\mathbb{Z}_2$ \textit{Ans\"atze}. In this section, we present and discuss the structure and classification of these short-ranged mean-field states.

Before proceeding further, it is useful to introduce an SU(2) flux operator, which provides additional gauge-invariant characterization of the \textit{Ans\"atze}, beyond the PSG labels. Defined for a closed loop $\mathcal{C}_i$ based at site $i$, the SU(2) flux or loop operator is given by
\begin{equation}
    \label{eq:loop_operator}
    \mathcal{P}^{\pdagger}_{\mathcal{C}_i} = u^{\pdagger}_{ij} u^{\pdagger}_{jl} \ldots u^{\pdagger}_{ki}.
\end{equation}
Under a local SU(2) gauge transformation $W_i$, this operator transforms as
\begin{equation}
    \mathcal{P}^{\pdagger}_{\mathcal{C}_i} \rightarrow W^\dagger_i \mathcal{P}^{\pdagger}_{\mathcal{C}_i} W^{\pdagger}_i,
\end{equation}
implying that the commutation or anticommutation relations between loop operators based at the same site remain invariant under such a rotation. Hence, the flux structure around various plaquettes becomes a powerful tool to distinguish between different \textit{Ans\"atze}.

In general, a loop operator around a $q$-sided plaquette takes the form~\cite{Bieri2016}
\begin{equation}
    P^{\pdagger}_{\mathcal{C}_{i}}(\varphi^{\pdagger}_{\mathcal{C}_{i}}) \propto g^{\pdagger}_i \mathcal{F}^{\pdagger}_z(i \varphi^{\pdagger}_{\mathcal{C}_{i}}) (\tau^z)^q g_i^\dagger, \quad g^{\pdagger}_i \in \mathrm{SU}(2),
\end{equation}
where $\varphi_{\mathcal{C}_i}$ represents the effective flux piercing the loop $\mathcal{C}_i$.

\begin{figure}[t]
\includegraphics[width=1.0\linewidth]{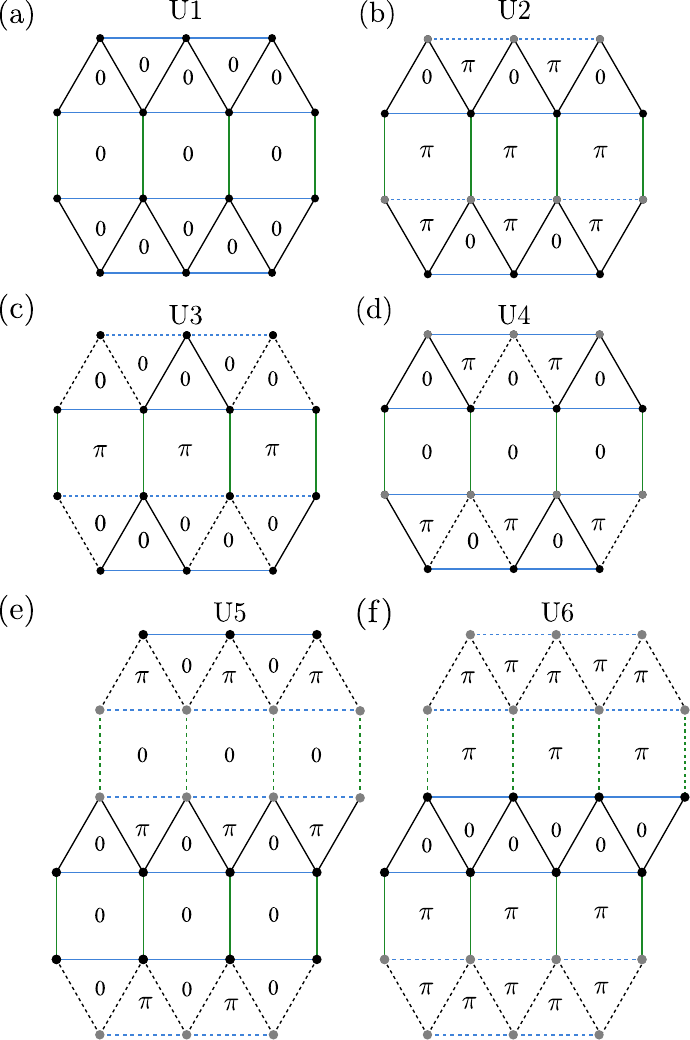}
\caption{Schematic representation of the class A, B, and C \textit{Ans\"atze}, described in Secs.~\ref{sec:u1_class_a}, \ref{sec:u1_class_b}, and \ref{sec:u1_class_c}, respectively. Solid (dashed) lines indicate hoppings with positive (negative) signs. The green, blue, and black lines correspond to $\chi_v\tau^z$, $\chi_h\tau^z$, and $\chi_z\tau^z$, respectively. Black (gray) points represent positive (negative) onsite hoppings. The associated PSGs of these \textit{Ans\"atze} are listed in Table ~\ref{tab:theta_rho_PSG}.}
\label{fig:u1_ansatz}
\end{figure}

On the trellis lattice, we consider a representative base site, labeled as ``1'' without loss of generality, as shown in Figs.~\ref{fig:flux}(a) and 2(b). Starting from this site, we define one hexagonal loop $P_h$, one square loop $P_s$, and two triangular loops $P_{t_1}$ and $P_{t_2}$, each associated with fluxes $\varphi_h$, $\varphi_s$, $\varphi_{t_1}$, and $\varphi_{t_2}$, respectively. In the case of a U(1) IGG, these fluxes commute, and their additive structure permits the identification of the total flux threading a hexagonal plaquette as the sum of the fluxes through its constituent triangular and square subloops. The corresponding flux configurations for each U(1) \textit{Ansatz} are illustrated in Fig.~\ref{fig:u1_ansatz}.

\begin{table*}
\caption{Symmetry-allowed mean-field parameters for different $\mathbb{Z}_2$ \textit{Ans\"atze}. The reference bonds are illustrated in Fig.~\ref{fig:flux}(c). The parent U(1) states corresponding to each $\mathbb{Z}_2$ \textit{Ansatz} are also indicated. $\mathbb{Z}_2$ \textit{Ans\"atze} that directly descend from a parent SU(2) state are labeled as class A, with an SU(2) flux of $\pi$ threading the hexagonal loop. The associated PSGs of these \textit{Ans\"atze} are listed in Table ~\ref{tab:z2_theta}.}
    \begin{ruledtabular}
    \begin{tabular}{cccccc}
     
   Label& $u^{\pdagger}_{v}$& $u^{\pdagger}_{h}$,$u_{h}^{'}$& $u^{\pdagger}_{z}$,$u_{z}^{'}$  &$a_{\mu,u}$& Parent U(1)\\
			\hline
Z1 & \{$\tau^z$\}  &\{$\tau^x,\tau^z$\},\{$\tau^x,\tau^z$\} & \{$\tau^x,\tau^z$\},\{$\tau^x,\tau^z$\}& \{$\tau^x,\tau^z$\},\{$\tau^x,\tau^z$\}&   U1,U5  \\
Z2  &\{$\tau^z$\}  &\{$\tau^x,\tau^z$\},\{$\tau^x,\tau^z$\} & \{$\tau^y$\},\{$\tau^y$\}& \{$\tau^x,\tau^z$\},\{$\tau^x,\tau^z$\}&U5   \\
Z3 &  \{$\tau^z$\}   &\{$\tau^x,\tau^z$\},\{$-\tau^x,-\tau^z$\} & \{$\tau^x,\tau^z$\},\{$-\tau^x,-\tau^z$\} & \{$\tau^x,\tau^z$\},\{$\tau^x,\tau^z$\}&U3 \\
Z4 &  \{$\tau^z$\}   &\{$\tau^x,\tau^z$\},\{$-\tau^x,-\tau^z$\} & \{$\tau^y$\},\{$-\tau^y$\} & \{$\tau^x,\tau^z$\},\{$\tau^x,\tau^z$\} & A\\
Z5 & \{$\tau^z$\}  &\{$\tau^x,\tau^z$\},\{$\tau^x,\tau^z$\} & $\{\dot \iota \tau^0\},\{-\dot \iota \tau^0\}$& \{$\tau^x,\tau^z$\},\{$\tau^x,\tau^z$\}   & U1 \\
Z6 &  \{$\tau^z$\}  &\{$\tau^x,\tau^z$\},\{$-\tau^x,-\tau^z$\} & \{$\dot \iota \tau^0$\},\{$\dot \iota \tau^0$\}  & \{$\tau^x,\tau^z$\},\{$\tau^x,\tau^z$\} &U3   \\
Z7 & \{$\tau^z$\}  &\{$\tau^x,\tau^z$\},\{$\tau^x,-\tau^z$\} & \{$\tau^z$\},\{$\tau^z$\} & \{$\tau^x,\tau^z$\},\{$\tau^x,-\tau^z$\}&  U2  \\
Z8&  \{$\tau^z$\}  &\{$\tau^x,\tau^z$\},\{$-\tau^x,\tau^z$\} & \{$\tau^z$\},\{$-\tau^z$\}& \{$\tau^x,\tau^z$\},\{$\tau^x,-\tau^z$\}&  U4\\
Z9 &\{$ \tau^z$\}  &\{$\tau^x,\tau^z$\},\{$-\tau^x,\tau^z$\} & \{$\tau^z$\},\{$\tau^z$\} & \{$\tau^x,\tau^z$\},\{$-\tau^x,\tau^z$\}  & U1\\
Z10 &\{$\tau^z$\}   &\{$\tau^x,\tau^z$\},\{$\tau^x,-\tau^z$\} & \{$ \tau^z$\},\{$-\tau^z$\} & \{$\tau^x,\tau^z$\},\{$-\tau^x,\tau^z$\}& U3\\
Z11& \{$\tau^z$\}  &\{$\tau^x,\tau^z$\},\{$\tau^x,-\tau^z$\} &\{$-\tau^x$\},\{$\tau^x$\}& \{$\tau^x,\tau^z$\},\{$\tau^x,-\tau^z$\}& U6 \\
Z12&\{$\tau^z$\}  &\{$\tau^x,\tau^z$\},\{$-\tau^x,\tau^z$\} & \{$\tau^x$\},\{$-\tau^x$\} & \{$\tau^x,\tau^z$\},\{$-\tau^x,\tau^z$\}&U5\\
Z13 &$\{ \tau^z$\}  &\{$\tau^x,\tau^z$\},\{$\tau^x,-\tau^z$\} & \{$\tau^x$\},\{$\tau^x$\} & \{$\tau^x,\tau^z$\},\{$-\tau^x,\tau^z$\}&A\\
Z14 &\{$\tau^z$\}  &\{$\tau^x,\tau^z$\},\{$-\tau^x,\tau^z$\} & \{$-\dot \iota \tau^0,-\tau^y$\},\{$\dot \iota \tau^0,\tau^y$\} & \{$\tau^x,\tau^z$\},\{$-\tau^x,\tau^z$\} & U1,U5\\
Z15 & \{$\tau^z$\}  &\{$\tau^x,\tau^z$\},\{$-\tau^x,\tau^z$\} & \{$\dot \iota \tau^0,-\tau^y$\},\{$-\dot \iota \tau^0,\tau^y$\}& \{$\tau^x,\tau^z$\},\{$\tau^x,-\tau^z$\}&   U4\\
Z16 & \{$\tau^z$\}  &\{$\tau^x,\tau^z$\},\{$\tau^x,-\tau^z$\} & \{$\dot \iota \tau^0,-\tau^y$\},\{$\dot \iota \tau^0,-\tau^y$\} & \{$\tau^x,\tau^z$\},\{$\tau^x,-\tau^z$\} &U2, U6\\
Z17 & \{$\tau^z$\}  &\{$\tau^x,\tau^z$\},\{$\tau^x,-\tau^z$\} & \{$\dot \iota \tau^0,\tau^y$\},\{$\dot \iota \tau^0,\tau^y$\}& \{$\tau^x,\tau^z$\},\{$-\tau^x,\tau^z$\}&U3\\
\hline
\hline
Z18 &\{$\tau^z$\}  &\{$\tau^z$\},\{$\tau^z$\} &\{$\dot \iota \tau^0,\tau^y$\} ,\{$-\dot \iota \tau^0,\tau^y$\}   & \{$\tau^z$\},\{$\tau^z$\} & U1\\
Z19 & \{$\tau^z$\} &\{$\tau^z$\},\{$-\tau^z$\} &\{$\dot \iota \tau^0,\tau^y$\} ,\{$\dot \iota \tau^0,-\tau^y$\}  & \{$\tau^z$\},\{$\tau^z$\}& U3,U5\\

Z20 & \{$\tau^z$\}&\{$\tau^z$\},\{$-\tau^z$\} &\{$\dot \iota \tau^0,\tau^y$\} ,\{$\dot \iota \tau^0,-\tau^y$\} & \{$\tau^z$\},\{$-\tau^z$\}&U2, U6\\
Z21 & \{$\tau^z$\}&\{$\tau^z$\},\{$\tau^z$\} &\{$-\dot \iota \tau^0,\tau^y$\} ,\{$\dot \iota \tau^0,\tau^y$\}  & \{$\tau^z$\},\{$-\tau^z$\} &  U4 \\
Z22 & \{$\tau^z$\}&\{$\tau^z$\},\{$-\tau^z$\} &\{$-\tau^x,\tau^z$\} ,\{$\tau^x,\tau^z$\}  & \{$\tau^z$\},\{$-\tau^z$\}&  U2, U6\\

Z23 &\{$\tau^z$\}&\{$\tau^z$\},\{$\tau^z$\} &\{$\tau^x,\tau^z$\} ,\{$\tau^x,-\tau^z$\}   & \{$\tau^z$\},\{$-\tau^z$\}&  U4\\
Z24 &\{$\tau^z$\}&\{$\tau^z$\},\{$\tau^z$\} &$\{\tau^x,\tau^z$\} ,\{$-\tau^x,\tau^z$\}   & \{$\tau^z$\},\{$\tau^z$\} & U1,U5\\
Z25 &\{$ \tau^z$\}&\{$\tau^z$\},\{$-\tau^z$\} &\{$\tau^x,\tau^z$\} ,\{$\tau^x,-\tau^z$\}   & \{$\tau^z$\},\{$\tau^z$\}&U3\\
    \end{tabular}
    \end{ruledtabular}
    \label{table:z2_modified}
\end{table*}

\subsection{U(1) \textit{Ans\"atze} }
\label{sec:u1_ansatze}

In this section, we present the different symmetric U(1) mean-field \textit{Ans\"atze} that are realizable under the constraint of having nonzero mean-field amplitudes only on the three types of nearest-neighbor bonds.

We classify the \textit{Ans\"atze} into four distinct classes---A, B, C, and D---based on the flux $\phi^{\pdagger}_{h}$ threading the hexagonal plaquette ($\theta = 0$ or $\pi$), and the spatial modulation of the mean-field amplitudes, which determines whether the unit cell must be enlarged. All the U(1) Ans\"atze are graphically illustrated in Fig.~\ref{fig:u1_ansatz} and the corresponding PSGs are enlisted in appendix (Table ~\ref{tab:theta_rho_PSG}).

\subsubsection{Class A}
\label{sec:u1_class_a}

Class A \textit{Ans\"atze} are characterized by a vanishing flux through the hexagonal plaquettes, i.e., $\phi_{h} = 0$. There are two U(1) \textit{Ans\"atze} in this class. The labeling of the bonds is defined in Fig.~\ref{fig:flux}(c), and the graphical representations of these \textit{Ans\"atze} are provided in Figs.~\ref{fig:u1_ansatz}(a) and \ref{fig:u1_ansatz}(b).

These \textit{Ans\"atze} can be further distinguished by the fluxes threading the square and triangular loops. In the U1 family, all loop fluxes vanish, i.e., $\phi_s = 0$ and $(\phi_{t_1}, \phi_{t_2}) = (0, 0)$. In contrast, the U2 states are characterized by $\phi_s = \pi$ and $(\phi_{t_1}, \phi_{t_2}) = (0, \pi)$.

\subsubsection{Class B}
\label{sec:u1_class_b}

Class B \textit{Ans\"atze} feature a $\pi$-flux through the hexagonal plaquettes, i.e., $\phi_{h} = \pi$. Their realization necessitates a doubling of the unit cell along the $T_1$ direction, due to sign-alternating mean-field amplitudes on the zigzag bonds. Explicitly, denoting the zigzag bonds within the $(x,y)$ unit cell as $u^{\pdagger}_z(x,y)$ and $u'_z(x,y)$, their modulation follows:
\begin{align}
    u^{\pdagger}_z(x,y) &= (-1)^x u^{\pdagger}_z, \label{eq:class_b_doubling} \\
    u'_z(x,y) &= (-1)^x u'_z.
\end{align}

This class contains two \textit{Ans\"atze}, labeled U3 and U4, illustrated in Fig.~\ref{fig:u1_ansatz}(c) and \ref{fig:u1_ansatz}(d), respectively. For U3, $\phi^{\pdagger}_s = \pi$ and $(\phi^{\pdagger}_{t_1}, \phi^{\pdagger}_{t_2}) = (0, 0)$, whereas for U4, $\phi^{\pdagger}_s = 0$ and $(\phi^{\pdagger}_{t_1}, \phi^{\pdagger}_{t_2}) = (0, \pi)$.

\subsubsection{Class C}
\label{sec:u1_class_c}

Class C \textit{Ans\"atze} also exhibit zero flux through the hexagonal plaquettes, but differ from class A in the spatial modulation of the amplitudes. In this case, the unit cell is doubled along the $T_2$ direction, with the mean-field amplitudes modulated as
\begin{alignat}{1}
    u^{\pdagger}_{v,h,z}(x,y) &= (-1)^y u^{\pdagger}_{v,h,z}, \label{eq:class_c_doubling} \\
    u'_{h,z}(x,y) &= (-1)^y u'_{h,z}, \\
    a'_{\mu,u} &= (-1)^y a_{\mu,u}.
\end{alignat}

This class includes two \textit{Ans\"atze}, labeled U5 and U6, depicted in Fig.~\ref{fig:u1_ansatz}(e) and \ref{fig:u1_ansatz}(f). For U5, the square flux vanishes and the triangular fluxes alternate with the unit cell's position along $T_2$: $(\phi^{\pdagger}_{t_1}, \phi^{\pdagger}_{t_2}) = (0, \pi)$ or $(\pi, 0)$. For U6, $\phi^{\pdagger}_s = \pi$, and $(\phi^{\pdagger}_{t_1}, \phi^{\pdagger}_{t_2}) = (0, 0)$ or $(\pi, \pi)$ depending on the $T_2$ position.

\subsubsection{Class D}

In Class D, the size of the unit cell depends on the value of the phase $\theta \notin \{0, \pi\}$. While the amplitudes on the $J_v$ and $J_h$ bonds remain translationally invariant, the zigzag bond amplitudes vary spatially as
\begin{align}
    u^{\pdagger}_z(x,y) &= \mathcal{F}^{\pdagger}_z(-x\theta) u^{\pdagger}_z, \label{eq:class_c_u1} \\
    u'_z(x,y) &= \mathcal{F}^{\pdagger}_z(-x\theta) u'_z,
\end{align}
where, as before, $\mathcal{F}_z(\xi) = \exp({i \xi \tau^z})$. This modulation is periodic only when $\theta = p\pi/q$ for integers $p$ and $q$. The corresponding \textit{Ansatz} then requires an enlarged unit cell of size $q$ along $T_1$. For example, $\theta = \pi/3$ implies a tripled unit cell, while $\theta = \pi/4$ requires quadrupling.

The associated flux through the hexagonal loop is $\phi_{h} = p\pi/q$, which is one realization of the U1$^p_q$ PSGs, leading to an infinite number of U1$^p_q$ {\it Ans\"atze}~\cite{Wen2002}. As the horizontal bond amplitudes vanish in these \textit{Ans\"atze}, no square or triangular fluxes exist. Although the resulting fluxes differ from $0$ or $\pi$ --- the usual values for time-reversal–invariant hoppings with real amplitudes--- these fractional fluxes, in appropriate gauges, alternate in sign between up and down spins, thereby preserving time-reversal symmetry~\cite{Bieri2016}.  It is worth noting that a chiral spin liquid is realized when nontrivial (not $0$ or $\pi$) fluxes thread odd-sided loops. 

\subsection{$\mathbb{Z}_2$ \textit{Ans\"atze} }
\label{sec:z2_ansatze}

All gauge-inequivalent \textit{Ans\"atze} with $\mathrm{IGG} \simeq \mathbb{Z}_2$ are enumerated in Table~\ref{table:z2_modified}, yielding a total of 25 distinct $\mathbb{Z}_2$ mean-field \textit{Ans\"atze}.

The emergence of $\mathbb{Z}_2$ \textit{Ans\"atze}, as descendants from different parent states, can occur through two mechanisms: (1) the first route involves adding symmetry-allowed pairing terms to an existing U(1) \textit{Ansatz}, thereby modifying the IGG from U(1)  to $\mathbb{Z}_2$;
(2) alternatively, $\mathbb{Z}_2$ \textit{Ans\"atze} can be obtained via directly breaking the symmetry down from an SU(2) IGG to $\mathbb{Z}_2$. The associated PSGs are tabulated in Table ~\ref{tab:z2_theta}. In this case, two distinct SU(2) parent states are possible, characterized by $0$- and $\pi$-flux threading the hexagonal plaquettes. These states feature vanishing mean-field amplitudes on the horizontal bonds and are continuously connected to the $0$- and $\pi$-flux SU(2) spin liquids previously studied on the honeycomb lattice~\cite{Vishwanath2012,Ying2011}.

Out of all the possible PSG solutions, No. 17 and No. 8 \textit{Ans\"atze} arise from the PSG classes corresponding to the second and fifth rows of Table~\ref{table:z2_psg1}, respectively. The remaining PSG classes do not permit a full reduction of the SU(2) IGG down to $\mathbb{Z}_2$, and thus do not yield viable $\mathbb{Z}_2$ spin liquids within our short-range framework. The Z1 state in Table~\ref{table:z2_modified} corresponds to the nonprojective (linear) representation of the space group with $s$-wave hoppings and pairings, and is a fully gapped $\mathbb{Z}_{2}$ QSL.

The last column of Table~\ref{table:z2_modified} documents the U(1) or SU(2) parent state from which each $\mathbb{Z}_2$ \textit{Ansatz} descends. Entries labeled with “A” refer to the SU(2) $\pi$-flux state, which are obtained via mechanism (2).

\begin{figure*}
\includegraphics[width=1.0\textwidth]{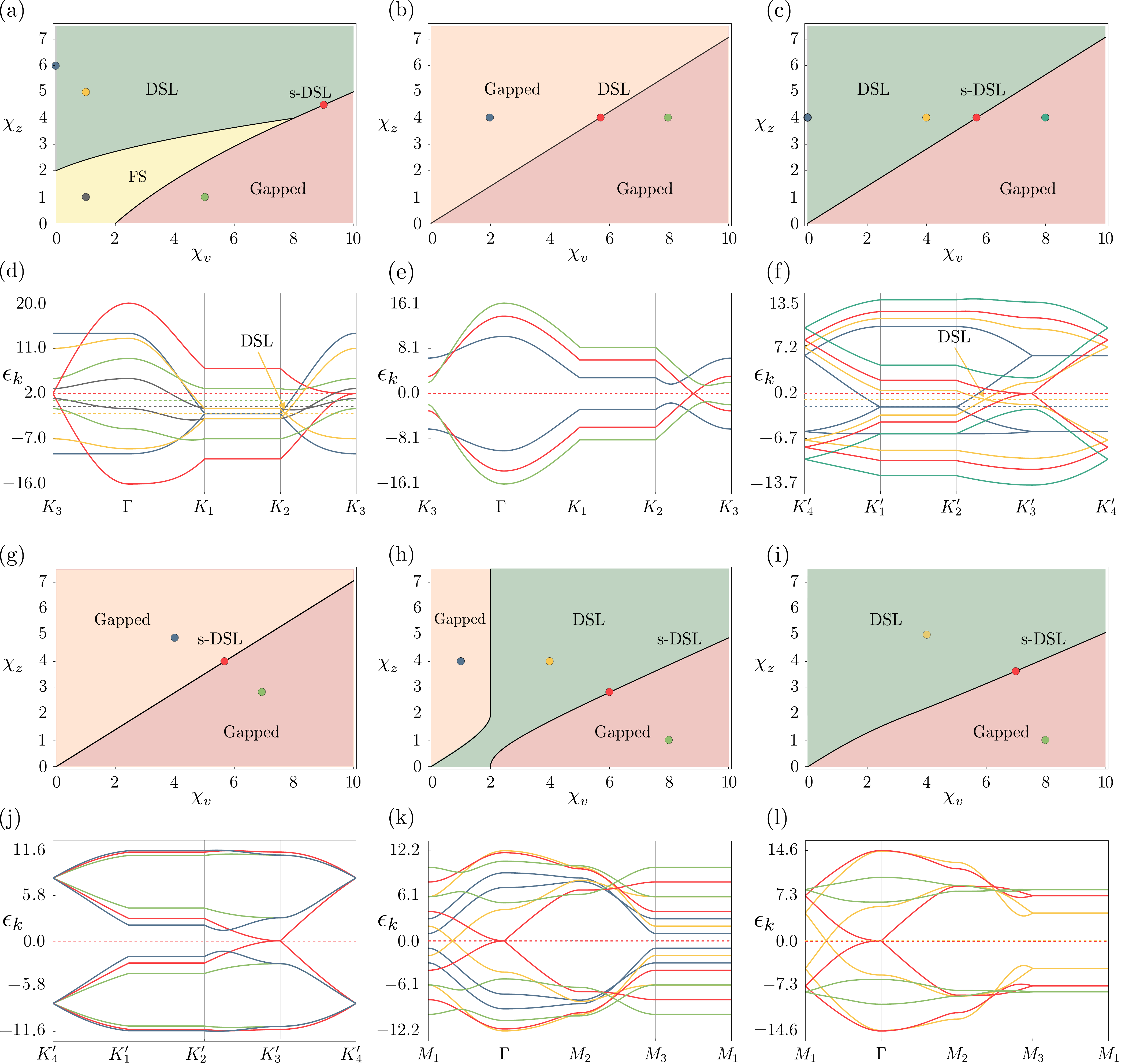}
\caption{Phase diagrams of the six U(1) \textit{Ans\"atze} 
(a) U1, (b) U2, (c) U3, (g) U4, (h) U5, and (i) U6 in the $(\chi_v, \chi_z)$ parameter space, with $\chi_h$\,$=$\,$1$ fixed. 
The corresponding band structures for representative points, indicated by colored circles, are shown in panels (d)–(f) and (j)–(l), respectively. 
(d), (e) The energy dispersions are plotted along a high-symmetry path $K_3 \rightarrow \Gamma \rightarrow K_1 \rightarrow K_2 \rightarrow K_3$ in the Brillouin zone, where $K_3 = (0,2\pi/3)$, $\Gamma = (0, 0)$, $K_1 = (2\pi/3, 0)$, and $K_2 = (2\pi/3, 2\pi/3)$. 
(f), (j) The band dispersions are plotted along a path that connects the momenta $K_4^\prime=(\pi/3, \pi/6)$, $K_1^\prime=(0, \pi/6)$, $K_2^\prime=(0,\pi/2)$, $ K_3^\prime =( \pi/3, \pi/2)$, and back to $K_4^\prime$. 
(k), (l) The spectra are displayed along the path $M_1 \rightarrow \Gamma \rightarrow M_2 \rightarrow M_3 \rightarrow M_1$, with $M_1 = (2\pi/3, 0)$, $\Gamma = (0, 0)$, $M_2 = (0, \pi/3)$, and $M_3 = (2\pi/3,\pi/3)$.
In all cases, dashed lines indicate the Fermi energy.
}
\label{fig:dispersion_all}
\end{figure*} 

\section{Quantum spin liquid phases}
\label{sec:dispersion}

\noindent
It is important to note that the PSG classification and the subsequent analysis of mean-field band structures address a question that is logically distinct from energetic stabilization in a specific microscopic Hamiltonian. The PSG framework enumerates all symmetry-allowed quantum spin liquid {\it Ans\"atze} on the trellis lattice, independent of their variational energies, and thereby identifies the full landscape of possible fractionalized phases. Within this space, the semi-Dirac spin liquid emerges as a distinct symmetry-allowed state characterized by a linear-quadratic spinon dispersion at high-symmetry points in the Brillouin zone. Whether such a state is realized as the ground state of a particular spin Hamiltonian is a separate, model-dependent question that requires energetic optimization or unbiased numerical methods. In the present work, we find that the nearest-neighbor Heisenberg model on the trellis lattice does not energetically stabilize the semi-Dirac spin liquid, which therefore remains a phase realized at the level of PSG classification and mean-field parameter space rather than in the optimized phase diagram of the physical spin model. This distinction is generic to PSG-based studies and does not diminish the conceptual significance of the semi-Dirac spin liquid as a symmetry-allowed quantum spin liquid with characteristic universal properties. This separation between PSG classification and energetic selection is standard in parton-based approaches~\cite{Wen2002,lu2011,Hermele-2008}.

The properties of the \textit{Ans\"atze} identified above depend on the choice of mean-field parameters, which are, in turn, determined by minimizing the energy with respect to a specific Hamiltonian. Consequently, the physical characteristics of each \textit{Ansatz} are inherently model dependent. In this section, however, instead of initially focusing on such microscopic details, we choose to first analyze the \textit{general} features of the six U(1) \textit{Ans\"atze}. Thereafter, we relate these findings to the physics of the nearest-neighbor Heisenberg model with $J_h-J_v-J_z$ couplings via a fully self-consistent calculation.

\subsection{Mean-field phase diagrams}

The unit cell of the trellis lattice contains three inequivalent bonds, each characterized by a hopping amplitude: $\chi_h$, $\chi_v$, and $\chi_z$. By independently varying these parameters, we uncover a rich variety of phases, including both gapped and gapless states exhibiting Fermi surfaces (FSs), nodal lines, semi-Dirac and Dirac points. We now present a detailed discussion of these phases for all six U(1) \textit{Ans\"atze} as well as their associated band structures. For clarity, we set $\chi_h = 1$ throughout the analysis.

\begin{figure*}[t]
    \centering
    \includegraphics[width=1\textwidth]{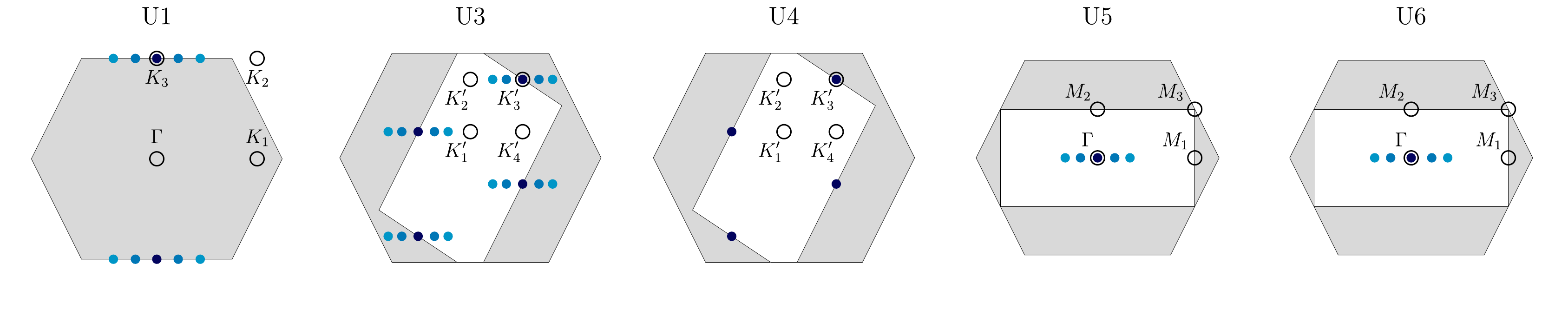}
    \caption{Evolution of the Dirac points as the system approaches the s-DSL phase from the DSL phase for the \textit{Ans\"atze} U1, U3, U4, U5, and U6 (see Fig.~\ref{fig:u1_ansatz}). The color intensity increases with progression toward the s-DSL, where two Dirac points merge to form the s-DSL. Each pair of dots with identical color corresponds to a particular value of the hopping parameter close to the s-DSL transition (see Fig.~\ref{fig:dispersion_all}). The U4 s-DSL state is separated from the others by two intermediate gapped phases. Hollow circles mark the momentum-space path along which the dispersions are displayed in Fig.~\ref{fig:dispersion_all}.}
    \label{fig:s-DSL_mechanism}
\end{figure*}

Among the six U(1) \textit{Ans\"atze}, the U1 family exhibits the most intricate phase diagram, as shown in Fig.~\ref{fig:dispersion_all}(a). Specifically, we observe extended windows of Dirac spin liquid (DSL), FS, and gapped phases, delineated in the parameter space defined by $\chi_z$ and $\chi_v$. Each phase is distinguished by its band structure and excitation spectrum near the Fermi energy, illustrated in Fig.~\ref{fig:dispersion_all}(d) for representative parameter values marked by colored dots in the phase diagram.
Here, the phase boundary between the DSL and FS phases is specified by 
$\chi^{}_v = -2\chi^{}_z + \chi_z^2,$ while the boundary between the FS and gapped phases is given by $\chi^{}_v = 2 + \chi^{}_z + {\chi_z^2}/{8}.$ 

Additional noteworthy features appear at the DSL's boundaries. For instance, the blue circle at $\chi_v = 0$ marks a nodal-line spin liquid phase. Moreover, along a segment of the line $\chi_v$\,$=$\,$2\chi_z$, we find a \emph{semi-Dirac} spin liquid (labeled s-DSL)~\cite{zhong2017}.
In this semi-Dirac phase, the bands touch at the point $K_3$\,$=$\,$(0,2\pi/3)$, at which the dispersion is linear along one momentum direction and quadratic along the orthogonal one\footnote{Interestingly, similar band structures also appear in other contexts~\cite{herrera2023tunable} including in studies of Kitaev spin liquids~\cite{PhysRevLett.133.146603}, and topological materials~\cite{Kruthoff-2017,wieder2018wallpaper,rudenko2024anisotropic,Isobe-2016}.}. Consequently, quasiparticles are effectively massive in one direction, while exhibiting massless Dirac-like behavior in the other. Notably, the semi-Dirac point acts as a precursor to the Dirac points that characterize the DSL phase (as discussed in Sec.~\ref{sec:s-DSL} below; see also Fig.~\ref{fig:s-DSL_mechanism}). Here, we note that this U1 \textit{Ansatz} includes the semi-Dirac dispersion observed in the tight-binding honeycomb model of Refs.~\cite{Bena-2011,Elsayed-2025a}. This connection is established by including further-neighbor amplitudes on the trellis lattice \textendash namely, the diagonal links within the square plaquettes and the longest vertical link within the hexagonal plaquette \textendash in addition to those shown in Fig.~\ref{fig:u1_ansatz}(a). The s-DSL nodes emerge when $\chi_v=\chi_z=\chi$, $\chi_h=0$, and the hopping amplitude on these further-neighbor links is set to $\chi/3$. The DSL, FS, s-DSL, and gapped phases all meet at a multicritical point $(\chi_v, \chi_z) = (8, 4)$.

Proceeding along the same lines, the phase diagram for the U2 \textit{Ansatz} is shown in Fig.~\ref{fig:dispersion_all}(b). It features two gapped phases separated by a gapless Dirac spin liquid phase, which lies along the line $\chi_v$\,$=$\,$\sqrt{2} \chi_z$. Representative band structures at selected points of this phase diagram are displayed in Fig.~\ref{fig:dispersion_all}(e). In this case, the Dirac point lies along the momentum path between the $K_2=(2\pi/3,2\pi/3)$ and $K_3$ points in reciprocal space.

Next, Fig.~\ref{fig:dispersion_all}(c) showcases the phase diagram for the U3 \textit{Ansatz}. At $\chi_v = 0$, this exhibits a nodal-line phase, with the nodal line itself stretching between $K_1' = (0,\pi/6)$ and $K_2'=(0, \pi/2)$. Upon increasing $\chi_v$, a Dirac spin liquid appears but with a tilted cone located along the path between $K_2'$ and $K_3'=(\pi/3, \pi/2)$, ~\cite{katayama2006,Amo2019}. As $\chi_v$ is increased even further, an s-DSL phase emerges that subsequently transitions into a gapped phase. The semi-Dirac spin liquid located along the line $\chi_v = \sqrt{2} \chi_z$ thus separates the DSL and gapped phases. It is characterized by a linear-quadratic dispersion centered at $K_3'$, and the tilted Dirac cone of the DSL phase in fact originates from the semi-Dirac point (see Sec.~\ref{sec:s-DSL} below). The corresponding band structures are all plotted in Fig.~\ref{fig:dispersion_all}(f).

Similarly, the U4 \textit{Ansatz}'s phase diagram is shown in Fig.~\ref{fig:dispersion_all}(g). It comprises two distinct gapped phases separated by a semi-Dirac spin liquid phase (black line) occurring, as before, along $\chi_v = \sqrt{2} \chi_z$. The associated band structures are presented in Fig.~\ref{fig:dispersion_all}(j). Unlike for the U1 \textit{Ansatz}, the semi-Dirac point here does not split into a pair of Dirac points and the spectrum remains gapped on either side of the s-DSL; however, such a splitting could be induced by incorporating further-neighbor hopping terms.

Continuing with our sequence of U(1) QSLs, Fig.~\ref{fig:dispersion_all}(h) presents the phase diagram for the U5 \textit{Ansatz}, wherein a DSL region is enclosed between two gapped phases. The DSL phase is bounded by the curves
\begin{alignat}{2}
\chi^{}_v &= \frac{\chi^{}_z}{2} \sqrt{8 - \chi_z^2} \quad&&(\text{for } \chi^{}_z < 2), \nonumber\\
\chi^{}_v &= 2,  \quad&&(\text{for } \chi^{}_z \geq 2), \nonumber\\
\chi^{}_v &= 2\sqrt{1 + \chi_z^2} .
\end{alignat}
At the boundary between the DSL and the gapped phase below it, we once again find semi-Dirac dispersion. The related band structures are illustrated in Fig.~\ref{fig:dispersion_all}(k).

Finally, the phase diagram for the U6 \textit{Ansatz} is shown in Fig.~\ref{fig:dispersion_all}(i). The topology of the phase diagram closely resembles that for the U3 case, and an s-DSL phase separates the DSL and gapped phases. The phase boundaries for the semi-Dirac region are defined by
\begin{alignat}{2}
\chi^{}_v &= \frac{\chi^{}_z}{2} \sqrt{8 + \chi_z^2} \quad&&(\text{for } \chi^{}_z < 2),\\
\chi^{}_z &= \frac{1}{2} \sqrt{4 + \chi_v^2} \quad&&(\text{for } \chi^{}_z \geq 2).
\end{alignat}
The associated spinon dispersions for this \textit{Ansatz} are sketched in Fig.~\ref{fig:dispersion_all}(l).

The s-DSL boundaries for all the  \textit{Ans\"atze} are tabulated in Table~\ref{tab:sdsl_boundaries}.

\begin{table}[h!]
    \centering
    \begin{ruledtabular}
    \begin{tabular}{cc}
       \textit{Ansatz} & s-DSL boundaries \\
        \hline
       U1 & $\chi_v = 2\chi_z$   \text{for } $\chi_z > 4$ \\
       U3 & $\chi_v = \sqrt{2} \chi_z$ \\
       U4 & $\chi_v = \sqrt{2} \chi_z$ \\
       U5 & $\chi_v = 2\sqrt{1 + \chi_z^2}$ \\
       U6 & $\begin{cases}
                 \chi_v = \dfrac{\chi_z}{2} \sqrt{8 + \chi_z^2} & \text{for }\chi_z < 2 \\[1pt]
                 \chi_z = \dfrac{1}{2} \sqrt{4 + \chi_v^2} &  \text{for } \chi_z \geq 2
              \end{cases}$ \\
    \end{tabular}
    \end{ruledtabular}
    \caption{The equations for the phase boundaries realizing s-DSL for each of the U(1) \textit{Ans\"atze}.}
    \label{tab:sdsl_boundaries}
\end{table}

While our analysis of the semi-Dirac \textit{Ansätze} has focused on a U(1) gauge structure, one can also realize $\mathbb{Z}_{2}$ semi-Dirac \textit{Ansätze} since their origin is rooted in the lattice (point group) symmetry.

\begin{table*}[t]
    \centering
    \begin{ruledtabular}
    \begin{tabular}{cccc}
    \textit{Ansatz} (sign structure) & No. of s-DSL points / BZ & Positions within EBZ & $\boldsymbol{q}$-vectors connecting s-DSL points \\
    \hline
     U1 (Fig.~\ref{fig:u1_ansatz}(a)) & 1 & $(0,\pm\frac{2\pi}{3}),(\pm\frac{4\pi}{3},0)$ & $(0,\pm\frac{4\pi}{3}),(\pm\frac{8\pi}{3},0),(\pm\frac{4\pi}{3},\pm\frac{2\pi}{3})$ \\
     U3 (Fig.~\ref{fig:u1_ansatz}(c)) & 2 & $\pm(\frac{\pi}{3},\frac{\pi}{2}),\ \pm(-\frac{\pi}{3},\frac{\pi}{6})$ & $\pm(\frac{2\pi}{3},{\pi}),\ \pm(-\frac{2\pi}{3},\frac{\pi}{3}),\ \pm(\frac{2\pi}{3},\frac{\pi}{3}),\ \pm(0,\frac{2\pi}{3})$ \\
     U4 (Fig.~\ref{fig:u1_ansatz}(d)) & 2 & $\pm(\frac{\pi}{3},\frac{\pi}{2}),\ \pm(-\frac{\pi}{3},\frac{\pi}{6})$ & $\pm(\frac{2\pi}{3},{\pi}),\ \pm(-\frac{2\pi}{3},\frac{\pi}{3}),\ \pm(\frac{2\pi}{3},\frac{\pi}{3}),\ \pm(0,\frac{2\pi}{3})$ \\
     U5 (Fig.~\ref{fig:u1_ansatz}(e)) & 1 & $(0,0)$ & $(0,\pm\frac{2\pi}{3}),(\pm\frac{4\pi}{3},0),(\pm\frac{4\pi}{3},\pm\frac{2\pi}{3})$ \\
     U6 (Fig.~\ref{fig:u1_ansatz}(f)) & 1 & $(0,0)$ & $(0,\pm\frac{2\pi}{3}),(\pm\frac{4\pi}{3},0),(\pm\frac{4\pi}{3},\pm\frac{2\pi}{3})$
    \end{tabular}
    \end{ruledtabular}
    \caption{Summary of the observations from Fig.~\ref{fig:s-DSL_mechanism}. For each U(1) \textit{Ansatz} shown in Fig.~\ref{fig:u1_ansatz}, we list the number of s-DSL points within the Brillouin zone, their positions in the extended Brillouin zone (EBZ), and the corresponding $\boldsymbol{q}$-vectors connecting them.}
    \label{tab:s-DSL_mechanism}
\end{table*}

\subsection{Properties of the semi-Dirac spin liquid state}
\label{sec:s-DSL}

\subsubsection{Origin and symmetry considerations}

To better understand the transition from the DSL to the s-DSL identified above, we analyze the motion of the Dirac points in the \textit{Ans\"atze} labeled U1, U3, U4, U5, and U6 in Fig.~\ref{fig:u1_ansatz}. As the hopping parameters are tuned, the Dirac points shift gradually in momentum space, as represented by the increasing intensity of color in Fig.~\ref{fig:s-DSL_mechanism}. This continuous motion ultimately drives two distinct Dirac points to merge, giving rise to the s-DSL. Each pair of identically colored dots in Fig.~\ref{fig:s-DSL_mechanism} corresponds to a specific choice of hopping parameter close to the critical regime where the s-DSL emerges (see Fig.~\ref{fig:dispersion_all}). The observations such as the number of s-DSL points within the extended Brillouin zone (EBZ) and the corresponding q-vectors connecting them (see Fig.~\ref{fig:s-DSL_mechanism}) are summarised in Table~\ref{tab:s-DSL_mechanism}. 

In general, as underscored by Fig.~\ref{fig:s-DSL_mechanism}, the merger of two Dirac nodes gives rise to a semi-Dirac node~\cite{Montambaux2009,Bena-2011}. Here, we demonstrate that a key requirement for this mechanism to work is that the {\it little} group at the band-touching momentum is the $C_{2v}$ point group; i.e., it has twofold rotational symmetry and two orthogonal reflection planes. Let us consider a two-band Hamiltonian in the sublattice basis, which can be parameterized as
\begin{equation}\label{eq:c2v_s-DSL_1}
H(\boldsymbol{k}) = a(\boldsymbol{k}) \varsigma^0 + \Vec{d}(\boldsymbol{k}) \cdot \Vec{\varsigma},
\end{equation}
where $\varsigma^0$ is the $2\times2$ identity matrix and $\Vec{\varsigma}=\{\varsigma^x,\varsigma^y,\varsigma^z\}$ are Pauli matrices that act on the two sublattices. The $C_{2v}$ point group consists of two mirror reflections, $\sigma_x$ and $\sigma_y$, and a $C_2$ rotation. Under these operations, the Hamiltonian transforms as
\begin{equation}
\mathcal{U}^{\pdagger}_\mathcal{O} H(\boldsymbol{k}) \mathcal{U}_\mathcal{O}^\dagger 
= H(\mathcal{O}(\boldsymbol{k})) \quad \text{for} \quad \mathcal{O}\in\{M^{\pdagger}_x,M^{\pdagger}_y,C^{\pdagger}_2\},
\end{equation}
with the corresponding sublattice transformation matrices
\begin{equation}
\mathcal{U}^{\pdagger}_{\sigma_x} = \varsigma^x, 
\quad \mathcal{U}^{\pdagger}_{\sigma_y} = \varsigma^0, 
\quad \mathcal{U}^{\pdagger}_{C_2} = \varsigma^x.
\end{equation}

The $\sigma_x$ symmetry constrains the Hamiltonian's coefficients as follows:
\begin{align}\label{eq:c2v_sdsl_3}
d_x(k_x,-k_y) &= d_x(k_x,k_y), \notag \\
d_y(k_x,-k_y) &= -d_y(k_x,k_y), \notag \\
d_z(k_x,-k_y) &= -d_z(k_x,k_y),
\end{align}
implying that $d_x(\boldsymbol{k})$ is even in $k_y$, while $d_y(\boldsymbol{k})$ and $d_z(\boldsymbol{k})$ are odd in $k_y$. Similarly, for $\mathcal{O}=\sigma_y$, Eq.~\eqref{eq:c2v_s-DSL_1} yields $H(-k_x,k_y)=H(k_x,k_y)$, implying that all components of $\Vec{d}(\boldsymbol{k})$ are even in $k_x$. The $C_2$ operation imposes no additional constraints since $C_2=\sigma_x\sigma_y$. The scalar function $a(\boldsymbol{k})$ is always even in both $k_x$ and $k_y$.

With these symmetry constraints in place, we construct a minimal model that yields a semi-Dirac nodal structure. Keeping only the leading symmetry-allowed terms consistent with a single mirror-protected linear dispersion along $k_y$, we choose
\begin{align}
d_x(\boldsymbol{k}) &= m - m_0 k_x^2, \notag \\
d_y(\boldsymbol{k}) &= v k_y, \notag \\
d_z(\boldsymbol{k}) &= a(\boldsymbol{k}) = 0.
\end{align}
The resulting minimal Hamiltonian is
\begin{equation}
H(\boldsymbol{k}) = (m - m_0 k_x^2)\sigma^x + v k_y \sigma^y,
\end{equation}
with spectrum
\begin{equation}
E(\boldsymbol{k}) = \pm \sqrt{v^2 k_y^2 + (m - m_0 k_x^2)^2}.
\end{equation}

For $m_0>0$, there exist three distinct regimes:
\begin{enumerate}
    \item $m$\,$>$\,$0$: Two Dirac nodes appear at $k_y$\,$=$\,$0$ and $k_x=\pm k_0=\pm\sqrt{m/m_0}$. Expanding around $k_x=\pm k_0$ yields $m - m_0(k_0+\delta k_x)^2 \approx -2m_0 k_0 \delta k_x$, showing that each node is a Dirac point with velocities $(v_x,v_y)=(2m_0k_0,v)$.
    \item $m=0$: The two Dirac nodes merge, producing the dispersion
    \begin{alignat}{2}
    E(\boldsymbol{k}) &= \pm \sqrt{v^2 k_y^2 +  m^2_0 k_x^4}\sim \pm v k_y, \,\,\pm m_0 k_x^2.
    \end{alignat}
    At $m=0$, the two Dirac cones merge into a semi-Dirac node, exhibiting quadratic dispersion along $k_x$ and linear dispersion along $k_y$.
    \item $m<0$: No nodal points exist, and a gap opens.
\end{enumerate}

This minimal model thus captures the essential mechanism underlying the transition from a Dirac spin liquid  to a gapped spin liquid through an intermediate semi-Dirac spin liquid. In particular, this mechanism accounts for the transitions observed in the U1, U3, U5, and U6 \textit{Ans\"atze}.

In contrast, higher little-group symmetries such as $C_{3v}$, $C_{4v}$, or $C_{6v}$ are incompatible with the realization of semi-Dirac dispersions. To illustrate this, let us focus on $C_{4v}$. In this case, the presence of a term such as $v k_y \sigma^y$ would, by virtue of $C_4$ rotational symmetry, require the existence of an equivalent linear term along the $k_x$ direction, thereby enforcing full Dirac-like dispersions in both momentum directions.
To formalize this argument, we generalize the matrices $\mathcal{U}_{\mathcal{O}}$ to pseudospin representations. These pseudospin matrices must be consistent with how the underlying physical degrees of freedom (e.g., spin, orbital, sublattice) transform under the operations of the point group, while simultaneously respecting the group multiplication rules (up to projective phases). For $C_{4v}$, the symmetry group is generated by $\sigma_x$ and $C_4$, for which we adopt the representation
\begin{equation}
\mathcal{U}_{\sigma_x} = \varsigma^x, \quad
\mathcal{U}_{C_4}= \dot\iota e^{\dot\iota\frac{\pi}{4}\varsigma^z}.
\end{equation}
Since $\mathcal{U}_{\sigma_x}$ coincides with the choice made in the $C_{2v}$ case, it imposes the same constraints as Eq.~\eqref{eq:c2v_sdsl_3}, leading to the low-energy expansion
\begin{align}
d_x(\boldsymbol{k}) &= a_x k_x + b_x k_y^2 + O(k_x^2) + O(k_y^4), \notag \\
d_y(\boldsymbol{k}) &= b_y k_y + O(k_y^3), \notag \\
d_z(\boldsymbol{k}) &= b_z k_y + O(k_y^3).
\end{align}
The additional constraints from $C_4$ symmetry read
\begin{align}
d_x(-k_y, k_x) &= -d_y(k_x, k_y), \notag \\
d_y(-k_y, k_x) &= d_x(k_x, k_y), \notag \\
d_z(-k_y, k_x) &= d_z(k_x, k_y).
\end{align}
The first two conditions enforce $a_x = b_y$, while simultaneously requiring $d_z = 0$. Consequently, the presence of a linear $b_y k_y$ term in $d_y$ necessitates the appearance of an equivalent $b_y k_x$ term in $d_x$. This enforces linear dispersions along both momentum directions, thereby ruling out the semi-Dirac scenario. An analogous argument applies to other little groups with higher symmetries, such as $C_{6v}$. 
We thus conclude that $C_{2v}$ constitutes a special case: It provides precisely the symmetry constraints necessary to stabilize semi-Dirac nodes, whereas higher point-group symmetries inevitably restore full Dirac dispersions.

\subsubsection{Spin-spin correlation functions}

\begin{figure}
\includegraphics[width=1.0\linewidth]{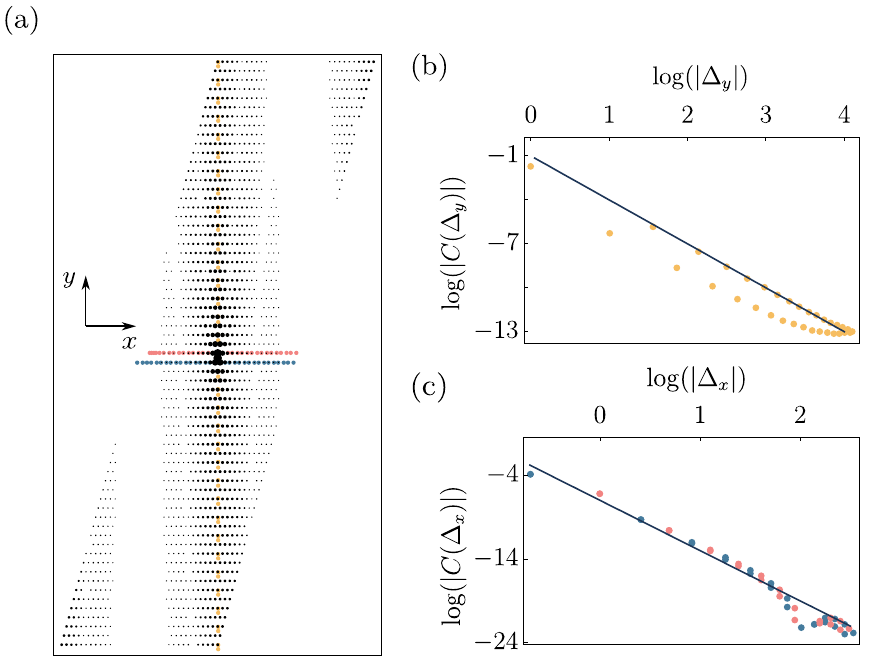}
\caption{(a) Real-space two-spin correlation decay for the U1 \textit{Ansatz} in the s-DSL phase at the mean-field level. The colored vertical and horizontal lines mark the directions analyzed in panels (b) and (c). The displacement variables are defined as $\Delta_y = y - y_0$ and $\Delta_x = x - x_0$, with $(x_0,y_0)$ labeling the reference site. The functions $C(\Delta_y)$ and $C(\Delta_x)$ denote the spin-spin correlations along the vertical and horizontal directions, respectively. (b) Along the vertical ($y$) direction, correlations decay with exponent $\approx 3$. (c) Along the two horizontal slices, correlations decay faster, with exponent $\approx 6$.}
\label{fig:decay_real_space}
\end{figure}

Having established the origin of the s-DSL state, we now turn to its correlation functions. To compute equal-time spin-spin correlations, it is convenient to rewrite the spins using the Abrikosov-fermion representation of Eq.~\eqref{spinon}. In this formulation, the Heisenberg exchange becomes a four-fermion term, which we decouple at the mean-field level in the hopping channel, yielding a quadratic spinon Hamiltonian. Diagonalizing this mean-field Hamiltonian gives the spinon eigenmodes $\psi_n(i)$ and their energies. The mean-field ground state is then obtained by filling the lowest half of these modes, from which we construct the single-particle Green's function $G^{}_{ij}=\sum_{n\in{\rm occ}}\psi_n^\ast(i)\psi^{}_n(j)$. 
Wick’s theorem allows us to express the physical spin–spin correlator entirely in terms of $G_{ij}$ as
$\langle \mathbf{\hat S}_i\!\cdot\!\mathbf{\hat S}_j\rangle \sim -|G_{ij}|^2$, so that the spatial decay of the spin correlations follows directly from the numerically evaluated $|G_{ij}|^2$.

Figure~\ref{fig:decay_real_space} presents the real-space decay of the spin-spin correlations for the U1 \textit{Ansatz} in the s-DSL phase with hopping parameters $\chi_h=1.0$, $\chi_v=9.0$, and $\chi_z=4.5$ (see Figs.~\ref{fig:u1_ansatz} and~\ref{fig:dispersion_all}). In panel (a), the radius of the black disks represents the correlation strength, while the colored vertical and horizontal lines indicate the directions analyzed in panels (b) and (c). Panel (b) shows that correlations along the vertical ($y$) direction decay with exponent $\approx 3$, whereas panel (c) demonstrates that correlations along the two horizontal cuts decay much more rapidly, with exponent $\approx 6$.

These exponents could be understood as follows. In the long-wavelength limit, the dispersion of the s-DSL can be approximated as
\begin{equation}
E(\boldsymbol{k}) =  m k^{2}_1+k^{}_2.
\end{equation}
The corresponding low-energy effective theory is obtained by adding a quadratic anisotropic term (mass) to the massless Dirac Lagrangian in $(1+1)$D:
\begin{equation}
\mathcal{L} = \bar{\psi} \left( i \gamma^\mu \partial^{}_\mu - m\, \partial_1^2 \right) \psi,
\end{equation}
where $\mu=0,2$ and $\bar{\psi} = \gamma^0 \psi^\dagger$. In $(2+1)$D, the Green's function satisfies
\begin{equation}
\left( i \gamma^\mu \partial_\mu - m\, \partial_1^2 \right) \mathcal{G}(\boldsymbol{r}-\boldsymbol{r}') = \delta^3(\boldsymbol{r}-\boldsymbol{r}').
\end{equation}
Here, we temporarily use the notation $\boldsymbol{r}\equiv(r^{}_0,r^{}_1,r^{}_2)$ to denote a three-vector in which the first component is a time coordinate and the other two are Cartesian coordinates. Using the Fourier transform
\begin{equation}
\mathcal{G}(\boldsymbol{r}-\boldsymbol{r}') = \int \frac{d^3\boldsymbol{k}}{(2\pi)^3} e^{i \boldsymbol{k}\cdot(\boldsymbol{r}-\boldsymbol{r}')} G(\boldsymbol{k}),
\end{equation}
we obtain the Green’s function in momentum space:
\begin{equation}
G(\boldsymbol{k}) = \frac{1}{\gamma^\mu k^{}_\mu + m k^{2}_1} 
= \frac{\gamma^\mu k^{}_\mu - m k^{2}_1}{k^{2}_\mu - m^2 k^{4}_1}.
\end{equation}
To analyze the spatial decay of correlations, we consider the real-space Green’s function $\mathcal{G}(\boldsymbol{r})$, obtained by Fourier transforming over spatial components:
\begin{equation}\label{eq:gx_1}
\mathcal{G}(\boldsymbol{r}) = \int \frac{d^2 q\, dk^{}_1}{(2\pi)^3} 
e^{i \boldsymbol{k}\cdot \boldsymbol{r}} 
\frac{\gamma^\mu q^{}_\mu - m k^{2}_1}{q^{2}_\mu - m^2 k^{4}_1}, 
\end{equation}
where $\boldsymbol{k}=(q_0,k_1,q_2)$ separates the $\mu=0,2$ components [the $(1+1)$D Dirac piece] from the anisotropic momentum $k_1$.

We can now carry out a simple scaling analysis that proves informative. In order to ascertain the asymptotic behavior of $\mathcal{G}(x,y)$ at large distances, we employ the anisotropic scaling: 
$r^{}_0 \sim \tilde{r}^{}_0 \Lambda$,  
$r^{}_1 \sim \tilde{r}^{}_1 \Lambda^{1/2}$, and
$r^{}_2 \sim \tilde{r}^{}_2 \Lambda$.  
With the rescaling $k^{}_0 \to \tilde{k}^{}_0/\Lambda$, $k^{}_1 \to \tilde{k}^{}_1/\Lambda^{1/2}$, and $k^{}_2 \to \tilde{k}^{}_2/\Lambda$, Eq.~\eqref{eq:gx_1} becomes
\begin{equation}\label{eq:gx_2}
\mathcal{G}(\Lambda \tilde{r}^{}_0, \Lambda^{1/2} \tilde{r}^{}_1, \Lambda \tilde{r}^{}_2) 
= \Lambda^{-3/2} \mathcal{G}(\tilde{r}^{}_0,\tilde{r}^{}_1,\tilde{r}^{}_2).
\end{equation}
For equal-time correlations, we set $r^{}_0=0$, yielding
\begin{equation}\label{eq:gx_3}
\mathcal{G}(\Lambda^{1/2}\tilde{r}^{}_1, \Lambda \tilde{r}^{}_2) 
= \Lambda^{-3/2} \mathcal{G}(\tilde{r}^{}_1, \tilde{r}^{}_2).
\end{equation}
The dependence along the two directions can then be straightforwardly extracted.
Along $r^{}_1$, we set $r^{}_2=0$ and choose $\Lambda=r^{2}_1$, such that $\tilde{r}^{}_1=1$. This gives $\mathcal{G}(r^{}_1,0)\sim r_1^{-3}$.
Likewise, along $r^{}_2$, we set $r^{}_1=0$ and choose $\Lambda=r^{}_2$, such that $\tilde{r}^{}_2=1$; this shows that $\mathcal{G}(0,r^{}_2)\sim r_2^{-3/2}$.

\begin{figure}[b]
\includegraphics[width=1.0\linewidth]{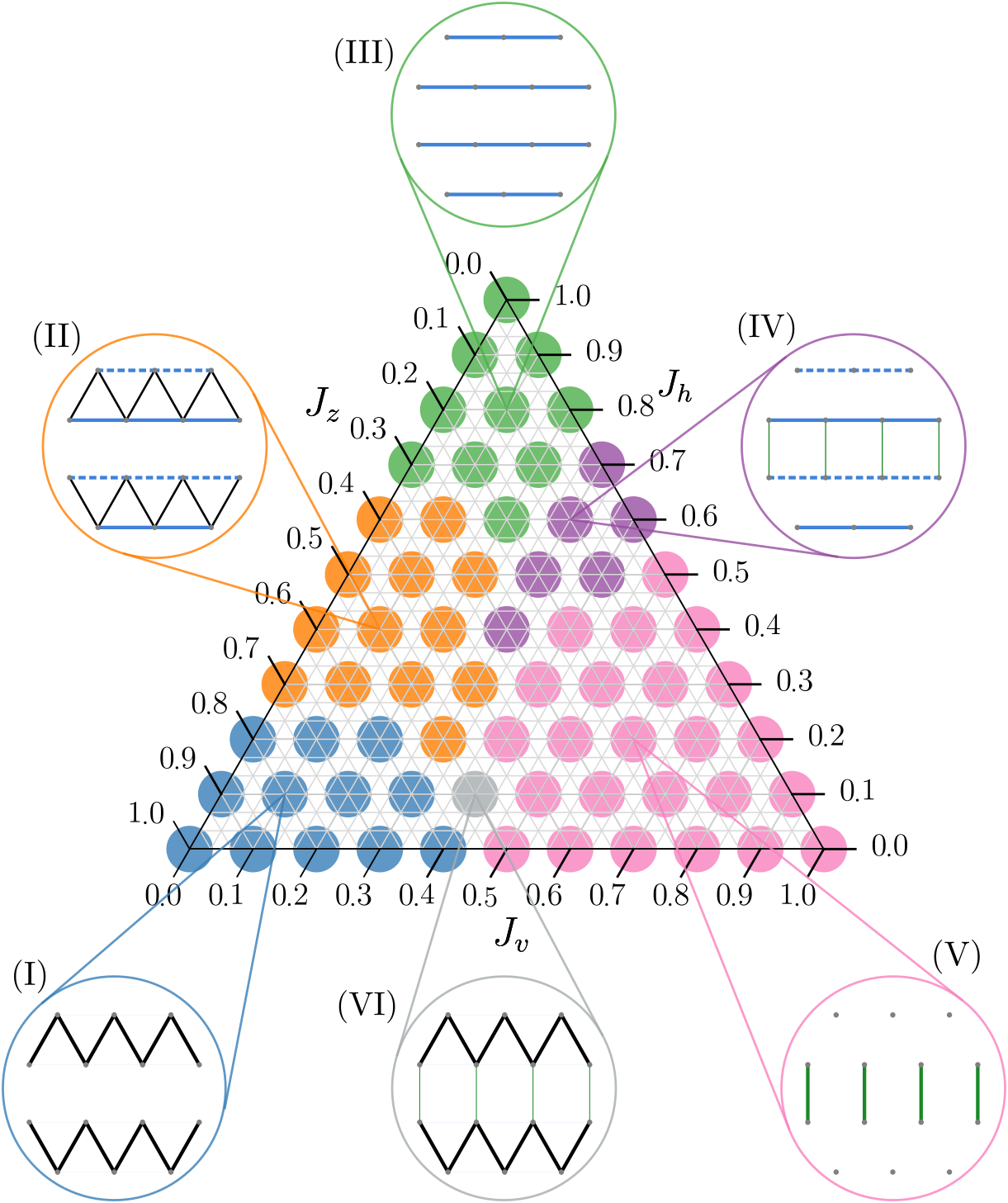}
\caption{The ternary phase diagram of the nearest-neighbor Heisenberg model on the trellis lattice with antiferromagnetic couplings $J_h$, $J_v$, and $J_z$ as illustrated in Fig.~\ref{fig:lattice}(a), subject to $J_h+J_v+J_z=1$. The six distinct regions colored here correspond to (I) zigzag-chain, (II) zigzag-ladder, (III) rung-chain, (IV) rung-ladder, (V) ladder-dimer, and (VI) honeycomb phases. Representative saddle-point structures are shown for selected points within each phase.}
\label{fig:phasediagram}
\end{figure}

\begin{figure}[b]
\includegraphics[width=1.0\linewidth]{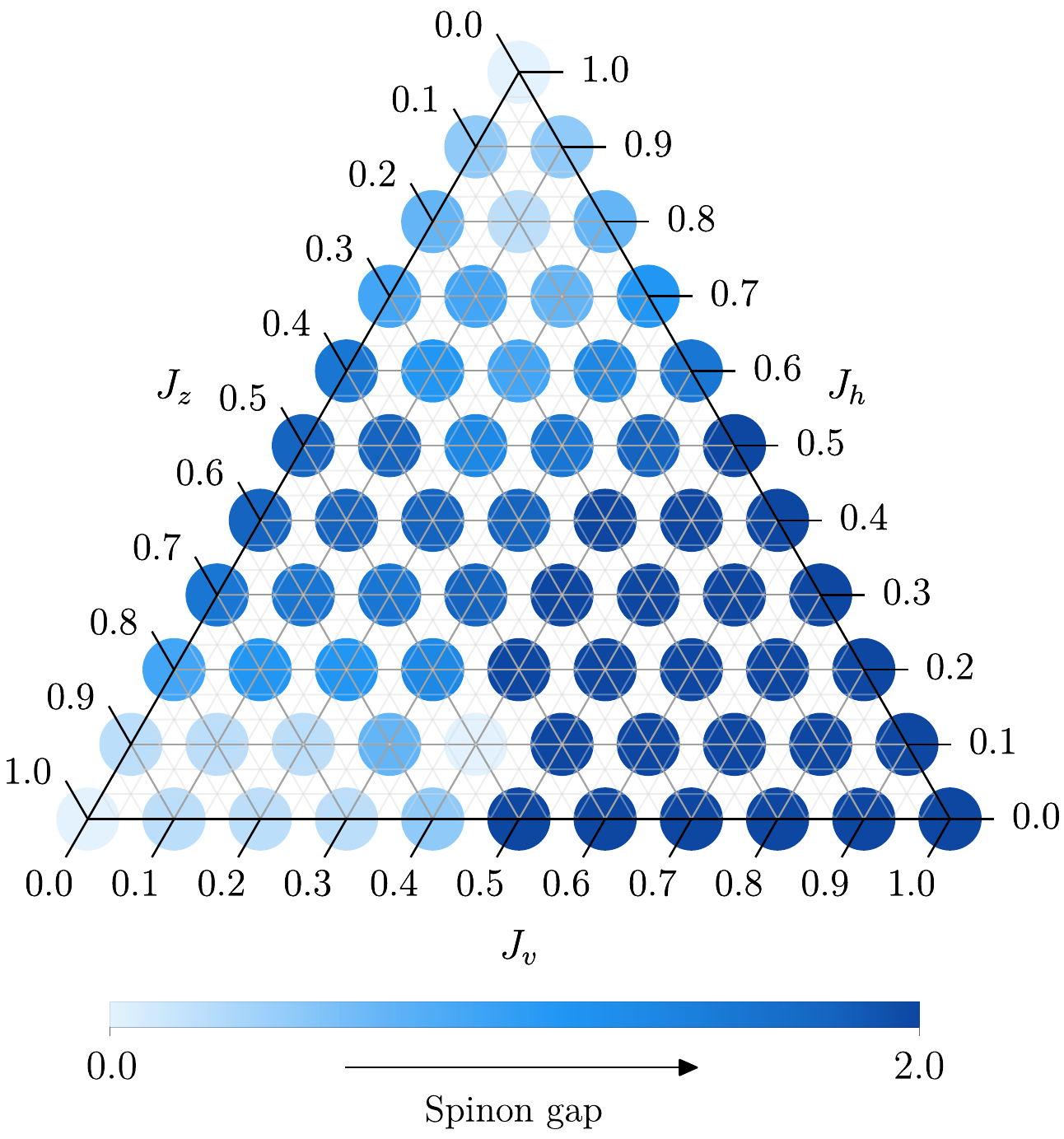}
\caption{Spinon excitation gap across the ternary phase diagram shown in Fig.~\ref{fig:phasediagram}. The color intensity represents the magnitude of the spinon gap, with the lowest intensity indicating gapless regions and higher intensity corresponding to an increasing excitation gap.}
\label{fig:gap}
\end{figure}

The equal-time correlator is defined as
\begin{align}
 \mathcal{S}^{}_{+-}(\boldsymbol{r}-\boldsymbol{r}') 
 &= \langle S^+(\boldsymbol{r}) S^-(\boldsymbol{r}') \rangle \notag \\
 &= \left\langle f^{\dagger}_\uparrow(\boldsymbol{r}) f^{\pdagger}_\downarrow(\boldsymbol{r}) 
 f^{\dagger}_\downarrow(\boldsymbol{r}') f^{\pdagger}_\uparrow(\boldsymbol{r}') \right\rangle \notag \\
 &= -\left\langle f^{\dagger}_\uparrow(\boldsymbol{r}) f^{\pdagger}_\uparrow(\boldsymbol{r}') \right\rangle 
 \left\langle f^{\dagger}_\downarrow(\boldsymbol{r}) f^{\pdagger}_\downarrow(\boldsymbol{r}') \right\rangle^\star,
\end{align}
where, in the last step, we use Wick’s theorem. Defining the equal-time two-point correlator as $\mathcal{G}^{}_\sigma(\boldsymbol{r}-\boldsymbol{r}') = \langle f^{\dagger}_\sigma(\boldsymbol{r}) f^{\pdagger}_\sigma(\boldsymbol{r}') \rangle$, we can write
\begin{equation}
\mathcal{S}^{}_{+-}(\boldsymbol{r}-\boldsymbol{r}') 
= \mathcal{G}^{}_\uparrow(\boldsymbol{r}-\boldsymbol{r}') \mathcal{G}^\star_\downarrow(\boldsymbol{r}-\boldsymbol{r}').    
\end{equation}
By virtue of spin-rotation symmetry, this reduces to
\begin{equation}
\mathcal{S}^{}_{+-}(\boldsymbol{r}-\boldsymbol{r}') 
= -|\mathcal{G}(\boldsymbol{r}-\boldsymbol{r}')|^2,
\end{equation}
implying
\begin{equation}
\mathcal{S}^{}_{+-}(\boldsymbol{r}-\boldsymbol{r}') 
\sim \begin{cases}
			|r^{}_1-r^{\prime}_1|^{-6}, & \text{if $r_2-r^{\prime}_2=0$;}\\
            |r^{}_2-r^{\prime}_2|^{-3}, & \text{if $r_1-r^{\prime}_1=0$.}
		 \end{cases}
\end{equation}
This reproduces the decay observed in Fig.~\ref{fig:decay_real_space}. To fully capture all the fine oscillatory features of the spin--spin correlation function, it is necessary to evaluate the complete integral over the Brillouin zone in Eq.~\eqref{eq:gx_1}. A detailed analysis of this behavior is left for future work. We note that all along the phase boundary separating the DSL and gapped state shown in Fig.\ref{fig:dispersion_all}(a), we realize a semi-Dirac spinon dispersion including at the multicritical point. These mean-field exponents should be compared with those in other gapless phases, such as a spinon Fermi surface state $(\sim r^{-3})$~\cite{Motrunich2005,Motrunich2011,Biswas2011} and Dirac spin liquid $(\sim r^{-4})$~\cite{Rantner-2002}. Since the semi-Dirac dispersion originates from a merger of two Dirac cones that introduces a length scale, i.e., the inverse of the distance between the two Dirac cones, it would be interesting to explore how the $\sim r^{-4}$ decay of the Dirac spin liquid transforms into the anisotropic exponents of the semi-Dirac state.

\subsection{Ternary phase diagram of the Heisenberg model}  
\label{sec:PD}

Building on our analysis of the generic features of the \textit{Ans\"atze}, we now investigate the phase diagram obtained from saddle-point solutions of a nearest-neighbor antiferromagnetic Heisenberg model with exchange couplings $J_v$ along the vertical ladder legs, $J_h$ along the horizontal rungs, and $J_z$ on the zigzag interladder bonds, as depicted in Fig.~\ref{fig:lattice}(a). The analysis is performed by self-consistently determining all mean-field amplitudes.

The saddle-point solutions of six U(1) \textit{Ans\"atze}, considered up to a doubling of the unit cell, are summarized in Fig.~\ref{fig:phasediagram}. Within the ternary parameter space, six distinct phases (I--VI) emerge. The graphical representation in Fig.~\ref{fig:phasediagram} directly highlights the saddle-point structures of the \textit{Ans\"atze}, with bond widths scaled according to the self-consistently determined mean-field amplitudes at representative points. 

\begin{table}[b]
    \centering
    \begin{ruledtabular}
    \begin{tabular}{lcc}
        Phase & Couplings & Hopping parameters  \\
         & $(J_z, J_h, J_v)$ & $(\chi_z, \chi_h, \chi_v)$  \\
        \hline
        Zigzag chain (I) & $(0.8, 0.1, 0.1)$ & $(0.642, -0.009, 0.003)$ \\
        Zigzag ladder (II) & $(0.5, 0.4, 0.1)$ & $(0.583, 0.352,-0.012)$ \\
        Rung chain (III) & $(0.1, 0.8, 0.1)$ & $(0.001, 0.636,0.013)$ \\
        Rung ladder (IV) & $(0.1, 0.6, 0.3)$ & $(0.000, 0.630, 0.192)$ \\
        Ladder dimer (V) & $(0.2, 0.2, 0.6)$ & $(-0.004, -0.009, 0.999)$ \\
        Honeycomb (VI) & $(0.5, 0.1, 0.4)$ & $(0.636,-0.009,-0.153)$ \\
    \end{tabular}
    \end{ruledtabular}
    \caption{Representative coupling strengths $(J_z, J_h, J_v)$ and self-consistent hopping parameters $(\chi_z, \chi_h, \chi_v)$ for the points marked in the ternary phase diagram of Fig.~\ref{fig:phasediagram}.}
    \label{tab:phases_hopping}
\end{table}

\begin{itemize}
\item
In phase I (blue), the amplitudes on the $J_v$ and $J_h$ bonds are negligible for all \textit{Ans\"atze}, resulting in an effective 1D zigzag chain structure. A representative point here is $(J_z,J_h,J_v) = (0.8, 0.1, 0.1)$ with optimized mean-field parameters $(\chi_z, \chi_h, \chi_v)$ $=$ $(0.642, -0.009, 0.003)$.  
\item
In phase II (orange), realized by the \textit{Ans\"atze} U2, U4, and U5 with $0$- and $\pi$-staggered fluxes through triangular plaquettes, the amplitudes on the legs dominate, producing an effective zigzag ladder structure. At the point $(J_z,J_h,J_v)=(0.5,0.4,0.1)$, the optimized values are $(0.583, 0.352,-0.012)$.  
\item
Phase III (green) corresponds to a rung-chain structure dominated by $J_h$ bonds, with negligible amplitudes on $J_v$ and $J_z$. For couplings $(J_z,J_h,J_v)$ $=$ $(0.1,0.8,0.1)$, the corresponding mean-field values are $(0.001, 0.636,0.013)$.  
\item
In phase IV (purple), a $\pi$-flux rung-ladder state appears, realized by the U2, U3, and U6 \textit{Ans\"atze}. For $(0.1,0.6,0.3)$, the optimized amplitudes are $(0.000, 0.630, 0.192)$.  
\item
In phase V (pink), all \textit{Ans\"atze} collapse to a ladder-dimer phase due to vanishingly small $\chi_z$ and $\chi_h$. At $(0.2,0.2,0.6)$, the optimized parameters are $(-0.004, -0.009, 0.999)$.  
\item
Finally, at the junction of phases I, II, and V, a zero-flux honeycomb resonating-valence-bond-like state (phase VI) governs the low-energy behavior. Here, $\chi_h \sim -0.009$ is nearly zero, while $(\chi_z, \chi_v) \sim (0.636,-0.153)$. The relevant parent states include U1, U2, U5, and U6.  
\end{itemize}
The isotropic point in the phase diagram, $J_z=J_h=J_v$ features a ladder-dimer state (phase V). In Fig.~\ref{fig:gap}, we show the spinon-gap profile across the ternary phase diagram in Fig.~\ref{fig:phasediagram}. Although the effective low-energy states in phases (I) and (III)—which reduce to isolated one-dimensional chains—are expected to be gapless, we find a finite gap except at the special points $J_v = J_h = 0$ and $J_z = J_v = 0$ in these two phases, respectively. This is because, in addition to the dominant mean-field parameters $\chi^{\pdagger}_z$ and $\chi^{\pdagger}_h$ in phases (I) and (III), respectively, the remaining mean-field parameters remain small but nonzero. As a result, a finite gap emerges.

In our self-consistent mean-field analysis, we also considered all $\mathbb{Z}_2$ \textit{Ans\"atze} but found them either to be higher in energy than the corresponding U(1) states or to exhibit vanishing optimized pairing amplitudes (which are responsible for the $\mathbb{Z}_{2}$ gauge structure in the first place), thereby flowing back to their parent U(1) states. It would be interesting to assess their variational energetics after Gutzwiller projection and investigate whether the $\mathbb{Z}_{2}$ states are stabilized in some region of the phase diagram.

\begin{figure}[tb]
\includegraphics[width=0.96\linewidth]{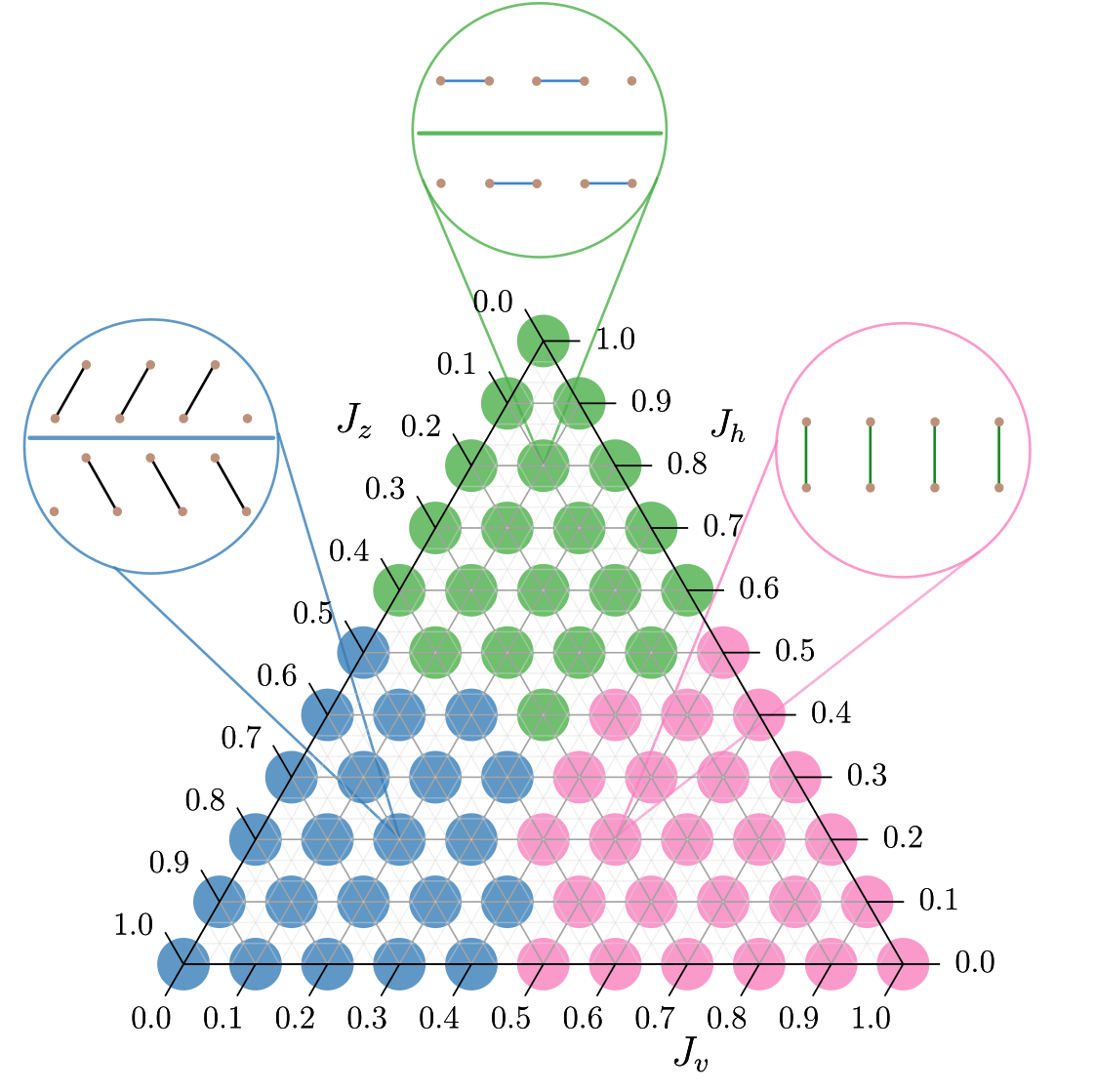}
\caption{Ternary phase diagram after symmetry reduction, consisting of three dimerized phases: Majumdar-Ghosh, rung dimer, and ladder-dimer phases, shown in blue, green, and pink, respectively. Realization of the Majumdar-Ghosh phase requires breaking $\sigma^{}_x$, while the rung-dimer phase requires breaking $T_1$.}
\label{fig:dimer_PD}
\end{figure} 

It is important to emphasize that the saddle-point phases described above assume full lattice and time-reversal symmetry. Certain paramagnetic phases, such as the Majumdar-Ghosh state~\cite{karlo1997}, fall outside this fully symmetric framework. To capture these, one must consider symmetry-reduced versions of the \textit{Ans\"atze}. For example, breaking $\sigma_x$ yields a dimerized state with singlets on the zigzag bonds, corresponding to the Majumdar-Ghosh phase, which is twofold degenerate per zigzag ladder and is indicated in blue in Fig.~\ref{fig:dimer_PD}. Similarly, breaking translational symmetry $T_1$ gives rise to a rung-dimer phase with singlets on the horizontal rungs, also twofold degenerate per rung, shown in green in Fig.~\ref{fig:dimer_PD}. The ladder-dimer phase remains stable within its domain. Overall, we identify three distinct dimerized quantum paramagnetic phases.

\begin{figure*}    \includegraphics[width=1.0\textwidth]{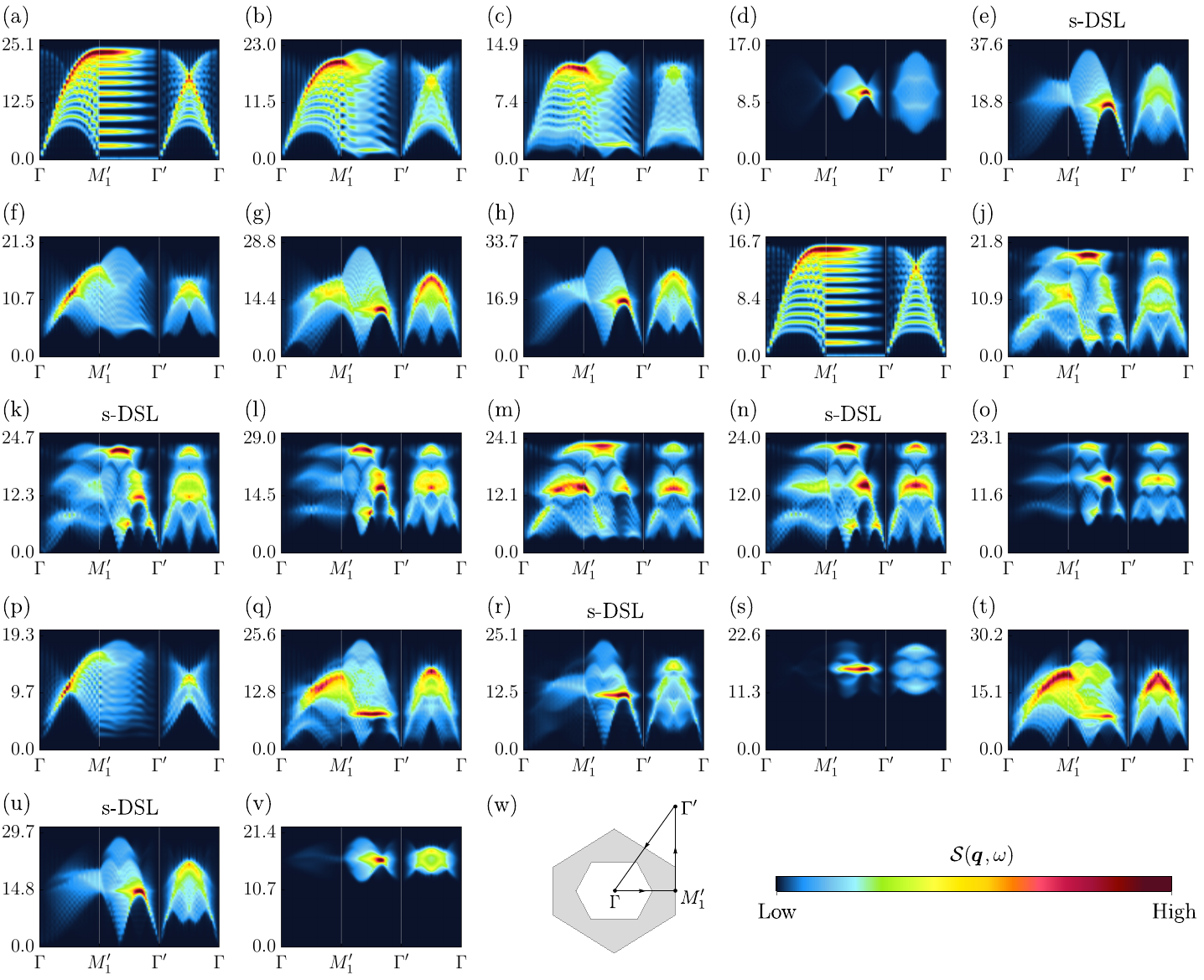}
   \caption{Dynamical spin structure factors plotted along the momentum-space path $\Gamma \rightarrow M'_1 \rightarrow \Gamma' \rightarrow \Gamma$, where the high-symmetry points are defined as $\Gamma = (0,0)$, $M'_1 = (4\pi/3, 0)$, and $\Gamma' = (4\pi/3, 2\pi)$. The calculations are performed on a system of size $24 \times 24 \times 2$. The panels correspond to the following \textit{Ans\"atze}: (a)–(e) U1, (f)–(h) U2, (i)–(l) U3, (m)–(o) U4, (p)–(s) U5, and (t)–(v) U6. The hopping amplitude $\chi_h$ is fixed to 1 in all cases.
For the U1 \textit{Ansatz}, shown in panels (a)–(e), the remaining mean-field parameters $(\chi_v, \chi_z)$ are set to
(a) $(0, 6)$, yielding a nodal-line spin liquid;
(b) $(1, 5)$, corresponding to a Dirac spin liquid;
(c) $(1, 1)$, giving a spinon Fermi surface;
(d) $(5, 1)$, resulting in a fully gapped spin liquid; and
(e) $(9, 4.5)$, which produces a semi-Dirac spin liquid.
These phases correspond to the band structures shown in Fig.~\ref{fig:dispersion_all}(d).
For the U2 \textit{Ansatz} [Fig.~\ref{fig:dispersion_all}(e)], panels (f)–(h) show results for
(f) $(2, 4)$: a gapped phase;
(g) $(4\sqrt{2}, 4)$: the DSL phase; and
(h) $(8, 4)$: the second gapped phase.
Here, the DSL phase occurs along a critical line separating two gapped regimes.
For the U3 \textit{Ansatz} [Fig.~\ref{fig:dispersion_all}(f)], panels (i)–(l) correspond to
(i) $(0, 4)$: a nodal-line phase;
(j) $(4, 4)$: a DSL phase;
(k) $(4\sqrt{2}, 4)$: a semi-Dirac spin liquid; and
(l) $(8, 4)$: a gapped phase.
For the U4 \textit{Ansatz} [Fig.~\ref{fig:dispersion_all}(j)], panels (m)–(o) show the DSF for
(m) $(4, 4\sqrt{3/2})$: a gapped phase;
(n) $(4\sqrt{2}, 4)$: an s-DSL phase; and
(o) $(4\sqrt{3}, 4/\sqrt{2})$: another gapped phase.
In this case, the s-DSL phase again separates two distinct gapped regions.
For the U5 \textit{Ansatz} [Fig.~\ref{fig:dispersion_all}(k)], panels (p)–(r) represent
(p) $(1, 4)$: a gapped phase;
(q) $(4, 4)$: a DSL phase; 
(r) $(6, 2.83)$: a state with s-DSL phase; and
(s) $(8, 1)$: the other gapped phase.
Finally, for the U6 \textit{Ansatz} [Fig.~\ref{fig:dispersion_all}(l)], panels (t)–(v) correspond to
(t) $(4, 5)$: the DSL phase;
(u) $(7, 3.64)$: the s-DSL phase;
(v) $(8, 1)$: a gapped phase and (w) shows the high-symmetry path along which the structure factors are plotted.}
   \label{fig:dsf_all}
\end{figure*}

\begin{figure*}
    \includegraphics[width=1.0\textwidth]{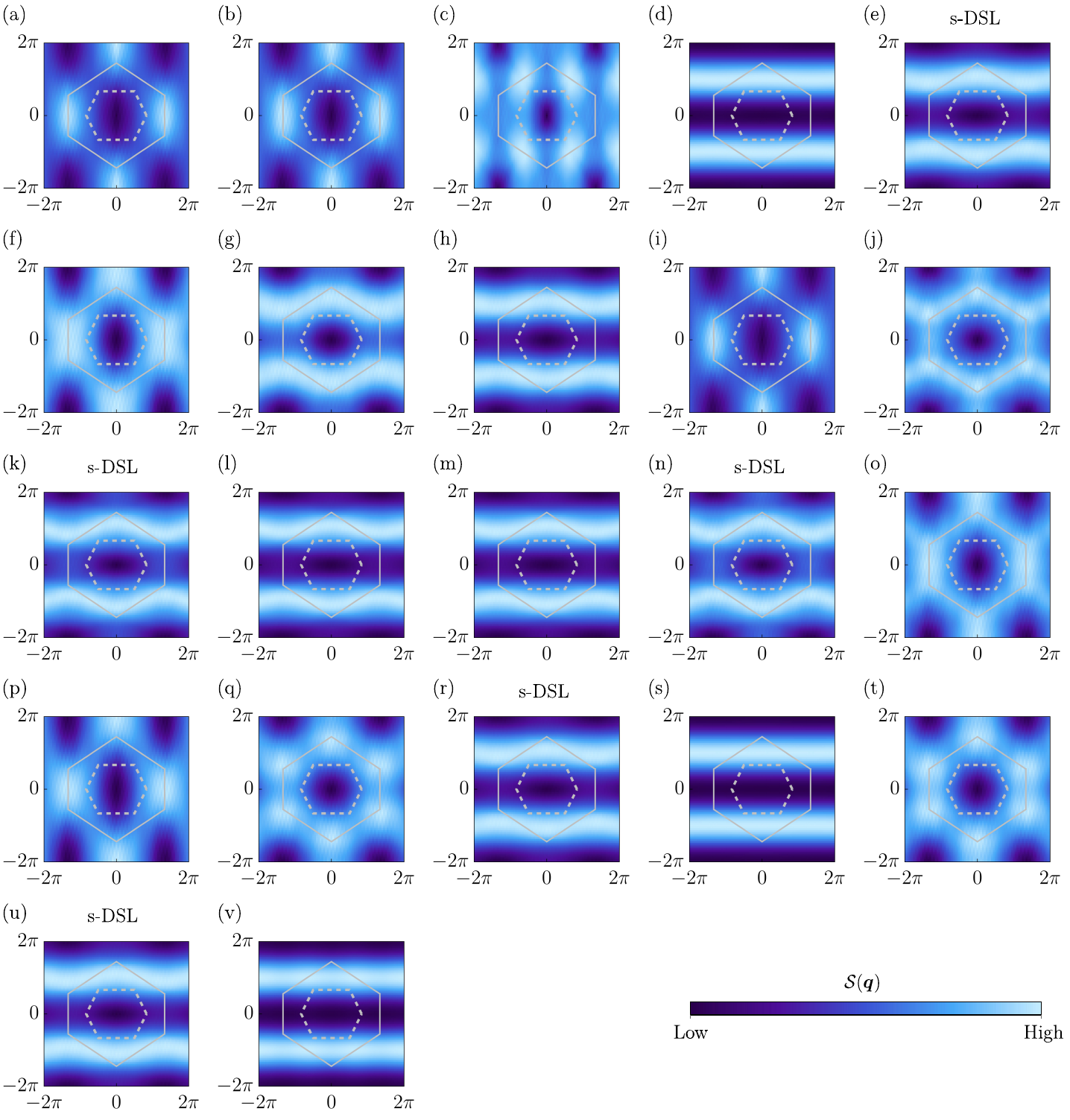}
    \caption{Equal-time structure factors shown for the same points as Fig.~\ref{fig:dsf_all}. The solid (dashed) hexagon marks the extended (first) Brillouin zones.}
    \label{fig:ssf_all}
\end{figure*}

\section{Spectral properties}
\label{sec:SF}

To further characterize the quantum spin liquid states, we examine the time- and momentum-resolved spin-spin correlation function, namely, the DSF, defined as  
\begin{equation}\label{eq:dsf_def}
    \mathcal{S}^{\mu\nu}(\boldsymbol{q},\omega)
    = \int \frac{d\tau \, e^{\dot\iota\omega\tau}}{2\pi N}
    \sum_{\langle i,j\rangle}
    e^{\dot\iota\boldsymbol{q}\cdot(\boldsymbol{r}_i-\boldsymbol{r}_j)}
    \langle \mathbf{S}^\mu_i(\tau)\mathbf{S}^\nu_j(0)\rangle,
\end{equation}
with $\mu,\nu = x,y,z$. The DSF is directly accessible in inelastic neutron scattering experiments and serves as a key probe of the spectral signatures of fractionalized excitations in QSLs.  

In the following, we compute the DSF (and, where relevant, its frequency-integrated equal-time counterpart) using both mean-field theory (Figs.~\ref{fig:dsf_all} and \ref{fig:ssf_all}) and beyond-mean-field approaches (Fig.~\ref{fig:keldysh_equal_time}), including Keldysh pf-FRG and DMRG.

Before presenting the results, we clarify the scope of the comparison between mean-field and numerical approaches. We emphasize that the dynamical structure factors computed from the fermionic mean-field Ans\"atze should be interpreted as qualitative probes of the underlying parton band structures rather than as quantitative predictions for the physical spin model \cite{Wen2002}. Although the microscopic exchange couplings $(J_v,J_h,J_z)$ are fixed, the mean-field excitation spectrum is governed by {\it Ansatz}-dependent bond parameters that set the spinon bandwidth and are not uniquely determined by the bare exchanges alone. More fundamentally, the unprojected Abrikosov-fermion mean-field state relaxes the single-occupancy constraint and therefore does not reside in the physical spin Hilbert space; as a consequence, its dynamical correlators do not, in general, satisfy the exact sum rules of the spin Hamiltonian, including frequency-moment sum rules that relate moments of the dynamical structure factors to equal-time correlators and exchange couplings \cite{auerbach2012interacting,HohenbergBrinkman1974}. By contrast, DMRG evaluates spin correlations strictly within the physical Hilbert space, while pf-FRG incorporates self-energy and vertex renormalizations derived from the microscopic Hamiltonian \cite{ReutherWolter2010}. For this reason, absolute energy scales, spectral weights, and detailed line shapes obtained at the mean-field level are not expected to quantitatively match those of pf-FRG or DMRG, even when expressed in units of $J$. Throughout, we therefore restrict comparisons to qualitative momentum-space features and trends, which nevertheless provide valuable insight into the symmetry properties and structure of the underlying fractionalized excitations.

\subsection{Mean-field results}

Owing to SU(2) spin-rotation symmetry, it suffices to compute only the longitudinal component of the DSF
\begin{equation}\label{eq:dsf}
	\mathcal{S}^{zz}(\boldsymbol{q},\omega)=\int^{+\infty}_{-\infty}\frac{d\tau\, e^{i\omega\tau}}{2\pi N} \sum_{i,j}e^{i\boldsymbol{q}\cdot\boldsymbol{r}_{ij}} \langle \hat{S}^z_{i}(\tau)\hat{S}^{z}_{j}(0)\rangle,
\end{equation}
where $\boldsymbol{r}_{ij}=\boldsymbol{r}_i-\boldsymbol{r}_j$ and $N$ denotes the total number of sites. Inserting the Heisenberg time evolution of $\hat{S}^z_{i}(\tau)=e^{i\hat{H}\tau}\hat{S}^z_{i}e^{-i\hat{H}\tau}$ and expressing the spin operators in terms of fermionic spinons recasts Eq.~\eqref{eq:dsf} into the form
\begin{align}
\mathcal{S}^{zz}(\boldsymbol{q},\omega)&=\int^{+\infty}_{-\infty}\frac{d\tau\, e^{i\omega\tau}}{8\pi N}\sum_{i,j}e^{i\boldsymbol{q}\cdot\boldsymbol{r}_{ij}}\tau^z_{\alpha\alpha}\tau^z_{\beta\beta} \notag\\
& \times \sum_{\alpha,\beta} \langle e^{i\hat{H}\tau} \hat{f}^\dagger_{i,\alpha} \hat{f}^{\pdagger}_{i,\alpha} e^{-i\hat{H}\tau} \hat{f}^\dagger_{j,\beta} \hat{f}^{\pdagger}_{j,\beta} \rangle. \label{eq:dsf_zz_f}
\end{align}

For U(1) \textit{Ans\"atze}, given the absence of pairing terms, the spin-up and spin-down sectors decouple, allowing us to work in a basis consisting solely of annihilation operators in each spin sector, i.e., $\hat{f}^\dagger_{\alpha} = (\hat{f}^\dagger_{1,\alpha}, \hat{f}^\dagger_{2,\alpha}, \cdots, \hat{f}^\dagger_{N,\alpha})$. Consequently, the Hamiltonian takes the form
\begin{equation}
    \hat{H} = \sum_{\alpha = \uparrow, \downarrow} \hat{f}^\dagger_{\alpha} \hat{\mathcal{H}} \hat{f}^{\pdagger}_{\alpha}.
    \label{eq:u1_ham}
\end{equation}

We now perform a unitary transformation $\hat{f}^{\pdagger}_{\alpha} = U \hat{c}^{\pdagger}_{\alpha}$ such that $U^\dagger \hat{\mathcal{H}} U = \mathrm{diag}(\epsilon_1, \epsilon_2, \dots, \epsilon_N)$. Substituting into Eq.~\eqref{eq:dsf_zz_f}, we obtain
\begin{widetext}
\begin{equation}\label{eq:dsf_u_1}
   \mathcal{S}^{zz}(\boldsymbol{q},\omega)=\int^{+\infty}_{-\infty}\frac{d\tau\, e^{i\omega\tau}}{8\pi N} \sum_{i,j,\mu,\mu',\nu,\nu'} e^{i\boldsymbol{q}\cdot \boldsymbol{r}_{ij}} \tau^z_{\alpha\alpha} \tau^z_{\beta\beta}
\, U^*_{i,\mu} U^{}_{i,\mu'} U^*_{j,\nu} U^{}_{j,\nu'} \sum_{\alpha,\beta} \langle e^{i\hat{H}\tau} \hat{c}^\dagger_{\mu,\alpha} \hat{c}^{\pdagger}_{\mu',\alpha} e^{-i\hat{H}\tau} \hat{c}^\dagger_{\nu,\beta} \hat{c}^{\pdagger}_{\nu',\beta} \rangle.
\end{equation}

The DSF involves contributions from processes in which a fermion is annihilated in an initially occupied state $(\nu', \beta)$ and a fermion is simultaneously created in an unoccupied state $(\nu, \beta)$ at $\tau = 0$. This particle-hole excitation persists until time $\tau$, when it is annihilated. This leads to the simplification
\begin{align}
\langle e^{i\hat{H}\tau} \hat{c}^\dagger_{\mu,\alpha} \hat{c}^{\pdagger}_{\mu',\alpha} e^{-i\hat{H}\tau} \hat{c}^\dagger_{\nu,\beta} \hat{c}^{\pdagger}_{\nu',\beta} \rangle &= e^{-i(\epsilon_{\nu}-\epsilon_{\nu'})\tau} \delta^{\pdagger}_{\nu',\mu} \delta^{\pdagger}_{\mu',\nu} \delta^{\pdagger}_{\alpha,\beta}\langle\hat{c}^\dagger_{\mu,\alpha}\hat{c}^{\pdagger}_{\nu',\beta}\rangle\langle \hat{c}^{\pdagger}_{\mu',\alpha}\hat{c}^\dagger_{\nu,\beta} \rangle, \label{eq:dsf_u_2}
\end{align}
which, upon substitution into Eq.~\eqref{eq:dsf_u_1}, yields
\begin{align}
\mathcal{S}^{zz}(\boldsymbol{q},\omega) &= \frac{1}{2N} \sum_{i,j,\mu,\nu} e^{i\boldsymbol{q}\cdot \boldsymbol{r}_{ij}} \delta(\omega - \epsilon^{}_{\nu} + \epsilon^{}_{\mu})   U^*_{i,\mu} U^{}_{i,\nu} U^*_{j,\nu} U^{}_{j,\mu} n^{}_{\mu}(1 - n^{}_{\nu}), \label{eq:dsf_u_3}
\end{align}
where the fermionic occupation number is given by $n_i = [\exp(\beta(\epsilon_i - \epsilon_F) + 1]^{-1}$, with $\epsilon_F$ denoting the Fermi energy.

In the zero-temperature limit, the Fermi distribution becomes a step function, and Eq.~\eqref{eq:dsf_u_3} simplifies to
\begin{align}
\mathcal{S}^{zz}(\boldsymbol{q},\omega) &= \frac{1}{2N} \sum_{i,j,\mu,\nu} e^{i\boldsymbol{q}\cdot \boldsymbol{r}_{ij}} \delta(\omega - \epsilon^{}_{\nu} + \epsilon^{}_{\mu})  U^*_{i,\mu} U^{}_{i,\nu} U^*_{j,\nu} U^{}_{j,\mu} \Theta(\epsilon^{}_F - \epsilon^{}_{\mu}) \Theta(\epsilon^{}_{\nu} - \epsilon^{}_F), \label{eq:dsf_u_4}
\end{align}
where $\Theta(x)$ is the Heaviside step function.
\end{widetext}

Following this calculation, Fig.~\ref{fig:dsf_all} presents the dynamical spin structure factors (DSFs) along high-symmetry momentum paths for the same representative points as in Fig.~\ref{fig:dispersion_all}. All the DSFs exhibit broad continua instead of sharp dispersive features, a hallmark of fractionalized excitations. 
The distribution of the dominant spectral intensity, however, varies significantly depending on the \textit{Ansatz} and the specific phase under consideration, appearing at different energies and momenta across the various cases. 
This sensitivity of the DSF profile offers a concrete experimental handle to distinguish between competing quantum spin liquid states and provides predictions for future experiments on, e.g., the candidate trellis lattice compounds discussed in Sec.~\ref{sec:mat}. It is worth mentioning that since our parton framework is strictly SU(2) spin-rotation invariant (as we do not include an antiferromagnetic order parameter), sharp modes can only appear beyond a mean-field treatment once fermion-fermion interactions are included, e.g., within the random phase approximation (RPA) that has recently been successful in accurately obtaining  dynamical spin structure factors~\cite{Willsher-2025,Rao-2025}.

\begin{figure*}[tb]
    \centering
    \includegraphics[width=1.0\textwidth]{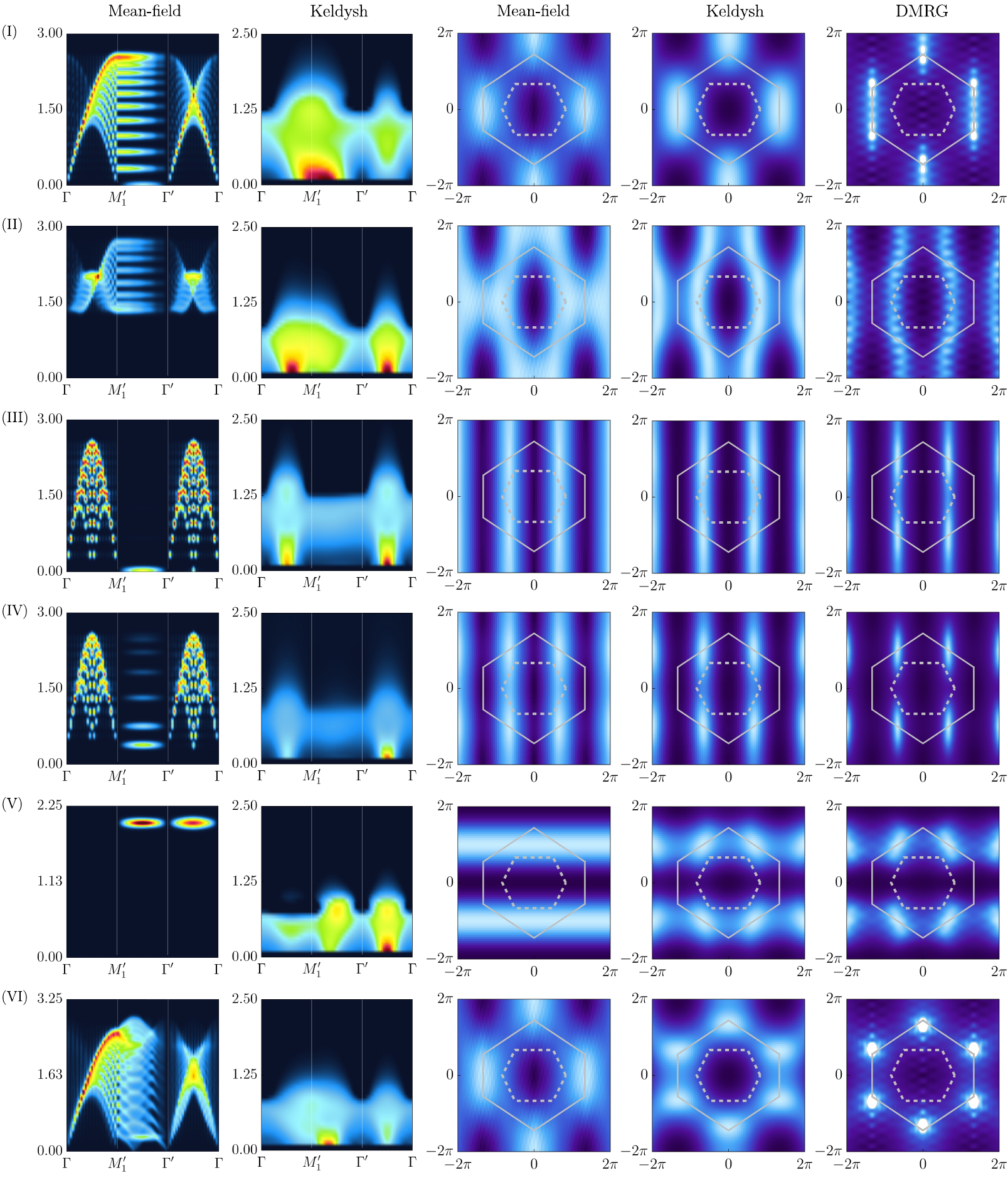}
    \caption{Equal-time and dynamical structure factors obtained within mean-field theory, the Keldysh pf-FRG formalism, and DMRG for the six phases (I)--(VI) defined in Fig.~\ref{fig:phasediagram}. For the mean-field results, the y-axis of the dynamical structure factors denotes the frequency $\omega$ whose scale depends on the $\chi$ parameters optimized with respect to $(J_z,J_h,J_v)$; for the Keldysh results, the y axis is labeled by $\omega/J$, where $J = J_z + J_h + J_v$. }
    \label{fig:keldysh_equal_time}
\end{figure*}

We also explicitly calculate the DSF for the six specific points in the mean-field phase diagram shown in Fig.~\ref{fig:phasediagram}. 
For these points, the self-consistent solutions yield vanishing hopping amplitudes in certain directions, which allows for a transparent interpretation of the $\mathcal{S}^{zz}(\boldsymbol{q},\omega)$ spectra in Fig.~\ref{fig:keldysh_equal_time}.
\begin{itemize}
\item
{Phase I:} Here, $\chi_v$ and $\chi_h$ are vanishingly small compared to $\chi_z$, resulting in effectively decoupled one-dimensional zigzag chains. The spectrum is dispersionless along $k_y$ between $M'$ and $\Gamma'$. Along $k_x$ ($\Gamma$--$M'$), $\mathcal{S}^{zz}(\boldsymbol{q},\omega)$ exhibits the characteristic particle-hole continuum of a one-dimensional cosinelike band.
\item
{Phase II:} In this regime, $\chi_z$ and $\chi_h$ dominate while $\chi_v$ is negligible. The triangular plaquettes carry staggered $0$ and $\pi$ fluxes, producing alternating $\chi_h$ hopping amplitudes. The system organizes into decoupled one-dimensional chains with uniform nearest-neighbor and alternating next-nearest-neighbor hoppings. This alternation opens a band gap, yielding a gapped $\mathcal{S}^{zz}(\boldsymbol{q},\omega)$. The spectrum remains dispersionless along $k_y$.
\item
{Phase III:} When $\chi_z$ and $\chi_v$ are small compared to $\chi_h$, the system reduces to decoupled horizontal chains with cosinelike parton dispersions. In this case, we observe a full particle-hole continuum along $k_x$ and a dispersionless response along $k_y$.
\item
{Phase IV:} This $\pi$-flux decoupled ladder phase arises when the hopping amplitudes along the upper and lower horizontal chains acquire opposite signs, leading to a band gap of $\Delta = 2\chi_v$. The horizontal chains thus exhibit gapped cosinelike dispersions, producing a $\mathcal{S}^{zz}(\boldsymbol{q},\omega)$ similar to phase III, but shifted by the gap. Additionally, finite dispersionless spectral weight emerges between $M'$ and $\Gamma'$ along $k_y$.
\item
{Phase V:} In this limit, horizontal and zigzag hoppings are negligible relative to vertical hopping, reducing the system to a lattice of fully decoupled dimers. The bonding-antibonding energy splitting of the dimers is $2\chi_v$, giving rise to a finite $\mathcal{S}^{zz}(\boldsymbol{q},\omega)$ only at $\omega = 2\chi_v$.
\item
{Phase VI:} With finite $\chi_z$ and $\chi_v$, the system forms a two-dimensional honeycomb lattice featuring Dirac points in the band structure. Consequently, $\mathcal{S}^{zz}(\boldsymbol{q},\omega)$ is gapless, with dispersive features along both $k_x$ and $k_y$.
\end{itemize}

Additionally, we calculate the equal-time momentum-resolved spin-spin correlation function, also known as the EQSF, which is relevant to elastic neutron scattering. This quantity can be obtained from Eq.~\eqref{eq:dsf_def} by summing the DSF over all frequencies:
$
\mathcal{S}^{zz}(\boldsymbol{q}) \equiv \sum_{\omega} \mathcal{S}^{zz}(\boldsymbol{q}, \omega),
$
leading to the final expression
\begin{align}
\mathcal{S}^{zz}(\boldsymbol{q}) &= \frac{1}{2N} \sum_{i,j,\mu,\nu} e^{i\boldsymbol{q}\cdot \boldsymbol{r}_{ij}} U^*_{i,\mu} U^{}_{i,\nu} U^*_{j,\nu} U^{}_{j,\mu} \notag\\
&\times \Theta(\epsilon^{}_F - \epsilon^{}_{\mu}) \Theta(\epsilon^{}_{\nu} - \epsilon^{}_F). \label{eq:ssf}
\end{align}
The resulting EQSFs are shown in Fig.~\ref{fig:ssf_all}, evaluated at the same representative parameter points used for the dispersions in Fig.~\ref{fig:dispersion_all} and the DSF plots in Fig.~\ref{fig:dsf_all}. Once again, we observe a smooth distribution of spectral weight in the Brillouin zone and an absence of sharp Bragg peaks, as expected for quantum paramagnetic states.

\subsection{Keldysh pf-FRG analysis}

To complement the mean-field results for the dynamical and equal-time spin structure factors, we employ the recently developed Keldysh extension~\cite{maity2024_3,Potten2025} of the pf-FRG~\cite{Mueller2024}. The core idea of this method is to formulate a renormalization group flow for the frequency-dependent effective interactions and onsite self-energies of the Abrikosov pseudofermions, as defined in Eq.~\eqref{spinon}, by introducing a cutoff scale $\Lambda$ that suppresses low-energy dynamics. At large $\Lambda$, the system is described by the bare microscopic Hamiltonian, while gradually lowering $\Lambda$ drives the flow toward the fully interacting regime, which is recovered at $\Lambda = 0$.

By utilizing the Keldysh formalism---originally developed for nonequilibrium systems---in an equilibrium setting at finite temperature $T$, the method provides direct access to the real-frequency structure of the interaction vertices. This circumvents the need for an ill-conditioned analytical continuation from imaginary (Matsubara) frequencies, as would be required in conventional equilibrium formulations.

From the flowing pf-FRG vertices and self-energies, we compute both the dynamical and equal-time spin structure factors for the full spin-$1/2$ system. It is important to note that within this formalism, excitations near the $\Gamma$ point are naturally suppressed, particularly in regimes close to long-range magnetic order~\cite{Potten2025}.

For the numerical calculations, we set the inverse temperature to $\beta = 1/T = 10/J$, where $J = J_z + J_h + J_v$ defines the overall energy scale of the system. Correlations are computed up to a spatial extent of 4 lattice spacings along the $\boldsymbol{T}_1$ direction and 2 along the $\boldsymbol{T}_2$ direction, resulting in a total of 41 symmetry-inequivalent lattice sites. The frequency axis is discretized using a linear-logarithmic mesh comprising 43 non-negative frequencies.

All pf-FRG flows remain smooth down to $\Lambda = 0$, except for the ladder phase, where a kink in the flow at $\Lambda_c = 0.05$ signals the onset of long-range magnetic order. Accordingly, for this case, we present the spin structure factor at $\Lambda_c$, while for all other cases the structure factors are shown in the fully renormalized ($\Lambda \to 0$) limit.

The DSFs and EQSFs thus obtained from our analysis are arrayed in Fig.~\ref{fig:keldysh_equal_time} for the six phases identified in the fully symmetric ternary phase diagram, alongside the corresponding mean-field results. In most cases, we find excellent qualitative agreement between the EQSFs derived from our PSG \textit{Ans\"atze} and those obtained from the beyond-mean-field Keldysh approach.  

For the DSFs, however, qualitative differences arise between the mean-field and Keldysh results. These discrepancies can be traced to two principal limitations of the Keldysh pf-FRG~\cite{Potten2025}: (1) Its inability to resolve spectral gaps when the associated order parameter is not explicitly included in the flow and (2) the absence of downward-dispersing branches toward $q=0$ in systems with antiferromagnetic order.  
The first limitation is generic to FRG approaches: If an order parameter is not explicitly incorporated into the flow equations, it cannot acquire a finite value, leading in practice to gapless spectra. Implementing a general order parameter that captures spectral gaps is nontrivial, as it cannot be introduced in a system-independent manner~\cite{Potten2025}. The second limitation is more severe and stems from the systematic suppression of long-range order within pf-FRG. In this framework, transitions from the initial paramagnetic state to an ordered phase are always accompanied by a diverging susceptibility, which would necessitate a divergence in the vertex function. Since the Keldysh pf-FRG relies on avoiding divergences to fully remove cutoff dependences, one is constrained to remain in the initial paramagnetic regime, preventing the correct reproduction of long-range features. Furthermore, the fermionic decomposition of spin operators introduces additional inaccuracies, as discussed in detail in Ref.~\onlinecite{Potten2025}.  
Despite these caveats, the Keldysh pf-FRG remains a powerful method for resolving the dominant low-energy excitations around the ground state. This makes it particularly well suited for spin liquid candidates, where long-range order is absent and alternative approaches may fail.  

Taking these limitations into account, we find that phases I and III show good agreement in terms of the principal low-energy excitations. Phases II and IV also display the expected qualitative features, aside from the absence of a spectral gap, which follows directly from limitation (1). More significant deviations from PSG-based expectations appear in phases V and VI. A comparison of the dynamical structure factors together with the equal-time spin structure factors with mean-field results suggests that (1) either beyond–mean-field corrections could account for these discrepancies, such as by incorporating fermion-fermion interactions, e.g., within an RPA framework~\cite{Willsher-2025,Rao-2025} or treating gauge fluctuations via Gutzwiller projection, or (2) these could be signatures of incipient magnetic orders that are captured in DMRG and Keldysh pf-FRG.  

In summary, while the Keldysh pf-FRG has intrinsic limitations, it remains a valuable tool for characterizing the main low-energy excitations beyond mean-field theory, provided that the above constraints are properly considered.

\subsection{DMRG calculations}

\begin{figure*}[tb]
\includegraphics[width=0.8\linewidth]{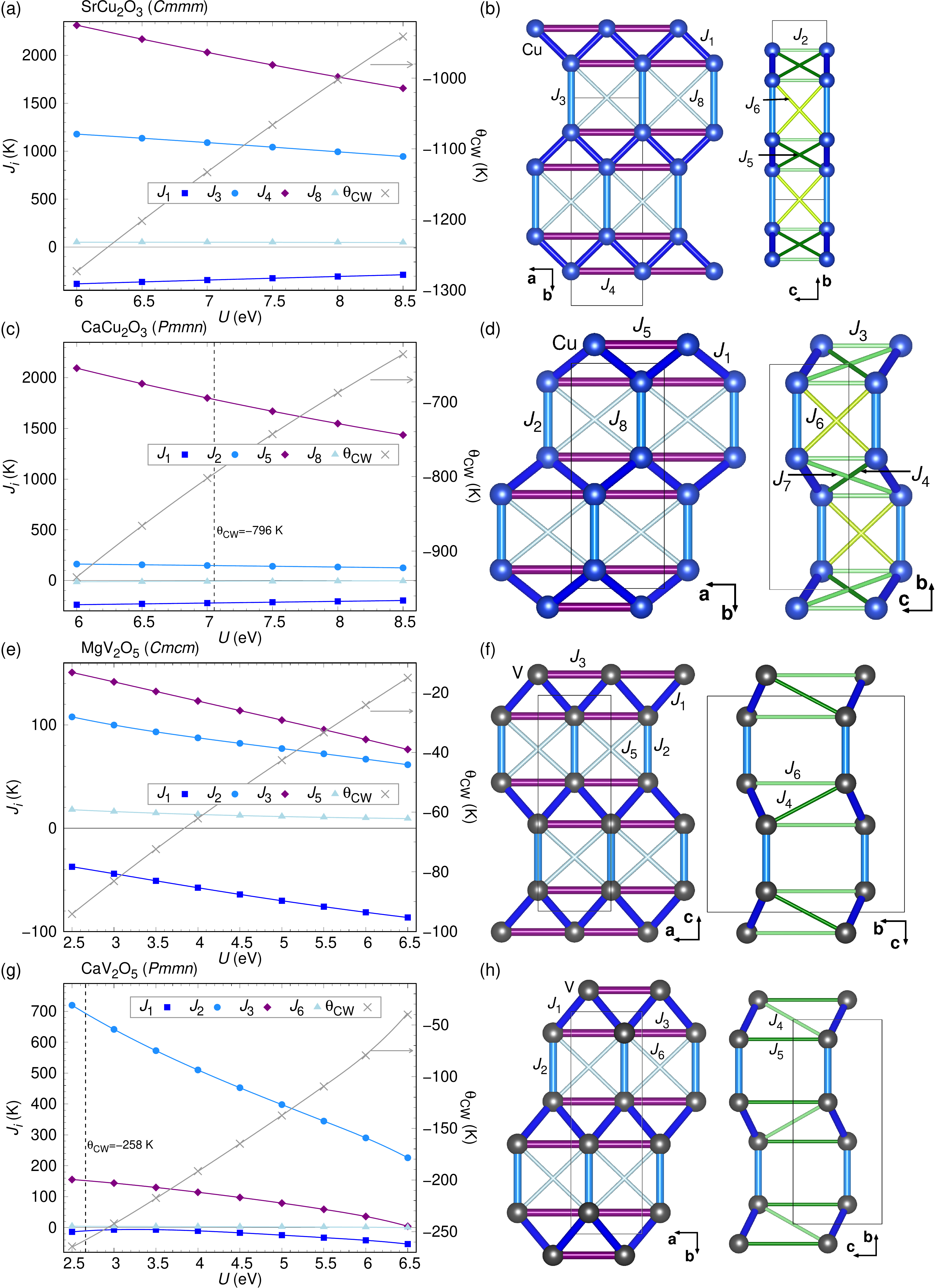}
\caption{DFT energy mapping results and Curie-Weiss fits for the trellis lattice compounds (a), (b) \ce{SrCu2O3}, (c), (d) \ce{CaCu2O3}, (e), (f) \ce{MgV2O5}, and (g), (h) \ce{CaV2O5}.}
\label{fig:mat}
\end{figure*}

Another unbiased approach for the calculation of spin correlation functions is DMRG~\cite{white1992density,white1993density,schollwock2005density}. Unlike Keldysh pf-FRG computations, which operate at small but finite temperatures, DMRG directly accesses the zero-temperature ground-state properties of the system.  

Our DMRG calculations are, in practice, based on the variational optimization of a matrix product state (MPS) wave function. Once such an MPS representation of the ground state is obtained for a given set of Hamiltonian parameters, the spin correlation functions can be readily computed. 
DMRG is particularly powerful for one-dimensional systems, for which it is known to provide extremely accurate results. In our context, this is advantageous for certain phases (e.g., I, III, and IV  in the mean-field phase diagram of Fig.~\ref{fig:phasediagram}), which become effectively quasi-one-dimensional due to the reduced couplings along some directions (refer to Table~\ref{tab:phases_hopping}).

Here, we simulate the trellis lattice system on a cylindrical geometry with open boundary conditions along $\boldsymbol{T}_1$ and periodic boundary conditions along the $\boldsymbol{T}_2$ direction\footnote{Fully periodic boundaries would require a quadratic increase in the number of states needed for the same accuracy, making the computation prohibitively expensive.}. A well-known limitation of two-dimensional DMRG is that the number of states required grows exponentially with the system width~\cite{schollwock2005density}, which restricts accessible system sizes. In this work, we consider a $10\times 5$ system (i.e., with up to 10 sites in the periodic direction), with a maximum MPS bond dimension of $2400$. Our sweeping strategy involves starting with many sweeps at relatively small bond dimensions and gradually increasing the bond dimension until convergence is achieved. Convergence is defined as the point where the truncation error falls below $10^{-5}$, which is typically satisfied after $\mathcal{O}(10^2)$ sweeps. To facilitate the buildup of long-range correlations, we also add a small noise term to the density matrix during the early sweeps, which is subsequently turned off~\cite{white2005density}.

The resulting equal-time structure factors, shown in the rightmost column of Fig.~\ref{fig:keldysh_equal_time}, display excellent agreement with the Keldysh pf-FRG results across all six phases (I--VI). In particular, the DMRG calculations capture features that lie beyond mean-field theory, which are also visible in the pf-FRG spectra, albeit broadened due to the finite temperatures intrinsic to the latter formalism.

\section{Material realizations}
\label{sec:mat}

The rich physics of magnetic frustration on the trellis lattice that we have discussed so far can also, in principle, be explored in solid-state materials. 
To this end, we introduce four compounds---\ce{SrCu2O3}, \ce{CaCu2O3}, \ce{MgV2O5}, and \ce{CaV2O5}---that are known to realize the trellis lattice geometry. Both Cu$^{2+}$ and V$^{4+}$ ions carry spin-$1/2$. To avoid confusion, we label the exchange couplings in order of strictly increasing distances between magnetic sites. In the various realizations of the trellis lattice, the three main couplings---namely, the chain ($J_h$), rung ($J_v$), and sawtooth ($J_z$) interactions---are realized through different intersite distances depending on the compound. In Fig.~\ref{fig:mat}, the colors purple, light blue, and blue are used to denote $J_h$, $J_v$, and $J_z$, respectively.

\subsection{Density functional theory}

For the four materials, we apply a DFT-based energy-mapping approach that has been successfully used to understand many low-dimensional copper-based quantum magnets~\cite{Yamamoto2021,Hering2022,Fujihara2022}. Our approach employs all electron electronic structure calculations using the full potential local orbital basis set~\cite{Koepernik1999} with a generalized gradient approximation exchange correlation potential~\cite{Perdew1996}. Strong electronic correlations on Cu$^{2+}$ and V$^{4+}$ $3d$ orbitals are treated with a DFT+$U$ correction~\cite{Liechtenstein1995}. This correction has two parameters, the on-site interaction $U$ and the Hund's coupling $J_{\rm H}$. As the latter is not expected to be strongly material dependent, we fix it to be $J_{\rm H}=1$\,eV for Cu~\cite{Jeschke2013} and $J_{\rm H}=0.68$\,eV for V~\cite{Mizokawa1996}. The interaction strength $U$ determines the energy scale of the magnetic interactions and is taken from experimental Curie-Weiss temperatures whenever possible. For the energy mapping, we classify all spin configurations in low-symmetry supercell structures and map large sets of DFT+$U$ total energies to the Heisenberg Hamiltonian written as $\hat{\mathcal{H}}= \sum_{i<j} J_{i j} \hat{\mathbf{S}}_i\cdot\hat{\mathbf{S}}_j$, where every bond is counted only once. Rather than making assumptions about an important subset of exchange couplings, we calculate between 9 and 12 interactions for each of the four materials, and the result flags the important couplings in an unbiased fashion; in most cases, we also gather information about exchange paths coupling the trellis layers, which are important for possible magnetic orders.

\ce{SrCu2O3} is a cuprate compound exhibiting a flat trellis lattice [Fig.~\ref{fig:mat}(b)], with couplings identified as $J_h \equiv J_4$, $J_v \equiv J_3$, and $J_z \equiv J_1$. We perform our calculations using the $T=100$\,K structure with $Cmmm$ space group, as determined by single-crystal x-ray diffraction~\cite{Sparta2006}. The DFT energy mapping reveals a system composed of robust two-leg $J_h$-$J_v$ ladders, with $J_h \approx 2J_v$. The zigzag coupling $J_z$ is found to be weaker and ferromagnetic. Reported spin gap values include $\Delta = 420$\,K based on susceptibility data~\cite{Azuma1994}, and $\Delta = 680$\,K from Cu NMR~\cite{Ishida1994}. The susceptibility presented in Ref.~\onlinecite{Azuma1994} is challenging to interpret via a traditional Curie-Weiss fit, as $\chi^{-1}$ exhibits a negative slope at high temperatures. Using the refined Curie-Weiss fitting procedure from Ref.~\onlinecite{Pohle2023}, we obtain $\theta_{\rm CW} = -626$\,K; however, this result is unreliable due to the limited temperature range (data only extend to 650\,K). Efforts to determine accurate Hamiltonian parameters have employed simplified models, such as the two-leg ladder approximation~\cite{Troyer1994,Sandvik1996,Johnston1996}. Among the computed Hamiltonians, the one obtained with $U = 8.5$\,eV appears most consistent with the expectation that $J_h$ should be approximately twice the spin gap, i.e., around 1400\,K. The extracted values are $J_h = J_4 = 1656(14)$\,K, $J_v = J_3 = 945(16)$\,K, and $J_z$\,$=$\,$J_1$\,$=$\,$-289(27)$\,K. Besides the three trellis lattice couplings, we obtain the diagonal coupling in the ladder $J_8=48(14)$\,K as well as interlayer couplings $J_2=41(22)$\,K, $J_5=-9(23)$\,K, $J_6=14(9)$\,K and $J_7=-17(13)$\,K that are small compared to the dominant interaction. For the rung-to-chain ratio $J_v/J_h$, we have $J_v/J_h=0.57$, which compares well to the value $J_v/J_h=0.5$ favored by \citet{Johnston1996} after weighing various experimental facts. 
Our result can be compared with an early {\it ab initio} study~\cite{deGraaf1999} that reports $J_h=1870$\,K, $J_v=1670$\,K, $J_1 = -145$\,K and thus a rather large rung-to-chain ratio of $J_v/J_h=0.89$. On the other hand, our results are in good agreement with local density approximation+U (LDA+\textit{U}) calculations reported in Ref.~\onlinecite{Johnston} yielding $J_h=1795$\,K, $J_v=809$\,K, $J_1 = -200$\,K. 

\begin{figure*}[tb]
    \includegraphics[width=1.0\textwidth]{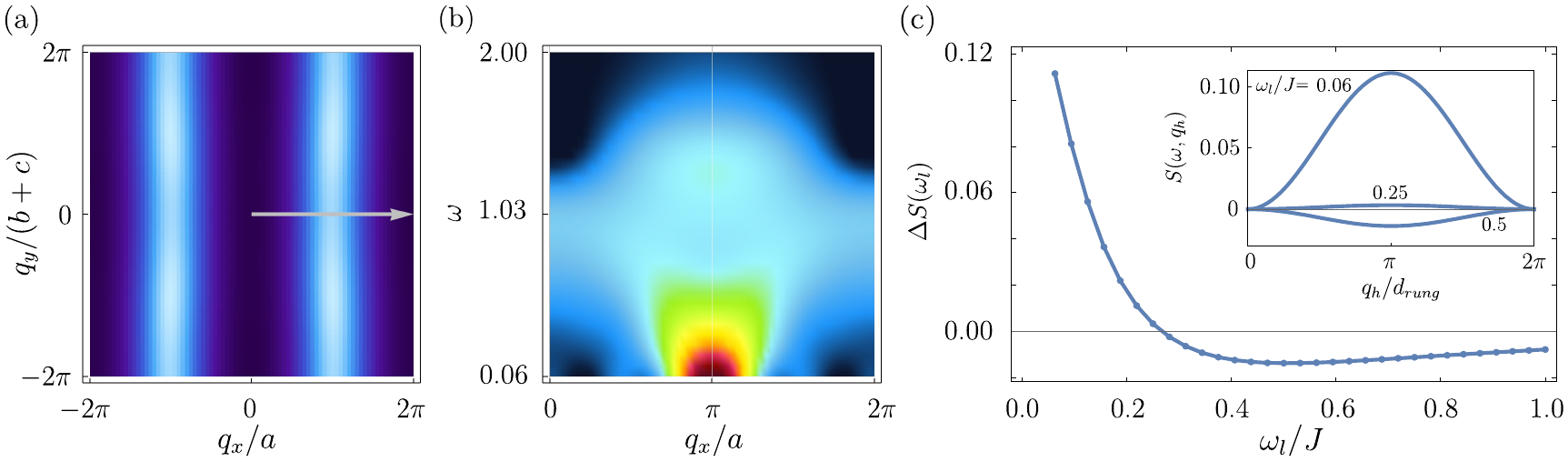}
    \caption{Keldysh pf-FRG results for \ce{CaCu2O3} obtained for the DFT model Hamiltonian with $J_h = 1786(9)$\,K, $J_v = 130(16)$\,K, and $J_z = -228(15)$\,K. 
    (a) Equal-time spin structure factor with $a,b$, and $c$ defined in Fig.~\ref{fig:lattice}. The nearly uniform behavior in the $q_y\equiv q_h$ (horizontal) direction supports a description in terms of one-dimensional chains along $\bm{T}_1$.  
    (b) Dynamical spin structure factor along the $q_x$ direction. The fractionalization dome of the spin chain is reproduced up to the branch at $q=0$, which the Keldysh pf-FRG cannot resolve.  
    (c) Confinement analysis following Ref.~\onlinecite{Lake2010}, calculated using Eq.~(\ref{eq:confinement}). The modulation is controlled by $1/q_\text{rung}$ and rapidly decays to zero. Inset: $q$-resolved modulation for selected data points. For $\omega_l>0.27$, the amplitude becomes negative, shifting the $q_y$ modulation to $q_y=0$, corresponding to an enhancement of the bonding susceptibility channel \cite{Lake2010,CuprateLadders}. }   
	\label{fig:CaCu2O3}           
\end{figure*}

\ce{CaCu2O3} exhibits a buckled trellis lattice [Fig.~\ref{fig:mat}(d)]. We use the $T=10$\,K structure with $Pnma$ space group obtained via neutron diffraction~\cite{Kiryukhin2001}. Here, the exchange couplings are given by $J_h \equiv J_5$, $J_v \equiv J_2$, and $J_z \equiv J_1$. Energy mapping shows that the antiferromagnetic chain coupling $J_h$ dominates over the other two. Using the Curie-Weiss fitting method of Ref.~\onlinecite{Pohle2023} to fit the susceptibility from Ref.~\onlinecite{Kiryukhin2001} [see Fig.~\ref{fig:mat}(c)], we find $\theta_{\rm CW} = -796$\,K. In contrast, a conventional $\chi^{-1}$ fit misleadingly suggests $\theta_{\rm CW} \approx 0$, inconsistent with the estimated spin chain scale $J_h\equiv J_{\parallel} = 2000(300)$\,K~\cite{Kiryukhin2001}. At this significant antiferromagnetic Curie-Weiss temperature, indicated by the dashed line in Fig.~\ref{fig:mat}(c), the extracted couplings are $J_h = J_5 = 1786(9)$\,K, $J_v = J_2 = 130(16)$\,K, and $J_z = J_1 = -228(15)$\,K. These results are consistent with the previous estimate of $J_{\parallel}$ within error bars. An inelastic neutron scattering study by~\citet{Lake2010} determined $J_h = 1880$\,K, supporting earlier quantum chemistry predictions~\cite{Calzado2003,Bordas2005} that $J_v$ is significantly smaller than $J_h$. To account for the incommensurate magnetic order observed at $T_{\rm N} = 25$\,K~\cite{Kiryukhin2001}, interlayer couplings would be important, but in the present study, at $J_3$\,$=$\,$-7(9)$\,K, $J_4 = -6(17)$\,K, and $J_6 = -12(9)$\,K, we have not been able to suppress the error bars sufficiently.

\ce{MgV2O5} is another material that realizes a buckled trellis lattice. Our calculations are based on the $T=83$\,K powder x-ray diffraction structure with $Cmcm$ space group~\cite{Millet1998}. In this case, the relevant couplings are $J_h \equiv J_3$, $J_v \equiv J_2$, and $J_z \equiv J_1$ [Fig.~\ref{fig:mat}(f)]. DFT energy mapping indicates a two-leg ladder model, with the rung coupling $J_v$ being comparable in strength to the chain coupling $J_h$, and with the sawtooth coupling $J_z$ being ferromagnetic. The Curie-Weiss temperature has been reported as $\theta_{\rm CW} = -174$\,K, although this value is sensitive to the choice of the constant van-Vleck contribution $\chi_0$. Varying $\chi_0$ between 0 and $0.7 \times 10^{-5}$\,emu/mol yields $\theta_{\rm CW}$ values ranging from $-205$\,K to $-77$\,K, with all $\chi^{-1}$ fits appearing equally satisfactory. Since onsite Coulomb interaction values much below $U = 2.5$\,eV are difficult to justify for V$^{4+}$, we consider the $U = 3$\,eV set of couplings: $J_h$\,$=$\,$J_3$\,$=$\,$147(1)$\,K, $J_v$\,$=$\,$J_2$\,$=$\,$100(1)$\,K, and $J_z $\,$=$\,$J_1$\,$=$\,$-44(1)$\,K (corresponding to $\theta_{\rm CW} = -83$\,K and a rung-to-chain ratio $J_v/J_h=0.68$). Alternatively, for $U = 2.5$\,eV, we find $J_h = J_3 = 151(1)$\,K, $J_v = J_2 = 108(1)$\,K, and $J_z = J_1 = -37(1)$\,K, with $\theta_{\rm CW} = -94$\,K and $J_v/J_h=0.72$. A perturbation-theory estimate of the couplings based on LDA+\textit{U} within the linear muffin-tin orbital basis~\cite{Korotin1999,Korotin2000} reported $J_h = 144$\,K, $J_v = 92$\,K, $J_z = 60$\,K which is very similar for the ladder but has a different sign for the zigzag coupling.

The fourth compound that we analyze is \ce{CaV2O5}, which also realizes a buckled trellis lattice. We base our calculations on the room-temperature x-ray powder diffraction structure with $Pmmn$ space group~\cite{Onoda1996} [Fig.~\ref{fig:mat}(h)]. Here, the relevant couplings are $J_h \equiv J_3$, $J_v \equiv J_2$, and $J_z \equiv J_1$. Fitting the susceptibility data from Ref.~\onlinecite{Onoda1996} using the Curie-Weiss approach of Ref.~\onlinecite{Pohle2023} [Fig.~\ref{fig:mat}(g)], we obtain $\theta_{\rm CW} = -258$\,K. This result suggests, similarly to \ce{MgV2O5}, that a relatively small $U$ is appropriate for this compound. The resulting couplings are $J_h = J_3 = 152(3)$\,K, $J_v = J_2 = 693(3)$\,K, and $J_z = J_1 = -11(3)$\,K. This places \ce{CaV2O5} in the regime of a two-leg ladder with very strong rungs, characterized by $J_v / J_h = 4.5$, consistent with the scenario proposed by~\citet{Millet1998}.
A perturbative estimate~\cite{Korotin1999,Korotin2000} was reported to be $J_h = 122$\,K, $J_v = 602$\,K, $J_z = -28$\,K which has a rung-to-chain ratio $J_v/J_h=4.9$. Various other models have been suggested, usually also in the strong-rung limit~\cite{Koo1999,deGraaf2004,Ming2010}. A QMC study~\cite{Miyahara_1998} arrived at a set of interactions $J_h = 135$\,K, $J_v = 665$\,K, and $J_z = -25$\,K which is quite close to our {\it ab initio} result. As far as subleading couplings are concerned, the diagonals in the ladder $J_6=18(4)$\,K are the most important, and interlayer couplings such as $J_4=1(2)$\,K and $J_7=4(2)$\,K are below 1{\%} of the dominant interaction. This makes \ce{CaV2O5} a good approximation of a two-dimensional material.

\subsection{Keldysh pf-FRG}

In this section, we employ the Keldysh formalism combined with the DFT-derived couplings to analyze the materials discussed above.

We begin with \ce{CaCu2O3}, which—due to its weak interchain couplings $J_2\equiv J_z$ and $J_3\equiv J_v$—is close to being a one-dimensional material.  This interpretation is supported by the equal-time spin structure factor in Fig.~\ref{fig:CaCu2O3}, which exhibits nearly uniform behavior along $q_y$. Consequently, the system effectively realizes one-dimensional chains along $\bm{T}_1$. The cyclic exchange $J_\text{cycl}$~\cite{TrellisStrongConfinement} cannot be modeled here, as it requires quartic spin interactions that lie beyond the truncation of the FRG. While this four-spin term frustrates the spin chains, its strength in \ce{CaCu2O3} is relatively weak ($J_\text{cycl}=4$\,meV~\cite{Lake2010}) compared to the approximate threshold $J_\text{cycl}/J_1 \approx 0.25$ for disrupting binding~\cite{TrellisStrongConfinement}, implying that neglecting it should introduce only minor discrepancies.
The dynamical spin structure factor in Fig.~\ref{fig:CaCu2O3}(b) qualitatively matches the experiment results~\cite{Lake2010}. The absence of the $q=0$ branch, a limitation of the Keldysh formalism, is not critical, since this feature is typically unresolved experimentally as well. Up to an expected mismatch in the frequency scale---intrinsic to qualitative methods such as pf-FRG---our description is broadly consistent with the data.  

\begin{figure*}[tb]
    \includegraphics[width=1.0\textwidth]{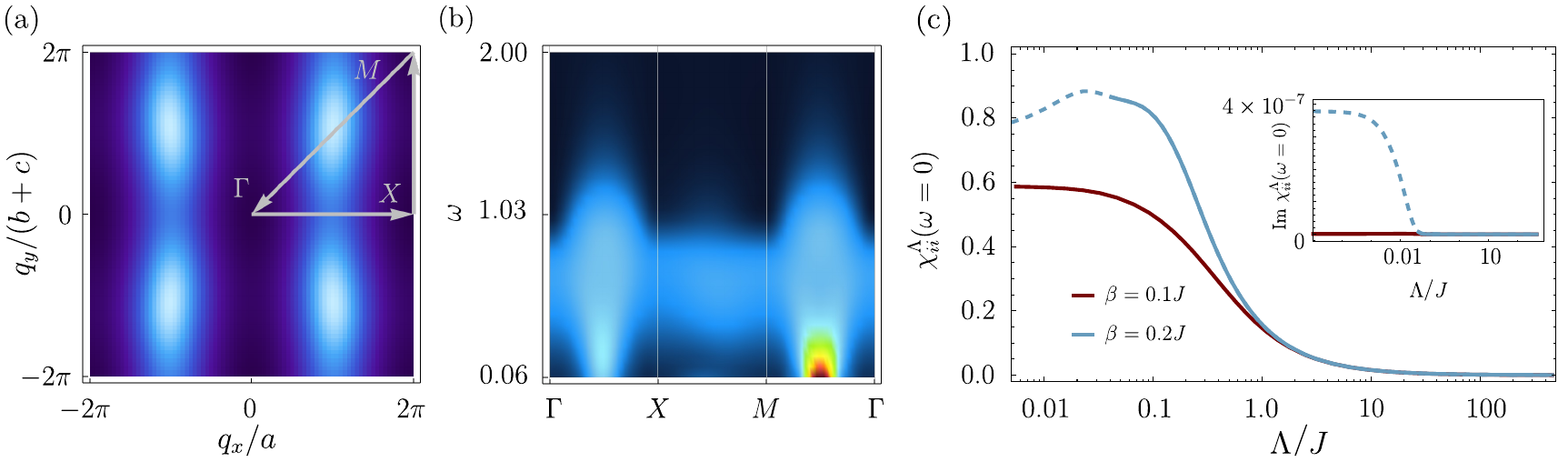}
    \caption{Keldysh pf-FRG results for \ce{SrCu2O3}. (a) Equal-time spin structure factor, with $a,b$, and $c$ defined in Fig.~\ref{fig:lattice}. The result captures the two-legged ladder structure and resembles phase (IV) in Fig.~\ref{fig:keldysh_equal_time}. (b) Dynamical spin structure factor, similar to that of the rung ladder. (c) FRG flow of the largest susceptibility component, showing a weak tendency toward magnetic ordering at $\beta=0.1J$. Both the kink in the flow and the symmetry violation in the imaginary part are small compared to magnetically ordered systems~\cite{Potten2025}.}   
	\label{fig:SrCu2O3}  
\end{figure*}

To further test this agreement, we investigate the measured difference between
strongly and weakly coupled regimes of the spin
chains. Strong coupling produces a modulation along the interchain direction at $1/Q_\text{rung}$, which should vanish in the weak-coupling case. \citet{Lake2010} introduced integrated frequency slices to quantify deviations from the expected $(1+\cos(Q_\text{rung}))\chi_{b}+(1-\cos(Q_\text{rung}))\chi_{ab}$ form, dominated experimentally by the anti-bonding susceptibility $\chi_{ab}$. Since we cannot disentangle bonding ($\chi_b$) and anti-bonding ($\chi_{ab}$) susceptibilities directly, we interpret the full excitation spectrum instead. 
Figure~\ref{fig:CaCu2O3}(c) shows the integrated dynamical susceptibility,
\begin{align}\label{eq:confinement}
	\Delta S(\omega_l) = \int_{\omega_l}^{\omega_u} \Big[ S(\omega,q^{}_{h,\text{max}}) - S(\omega,q^{}_{h,\text{min}}) \Big] \, \mathrm{d}\omega ,
\end{align}
with a constant offset removed. Here, we use $q_{h,\text{max}}=(\pi,\pi/d_\text{rung})^T$, $q_{h,\text{min}}=(\pi,0)^T$, and $\omega_u = 5J$ sets the upper frequency cutoff. This choice captures the essential modulation falloff without relying on arbitrary frequency windows. The correlations decay rapidly to zero around $\omega \approx 0.27$, followed by a sign reversal at larger $\omega/J$, indicating that $\chi_b$ eventually dominates over $\chi_{ab}$. We interpret this as a remnant of strong chain interactions persisting at higher frequencies, with a small amplitude, potentially enhancing the singlet channel. Nevertheless, the Keldysh pf-FRG is, in principle, capable of capturing triplons that induce $\chi_b$-dominated modulation at low frequencies, extending beyond mean-field analyses.

For the remaining trellis lattice materials, no experimental benchmarks for the dynamical spin structure factor $\mathcal{S}(\boldsymbol{q},\omega)$ yet exist. We therefore use the Keldysh pf-FRG calculations to make predictions. As discussed in Ref.~\onlinecite{Potten2025}, it should be kept in mind that potential gaps and exponentially suppressed branches at $q=0$ are not resolved, restricting our predictions to the low-energy sector.  

For \ce{SrCu2O3}, the strongest couplings are again along the chains, but significant interchain couplings convert the system into coupled two-leg ladders \cite{SrCuOLadder}.  This is reflected in the equal-time spin structure factor shown in Fig.~\ref{fig:SrCu2O3}(a), which exhibits two distinct maxima along $q_y$, consistent with phase IV in Fig.~\ref{fig:keldysh_equal_time}. The experimentally known gap of $E_\text{gap}\sim 36$\,meV~\cite{Azuma1994} lies far below the Keldysh resolution ($E_\text{gap}/J \approx 0.015$). Moreover, doping rapidly introduces in-gap states and suppresses singlet-triplet excitations~\cite{SrCuOGapDoping}. We therefore consider the influence of the gap to be small and expect the remaining excitation features to be well captured by the Keldysh pf-FRG. The dynamical spin structure factor Fig.~\ref{fig:SrCu2O3}(b) closely resembles that of the rung-ladder phase (phase IV in Fig.~\ref{fig:phasediagram}). Its mean-field counterpart corresponds to a quadratic hopping Hamiltonian with $\pi$-flux per square plaquette, obtained as a saddle point of the parent U(1) \textit{Ans\"atze} U2, U3, and U6 in Fig.~\ref{fig:u1_ansatz}.

It is worth noting that the Keldysh FRG flow does not monotonically continue to vanishing cutoff scales. Instead, as shown in Fig.~\ref{fig:SrCu2O3}(c), the imaginary part of the susceptibility begins to grow precisely at the position of the kink, which we regard as a signal to terminate the flow, since it should vanish by symmetry.
In the pf-FRG framework, such a breakdown serve as an indicator for the onset of magnetic ordering: The FRG cannot flow into a symmetry-broken phase and therefore, the susceptibility diverges at the corresponding phase boundary \cite{Mueller2024,ReutherFirstFRG}. This behavior can also occur in the vicinity of symmetry-broken phases---even though a purely two-dimensional model should not order at finite temperature as dictated by the Hohenberg-Mermin-Wagner theorem---which is why we observe these ordering tendencies. 
Compared to strongly ordered systems, however, the kink seen here is relatively small. This interpretation is supported by the fact that raising the temperature removes the kink entirely, allowing the flow to proceed down to $\Lambda = 0$. Given that the cutoff at the divergence is already small, we use the $\beta = 0.1 J$ data above the nonmonotonicity for the equal-time and dynamic spin structure factors, thereby capturing both the incipient ordering tendencies and the underlying spin liquid behavior, which lie beyond the scope of a pure mean-field analysis.

\begin{figure}[tb]
    \includegraphics[width=1.0\textwidth]{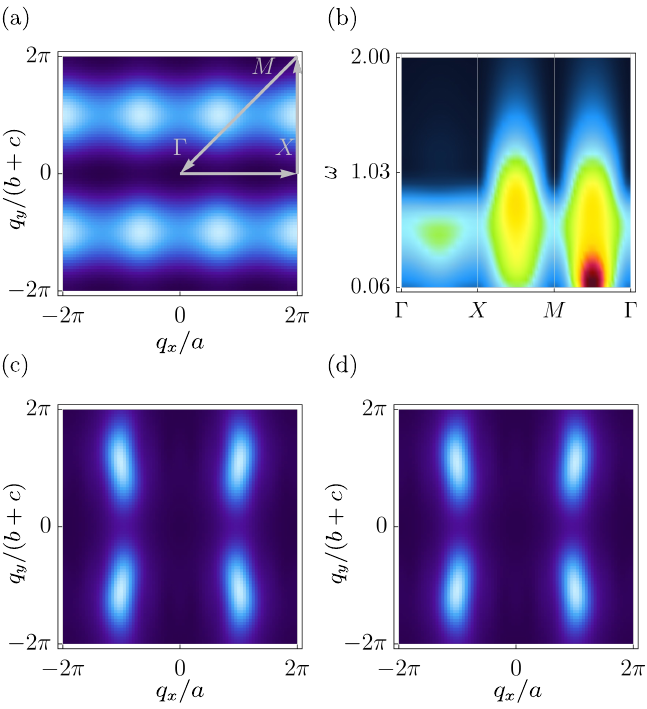}
    \caption{(a), (b) Keldysh pf-FRG results for \ce{CaV2O5}. (a) Equal-time spin structure factor, with $a,b$, and $c$ defined in Fig.~\ref{fig:lattice}. Strong rung coupling ($J_v/J_h = 4.5$) leads to dominant modulation along $q_y$, consistent with ladder-dimer phase (V) in Fig.~\ref{fig:phasediagram}. (b) Dynamical spin structure factor, displaying a broader, more featureless spectrum compared to \ce{SrCu2O3}, consistent with a higher breakdown scale $\Lambda \approx 0.08$. This broadening reflects the finite cutoff rather than fractionalization. Equal-time spin structure factors for \ce{MgV2O5} at (c) $U=3$\,eV and (d) $U=2.5$\,eV. Both exhibit strong ordering tendencies persisting at finite temperature, precluding dynamical analysis. While the equal-time correlations appear similar, the flow shows much stronger growth of the symmetry-breaking channel at $U=2.5$\,eV, indicating enhanced ordering that cannot be resolved in the equal-time limit.}
    \label{fig:remainingMaterials}
\end{figure}

For \ce{CaV2O5}, DFT results identify a two-legged ladder, similar to \ce{SrCu2O3}, but with significantly stronger rung couplings ($J_v/J_h=4.5$). Accordingly, the equal-time spin structure factor [Fig.~\ref{fig:remainingMaterials}(a)] shows dominant modulation along $q_y$, consistent with the ladder-dimer phase V in Fig.~\ref{fig:phasediagram}. The dynamical spin structure factor [Fig.~\ref{fig:remainingMaterials}(b)] shares similarities with that of \ce{SrCu2O3}, though the spectral weight is more broadly distributed beyond $\omega=0$. This broadening stems from the higher breakdown scale $\Lambda \approx 0.08$, about twice that of \ce{SrCu2O3}. While there is no direct correspondence between the cutoff and a transition temperature or an energy scale~\cite{Mueller2024}, a larger divergence scale qualitatively reflects stronger ordering tendencies. Consequently, this finite cutoff still manifests in the dynamical spin structure factor and leads to an exponential broadening of all $\omega \neq 0$ excitations, unrelated to any intrinsic signatures of spin liquid behavior.

Finally, for \ce{MgV2O5}, we consider two viable DFT configurations, with $U$\,$=$\,$3$\,eV and $U$\,$=$\,$2.5$\,eV. Both exhibit strong ordering tendencies within pf-FRG that persist even at elevated temperatures. We therefore refrain from presenting the dynamical spin structure factor, as ordering effects and the cutoff-related broadening reduce the applicability of the Keldysh pf-FRG beyond the typical ground-state analysis known from standard Matsubara pf-FRG (and the pseudo-Majorana functional renormalization group). The corresponding equal-time spin structure factors [Figs.~\ref{fig:remainingMaterials}(c) and ~\ref{fig:remainingMaterials}(d)] appear nearly identical, suggesting similar ground-state properties. However, the flow analysis reveals that the imaginary part of susceptibility, signifying the symmetry-breaking tendency, grows 30 times larger for $U=2.5$\,eV than for $U=3$\,eV, indicating a stronger ordering propensity. In both cases, the systems appear magnetically ordered, and further analysis should therefore employ methods such as linear spin-wave theory, which are suited for dynamic correlations on top of magnetically ordered states. 

\section{Discussion and outlook}
\label{sec:end}

Our work introduces a unique class of QSLs, namely, semi-Dirac spin liquids. The fermionic parton constructions presented here provide a foundation for studying the role of U(1) gauge fluctuations beyond the mean-field approximation. For conventional Dirac spin liquids, the impact of gauge fluctuations has been extensively analyzed, particularly on triangular and kagome lattices, where such states are regarded as excellent variational candidates for the ground state of the spin-$1/2$ Heisenberg antiferromagnet~\cite{Hermele-2008,yasir2013,yasir2016}. In these systems, gauge fluctuations significantly modify spectroscopic signatures~\cite{Ferrari-2019,Kiese-2023,Ferrari-2023,Niggemann-2023}, while preserving the gapless nature of excitations~\cite{yasir2011,yasir2014,yasir2016,Iqbal-2018,Iqbal-2021_j1j2} and their disordered character without lattice or spin-rotation symmetry breaking~\cite{Iqbal-2011_vbc,Iqbal_2012_vbc}. Moreover, singlet and triplet monopoles remain gapless on kagome and triangular lattices, without destabilizing the U(1) Dirac state~\cite{Budaraju-2023,Budaraju-2025}. By contrast, the influence of gauge fluctuations on U(1) semi-Dirac states with neutral fermions remains unexplored, in constrast to effects of Coulomb interactions on charged fermions~\cite{Elsayed-2025b}. It is therefore important to investigate how Gutzwiller projection modifies the long-distance behavior of spin-spin correlations and the spin structure factor. Possible outcomes include renormalization of the mean-field exponents, as occurs in Dirac spin liquids~\cite{Hermele-2008,Ferrari-2021_gapless,Kim-2025}, or more drastic effects such as gap opening~\cite{Rantner-2002} or spontaneous symmetry breaking~\cite{Li_2013}. Since, Gutzwiller projection incorporates spinon interactions in a nonperturbative manner, substantial modifications of low-energy dynamical structure factors are likely. A complementary direction is the symmetry analysis of singlet and triplet monopoles and their condensation patterns into valence-bond crystal and magnetic orders, respectively, which would yield crucial insights into the spectroscopic fingerprints of low-energy monopole excitations~\cite{Song-2019}. Constructing variational wave functions for monopole excitations and estimating their scaling dimensions would provide further guidance on their stability. Another interesting avenue would be to identify the key ingredients necessary to construct tight-binding models that yield generalized semi-Dirac dispersions, i.e., where the leading term is $\propto |\mathbf{k}|^{2N}$ (with $N>1$) or host type-II semi-Dirac fermions wherein the spinons disperse with a combination of different momentum components away from the reciprocal lattice vectors~\cite{Huang2015}.

While not stabilized for the nearest-neighbor Heisenberg model on the trellis lattice, the semi-Dirac spin liquid represents a distinct symmetry-allowed quantum spin liquid, motivating the search for microscopic spin models in which it may emerge as a viable variational ground state. At the mean-field level, these states appear only at fine-tuned phase boundaries, and it remains to be seen whether upon inclusion of quantum fluctuations the sDSL is robust against its many potential instabilities such as monopole proliferation and confinement, Higgsing of the U(1) gauge boson to $\mathbb{Z}_2$, or spontaneous symmetry breaking driven by (strong) fermion interactions~\cite{Uryszek2019}. In the scenario that the sDSL happens to be stable to these perturbations and is realized as a phase while being the variational ground state over a region of parameter space of the nearest-neighbor antiferromagnetic model, it would be interesting to identify additional interactions that could potentially enhance its region of stability. These include terms such as ring-exchange, XXZ anisotropy, etc., and to this effect, it would be important to map out the corresponding ground state phase diagrams using state-of-the-art many-body methods, including variational Monte Carlo, DMRG, and tensor-network approaches. It is also valuable to systematically determine which lattice and Hamiltonian symmetries are required to host semi-Dirac spectra~\cite{Hou-2013,Hou-2015,Das-2020}, and to examine their feasibility across other wallpaper groups and in three-dimensional lattices. A particularly intriguing direction is the realization of semi-Dirac QSLs in tight-binding models on two-dimensional hyperbolic lattices~\cite{Dusel-2025,Mosseri-2025,Lenggenhager,vidal2025}, a setting that remains largely unexplored.

Given the delicate competition between quantum spin liquid and weak magnetic order in the spin-$1/2$ nearest-neighbor Heisenberg antiferromagnet, it would be instructive to study this interplay using variational Monte Carlo. Jastrow-Slater wave functions have previously provided accurate estimates of order parameters and phase boundaries in frustrated magnets. Should the magnetic order parameter vanish in the thermodynamic limit, the U(1) and $\mathbb{Z}_{2}$ states classified here would constitute strong candidates for the true ground state. Comparing the variational energies of their Gutzwiller-projected wave functions would then allow one to identify the lowest-energy phase and thereby elucidate the microscopic nature of the ground state. In particular, the quantum melting of classical incommensurate spirals has been argued to give rise to gapped $\mathbb{Z}_{2}$ QSLs~\cite{Bernier-2004,Capriotti-2004}, providing a concrete starting point for future investigations.

Finally, we note that recent advances in quantum simulation have enabled the realization of tunable Heisenberg and XXZ models on arbitrary lattice geometries, including the trellis lattice~\cite{bintz2024dirac,tian2025engineering}. Such synthetic quantum systems offer yet another pathway to probe the correlated phases discussed in this work, complementing solid-state materials. In particular, these platforms' capability for site-resolved measurements provides direct microscopic access to quantities that are typically inaccessible in bulk experiments~\cite{semeghini2021-lukin}. This includes not only local spin correlations but also more refined probes such as the entanglement entropy~\cite{teng2024learning,ott2024probing}, thereby offering unique insights into the underlying quantum many-body physics.

\newpage

\section*{Acknowledgments}
We thank Subhro Bhattacharjee, Titas Chanda, Andrey Chubukov, Johannes Knolle, Zohar Nussinov, Sri Raghu, and Subir Sachdev for helpful discussions. 
The work of K.P., R.S., and Y.I. was performed in part at the Aspen Center for Physics, which is supported by a grant from the Simons Foundation (1161654, Troyer). This research was also supported in part by grant NSF PHY-2309135 to the Kavli Institute for Theoretical Physics. S.C. acknowledges the use of the computing resources at HPCE, IIT Madras and support from the Anusandhan National Research Foundation (ANRF), India in the form of a Junior Research Fellowship via the Prime Minister Early Career Research Grant Scheme ANRF/ECRG/2024/001198/PMS. K.P, R.T., and H.O.J. thank IIT Madras for a Visiting Faculty Fellow position under the IoE program that enabled completion of this work. H.O.J. acknowledges support through JSPS KAKENHI Grants No.~24H01668 and No.~25K0846007. The work in W\"urzburg was supported by the Deutsche Forschungsgemeinschaft (DFG, German Research Foundation) through Project ID 258499086 --- SFB 1170 and through the W\"urzburg-Dresden Cluster of Excellence on Complexity and Topology in Quantum Matter - ct.qmat Project ID 390858490-EXC 2147. K.P. was supported by Hungarian NKFIH OTKA Grant No.~K142652. Y.I. acknowledges support from the Abdus Salam International Centre for Theoretical Physics through the Associates Programme, from the Simons Foundation through Grant No.~284558FY19, from IIT Madras through the Institute of Eminence program for establishing QuCenDiEM (Project No. SP22231244CPETWOQCDHOC), and the International Centre for Theoretical Sciences for participation in the Discussion Meeting --- Fractionalized Quantum Matter (code: ICTS/DMFQM2025/07). R.S. was supported by the Princeton Quantum Initiative Fellowship. The DMRG calculations presented in this article were implemented using the \textsc{ITensor} library~\cite{itensor} and performed on computational resources managed and supported by Princeton Research Computing, a consortium of groups including the Princeton Institute for Computational Science and Engineering (PICSciE) and the Office of Information Technology's High Performance Computing Center and Visualization Laboratory at Princeton University. 

\section*{Data Availability Statement} The data generated during the current study are available from the corresponding author upon reasonable request.

\clearpage
\pagebreak
\appendix

\section{Algebraic symmetry relations}
\label{sec:genric_gauge_con}

The projective realizations $W^{\pdagger}_\mathcal{O}$ of the symmetry elements $\mathcal{O}$ must satisfy the algebraic constraints specified in Eqs.~\eqref{eq:translations}–\eqref{eq:TRO}. These constraints are enforced by replacing each occurrence of the identity symmetry element $\mathcal{I}$ on the right-hand sides with an element of the IGG, since the PSG associated with $\mathcal{I}$ coincides with the IGG. Consequently, the conditions on the projective gauge transformations $W^{\pdagger}_\mathcal{O}$ are
\begin{widetext}
\begin{align}
 W^{\pdagger}_{T_1}(x,y,u) W^{\pdagger}_{T_{2}}(x-1,y,u)W^{-1}_{T_1}(x,y-1,u)W^{-1}_{T_{2}}(x,y,u) &\in \mathrm{IGG} \label{eq:g_translations} \\
 W^{\pdagger}_{T_1}(x,y,u) W_{C_2}^{-1}(-x+1,-y,\Bar{u}) W^{\pdagger}_{T_1}(-x+1,-y,\Bar{u}) W^{\pdagger}_{C_2}(-x,-y,\Bar{u})&\in \mathrm{IGG}, \label{eq:g_C2_T1}\\
 W^{\pdagger}_{T_2}(x,y,u)W_{C_2}^{-1}(-x,-y+1,\Bar{u})W^{\pdagger}_{T_2}(-x,-y+1,\Bar{u}) W^{\pdagger}_{C_2}(-x,-y,\Bar{u}) &\in \mathrm{IGG},\label{eq:g_C2_T2}\\
W_{T_1}^{-1}(x+1,y,u)W_{\sigma_x}^{-1}(x+y+1,-y,\Bar{u}) W^{\pdagger}_{T_1}(x+y+1,-y,\Bar{u})W^{\pdagger}_{\sigma_x}(x+y,-y,\Bar{u}) & \in \mathrm{IGG}, \label{eq:g_sigma_T1}\\
 W_{T_1}^{-1}(x+1,y,u)W^{\pdagger}_{T_2}(x+1,y,u)W_{\sigma_x}^{-1}(x+y,-y+1,\Bar{u}) W^{\pdagger}_{T_2}(x+y,-y+1,\Bar{u}) W^{\pdagger}_{\sigma_x}(x+y,-y,\Bar{u})&\in \mathrm{IGG}, \label{eq:g_sigma_T2}\\
W^{\pdagger}_{C_2}(x,y,u) W^{\pdagger}_{\sigma_x}(-x,-y,\Bar{u}) W^{\pdagger}_{C_2}(-x-y,y,u) W^{\pdagger}_{\sigma_x}(x+y,-y,\Bar{u}) & \in \mathrm{IGG}, \label{eq: G_C2_s2}\\
 W^{\pdagger}_{C_2}(x,y,u)W^{\pdagger}_{C_2}(-x,-y,\Bar{u}) &\in \mathrm{IGG}, \label{eq:G_C22}\\
 W^{\pdagger}_{\sigma_x}(x,y,u)W^{\pdagger}_{\sigma_x}(x+y,-y,\Bar{u})&\in \mathrm{IGG}, \label{eq:G_sigma2} \\
 W_{T_1}^{-1}(x+1,y,u) W^{\pdagger}_{T_2}(x+1,y,u)W^{\pdagger}_{\sigma_x}(x+1,y-1,u) W^{\pdagger}_{T_1}(x+y,-y+1,\Bar{u})\notag\\ \times W_{\sigma_x}^{-1}(x,y-1,u) W_{T_2}^{-1}(x,y,u)&\in \mathrm{IGG}, \label{eq:G_sigma_T2T1} \\
W^{\pdagger}_{C_2}(x,y,u) W^{\pdagger}_{T_1}(-x,-y,\Bar{u})W_{T_2}^{-1}(-x-1,-y+1,\Bar{u})W^{\pdagger}_{\sigma_x}(-x-1,-y+1,\Bar{u})  \notag \\
\times  W^{\pdagger}_{C_2}(-x-y,y-1,u) W^{\pdagger}_{T_2}(x+y,-y+1,\Bar{u})W^{\pdagger}_{\sigma_x}(x+y,-y,\Bar{u})&\in \mathrm{IGG}, 
\label{eq:G_C2_T1_sigma_x}\\
    W^{\pdagger}_{\mathcal{T}}(x,y,u)W^{\pdagger}_{\mathcal{T}}(x,y,u)&\in \mathrm{IGG}, \label{eq:U1_g_T2} \\
    W^{\pdagger}_{\mathcal{T}}(x,y,u) W^{\pdagger}_{T_1}(x,y,u)    W_{\mathcal{T}}^{-1}(x-1,y,u) W_{T_1}^{-1}(x,y,u) &\in \mathrm{IGG}, \label{eq:U1_g_T_T1} \\
    W^{\pdagger}_{\mathcal{T}}(x,y,u) W^{\pdagger}_{T_2}(x,y,u)    W_{\mathcal{T}}^{-1}(x,y-1,u) W_{T_2}^{-1}(x,y,u) &\in \mathrm{IGG}, \label{eq:U1_g_T_T2} \\
  W^{\pdagger}_{\mathcal{T}}(x,y,u) W^{\pdagger}_{C_2}(x,y,u)    W_{\mathcal{T}}^{-1}(-x,-y,\Bar{u}) W_{C_2}^{-1}(x,y,u) &\in \mathrm{IGG},\label{eq:U1_g_T_C2} \\
    W^{\pdagger}_{\mathcal{T}}(x,y,u) W^{\pdagger}_{\sigma_x}(x,y,u)     W_{\mathcal{T}}^{-1}(x+y,-y,\Bar{u}) W_{\sigma_x}^{-1}(x,y,u) &\in \mathrm{IGG}.\label{eq:U1_g_T_sigma}
\end{align}    
\end{widetext}

In the following sections, we explicitly demonstrate how the above equations are applied to determine the algebraic PSGs corresponding to different IGGs, namely U(1) and $\mathbb{Z}_2$.  
Before proceeding, we clarify the notation employed in the subsequent analysis. For the U(1) gauge group, the IGG elements are parametrized as ${\exp({\dot\iota \theta \tau^z})}$ with $\theta \in [0, 2\pi)$. For the $\mathbb{Z}_2$ gauge group, the IGG elements are given by $\eta \in \{+1, -1\}$. In addition, any parameter labeled by ``$n_{\ldots}$'' is restricted to take values in $\{0,1\}$.

\section{U(1) projective symmetry groups}
\label{app:u1_PSG_derivation}

In this section, we derive the algebraic solutions of the PSG with IGG U(1). Any \textit{Ansatz} with this IGG can be expressed solely in terms of hopping mean-field parameters,
\begin{equation}
    u^{\pdagger}_{ij} = \dot{\iota}\,\chi_{ij}^{0} \tau^0 + \chi_{ij}^{3} \tau^z ,
\end{equation}
which corresponds to the canonical gauge where the U(1) IGG is explicit. In this gauge, elements of the IGG are global U(1) matrices of the form
\begin{equation}
    W = \{\exp\{\dot{\iota}\,\theta\,\tau^z\} \mid 0 \leq \theta < 2\pi\}.
\end{equation}
The gauge transformations preserving this canonical form can be written as
\begin{equation}
    W^{\pdagger}_{\mathcal{O}}(r,u) = \mathcal{F}^{\pdagger}_z(\phi^{\pdagger}_{\mathcal{O}}(r,u)) (\dot{\iota}\,\tau^x)^{w^{\pdagger}_{\mathcal{O}}}, \label{eq:U1_G}
\end{equation}
where $\mathcal{F}_z(\xi) \equiv \exp({i \xi\, \tau^z})$ and $w^{\pdagger}_{\mathcal{O}} \in \{0,1\}$ for the symmetry operations $\mathcal{O} \in \{T^{\pdagger}_1, T^{\pdagger}_2, C^{\pdagger}_2, \sigma^{\pdagger}_x, \mathcal{T}\}$. The reason behind including the factor $(\dot{\iota}\,\tau^x)^{w^{\pdagger}_{\mathcal{O}}}$ on the right-hand side can be understood from the SU(2) flux operator $\mathcal{P}_{\mathcal{C}_i}(\varphi_i)=\mathcal{F}_z(\varphi_i)$, which transforms under projective operations as follows:
\begin{align}
 \mathcal{P}_{\mathcal{C}_i}(\varphi_i)&=W^\dagger_{\mathcal{O}(i)}\mathcal{P}^{\pdagger}_{\mathcal{C}_{\mathcal{O}(i)}}(\varphi_{\mathcal{O}(i)})W^{\pdagger}_{\mathcal{O}(i)}  \notag \\
 &=(\tau_x)^{w_\mathcal{O}}\mathcal{P}_{\mathcal{C}_{\mathcal{O}(i)}}(\varphi_{\mathcal{O}(i)})(\tau_x)^{w_\mathcal{O}}\notag\\
&=\mathcal{P}_{\mathcal{C}_{\mathcal{O}(i)}}((-)^{w_\mathcal{O}}\varphi_{\mathcal{O}(i)})
\end{align}
Notice, that for $w_\mathcal{O}=0$ ($w_\mathcal{O}=1$), the flux $\varphi_{\mathcal{O}(i)}=\varphi_i$ ($\varphi_{\mathcal{O}(i)}=-\varphi_i$). Thus, if we do not consider the $w_\mathcal{O}=1$ case, we would not be able to include those \textit{Ansätze} for which the flux operators connected by the symmetry operation $\mathcal{O}$ have antiparallel orientations.

Specifically, for translations $\mathcal{O} \in \{T^{\pdagger}_1, T^{\pdagger}_2\}$, Eq.~\eqref{eq:U1_G} gives
\begin{align}
     W^{\pdagger}_{T_1}(x,y,u) &= \mathcal{F}^{\pdagger}_z (\phi^{\pdagger}_{T_1}(x,y,u)) (\dot{\iota}\,\tau^x)^{w^{\pdagger}_{T_1}}, \notag \\
     W^{\pdagger}_{T_2}(x,y,u) &= \mathcal{F}^{\pdagger}_z (\phi^{\pdagger}_{T_2}(x,y,u)) (\dot{\iota}\,\tau^x)^{w^{\pdagger}_{T_2}}.
\end{align}
Loop operators connected by translations $T_i$ differ by a factor $(-1)^{w^{\pdagger}_{T_i}}$ in order to maintain translation invariance. Consequently, there are four possible choices for $(w^{\pdagger}_{T_1}, w^{\pdagger}_{T_2})$:
\begin{align}
    w^{\pdagger}_{T_1} &= 0, \quad w^{\pdagger}_{T_2} = 0, \label{eq:w_T11}\\
    w^{\pdagger}_{T_1} &= 0, \quad w^{\pdagger}_{T_2} = 1, \label{eq:w_T14}\\
    w^{\pdagger}_{T_1} &= 1, \quad w^{\pdagger}_{T_2} = 0, \label{eq:w_T12}\\
    w^{\pdagger}_{T_1} &= 1, \quad w^{\pdagger}_{T_2} = 1. \label{eq:w_T13}
\end{align}
However, the cases in Eqs.~\eqref{eq:w_T12} and~\eqref{eq:w_T13} fail to satisfy condition~\eqref{eq: T_2}, and are therefore excluded. Thus, we restrict our attention to cases~\eqref{eq:w_T11} and~\eqref{eq:w_T14}, which we analyze in the subsequent sections.

\subsection{U(1) class with $w^{\pdagger}_{T_1}=w^{\pdagger}_{T_2}=0$}

\subsubsection{Lattice symmetries}

Due to the SU(2) gauge redundancy, under a local gauge transformation $W(r,u)$, the projective symmetry operator $W^{\pdagger}_{\mathcal{O}}(r,u)$ transforms as
\begin{equation}
    W^{\pdagger}_{\mathcal{O}}(r,u) \;\rightarrow\; W^{\dagger}(r,u) \, W^{\pdagger}_{\mathcal{O}} \, W[\mathcal{O}^{-1}(r,u)].
\end{equation}
This transformation law allows significant simplification of the PSG solutions. To wit, by choosing $W(x,y,u)=\mathcal{F}_z(\phi(x,y,u))$, we obtain
\begin{equation}
    W^{\pdagger}_{T_2}(x,y,u) = W^{\pdagger}_{T_1}(x,0,u) = \tau^0. \label{eq:U1_T2a}
\end{equation}
Imposing the translation symmetry condition~\eqref{eq:g_translations}, we find the translational PSGs
\begin{equation}\label{eq:g_U1_T1T2}
     W^{\pdagger}_{T_1}(x,y,u)=\mathcal{F}^{\pdagger}_z(y\, \theta),\quad W^{\pdagger}_{T_2}(x,y,u)=\tau^0.
\end{equation}

Next, we analyze the point-group symmetries. For $C_2$, the PSG takes the form
\begin{equation}
    W^{\pdagger}_{C_2}(x,y,u) = \mathcal{F}^{\pdagger}_z(\phi^{\pdagger}_{C_2}(x,y,u)) (\dot{\iota}\,\tau^x)^{w^{\pdagger}_{C_2}}.
\end{equation}
Defining $\Delta^{\pdagger}_i \phi(x,y,u)=\phi(x,y,u)-\phi(x-1,y,u)$, and using Eq.~\eqref{eq:g_U1_T1T2} in Eqs.~\eqref{eq:g_C2_T1} and~\eqref{eq:g_C2_T2}, we obtain
\begin{align}
\Delta^{\pdagger}_{1} \phi^{\pdagger}_{C_2}(x,y,u) &= -(-1)^{w^{\pdagger}_{C_2}}\theta^{\pdagger}_{C_2 T_1} 
+ y \theta [1-(-1)^{w^{\pdagger}_{C_2}}], \notag \\
\Delta^{\pdagger}_{2} \phi^{\pdagger}_{C_2}(x,y,u) &= -(-1)^{w^{\pdagger}_{C_2}} \theta^{\pdagger}_{C_2 T_2}. \label{eq:U1_C2a}
\end{align}
These must further satisfy the consistency condition
\begin{align}\label{eq:c2_consis}
    &\Delta^{\pdagger}_i \phi^{\pdagger}_{C_2}(x,y,u) + \Delta^{\pdagger}_{i+1} \phi^{\pdagger}_{C_2}(x-1,y,u) \notag \\
    &= \Delta^{\pdagger}_{i+1} \phi^{\pdagger}_{C_2}(x,y,u)+\Delta^{\pdagger}_i \phi^{\pdagger}_{C_2}(x,y-1,u).
\end{align}
For $i=1$, inserting Eq.~\eqref{eq:U1_C2a} into Eq.~\eqref{eq:c2_consis} yields the constraint
\begin{align}
\big[1 - (-1)^{w^{\pdagger}_{C_2}}\big] \theta = 0, \quad 
\begin{cases} 
\theta \in [0, 2\pi) & \text{if } w^{\pdagger}_{C_2} = 0, \\
\theta = 0, \pi & \text{if } w^{\pdagger}_{C_2} = 1.
\end{cases}
\label{eq:w_c2_constraint}
\end{align}
Substituting back into Eq.~\eqref{eq:U1_C2a}, we find that the general solution is
\begin{equation}
     \phi^{\pdagger}_{C_2}(x,y,u)=-(-1)^{w^{\pdagger}_{C_2}}(x \theta^{\pdagger}_{C_2 T_1}+y \theta^{\pdagger}_{C_2 T_2}) + \rho^{\pdagger}_{C_2,u}, \label{eq:U1_C2}
\end{equation}
where $\rho_{C_2,u} \equiv \phi_{C_2}(0,0,u)$.

We now consider the PSG associated with $\sigma_x$. Using Eqs.~\eqref{eq:g_sigma_T1} and~\eqref{eq:g_sigma_T2} together with Eq.~\eqref{eq:g_U1_T1T2}, we obtain
\begin{align}
&    \Delta^{\pdagger}_1 \phi^{\pdagger}_{{\sigma_x}}(x,y,u)=-(-1)^{w^{\pdagger}_{\sigma_x}} \theta^{\pdagger}_{{\sigma_x T_1}} +y \theta\big(1+(-1)^{w^{\pdagger}_{\sigma_x}}\big), \notag \\
&    \Delta^{\pdagger}_2 \phi^{\pdagger}_{\sigma_x}(x,y,u)=-(-1)^{w^{\pdagger}_{\sigma_x}} \theta^{\pdagger}_{{\sigma_x T_2}} +(-1)^{w^{\pdagger}_{\sigma_x}}(y-1) \theta \, , \label{eq:sigma_delta}
\end{align}
which must satisfy the consistency relation analogous to Eq.~\eqref{eq:c2_consis},
\begin{align}
   & \Delta^{\pdagger}_i \phi^{\pdagger}_{\sigma_x}(x,y,u)+\Delta^{\pdagger}_{i+1} \phi^{\pdagger}_{\sigma_x}(x-1,y,u) \notag\\
    &= \Delta^{\pdagger}_{i+1} \phi^{\pdagger}_{\sigma_x}(x,y,u)+\Delta^{\pdagger}_i \phi^{\pdagger}_{\sigma_x}(x,y-1,u) \, .\label{eq:sigma_consis}
\end{align}
For $i=1,2$ in Eq.~\eqref{eq:sigma_consis}, substituting Eq.~\eqref{eq:sigma_delta} imposes the following constraints:
\begin{align}
\label{eq:w_sigma_constraint} 
\left[1+(-1)^{w^{\pdagger}_{\sigma_x}}\right] \theta =0,\quad
\begin{cases}
\theta = 0, \pi & \text{if } w^{\pdagger}_{\sigma_x} = 0, \\
\theta \in [0, 2\pi) & \text{if } w^{\pdagger}_{\sigma_x} = 1\,.
\end{cases}
\end{align}
Combining these constraints with Eq.~\eqref{eq:sigma_delta}, the general solution for $\phi^{\pdagger}_{\sigma_x}(x,y,u)$ is
\begin{align}
    \phi^{\pdagger}_{\sigma_x}(x,y,u)&=-(-1)^{w^{\pdagger}_{\sigma_x}}\big[x \theta^{\pdagger}_{{\sigma_x T_1}}+y \theta^{\pdagger}_{{\sigma_x T_2}}\big] \notag \\
&\quad +(-1)^{w^{\pdagger}_{\sigma_x}}\frac{1}{2} y(y-1) \theta +\rho^{\pdagger}_{\sigma_x,u}\,, \label{eq:U1_sigma}
\end{align}
where $\rho_{\sigma_x,u}=\phi_{{\sigma_x}}(0,0,u)$.  
To summarize, the constraints obtained from Eqs.~\eqref{eq:w_c2_constraint} and~\eqref{eq:w_sigma_constraint} are compiled in Table~\ref{tab:U1_theta}.

\begin{table}[h!]
    \centering
    \begin{ruledtabular}
    \begin{tabular}{c ccc}
        Case & $w^{\pdagger}_{C_2}$ (rotations) & $w^{\pdagger}_{\sigma_x}$ (reflections) & Constraints on $\theta$  \\
        \hline
        (i) & 0 & 0& $0,\pi$\\
        (ii) &1 & 0& $0,\pi$\\
        (iii) &0 & 1& $[0, 2\pi)$\\
        (iv)& 1 & 1& $0,\pi$\\
    \end{tabular}
    \end{ruledtabular}
    \caption{Allowed values of $\theta$ for different combinations of $(w^{\pdagger}_{C_2},w^{\pdagger}_{\sigma_x})$.}
    \label{tab:U1_theta}
\end{table}

Continuing with our derivation, substituting Eq.~\eqref{eq:U1_C2} into Eq.~\eqref{eq:G_C22} yields the conditions
\begin{align}\label{eq:C2_cyclic}
    &\rho^{\pdagger}_{C_2,u}+(-1)^{w^{\pdagger}_{C_2}} \rho^{\pdagger}_{{C_2},\bar{u}} =\theta^{\pdagger}_{C_2} \notag\\
    &w^{\pdagger}_{C_2}=1 : \theta^{\pdagger}_{{C_2 T_i}} =0,\pi \, .
\end{align}
Similarly, substituting the solution~\eqref{eq:U1_sigma} into Eq.~\eqref{eq:G_sigma2} leads to
\begin{align}
& w^{\pdagger}_{\sigma_x}=0: \theta^{\pdagger}_{\sigma_x T_1} =\theta,\quad \rho^{\pdagger}_{\sigma_x,u}+ \rho^{\pdagger}_{\sigma_x,\Bar{u}}=\theta^{\pdagger}_{{\sigma_x}}, \label{eq:sigma_cyc_1}\\
& w^{\pdagger}_{\sigma_x}=1: \theta^{\pdagger}_{\sigma_x T_1} =2 \theta^{\pdagger}_{\sigma_x T_2}+ \theta,\quad \rho^{\pdagger}_{\sigma_x,u}- \rho^{\pdagger}_{\sigma_x,\Bar{u}}=\theta^{\pdagger}_{{\sigma_x}} \, . \label{eq:sigma_cyc_2}
\end{align}
Furthermore, one can use Eq.~\eqref{eq:G_sigma_T2T1} to obtain
\begin{equation}
    \rho^{\pdagger}_{\sigma_x,u} -\rho^{\pdagger}_{\sigma_x,u}=\theta^{\pdagger}_{1}+(-1)^{w^{\pdagger}_{\sigma_x}}[\theta^{\pdagger}_{\sigma_x T_1}-\theta]\, ,
\end{equation}
 while Eq.~\eqref{eq:G_C2_T1_sigma_x} leads to the following constraints depending on the different choices of $(w^{\pdagger}_{C_2},w^{\pdagger}_{\sigma_x})$:
\begin{align}
\mbox{(i) }&   w^{\pdagger}_{C_2}=0,w^{\pdagger}_{\sigma_x}=0\;: \notag \\
&   2 \theta^{\pdagger}_{{C_2 T_2}} + \theta^{\pdagger}_{\sigma_x T_1} =2 \theta^{\pdagger}_{\sigma_x T_2} + \theta^{\pdagger}_{C_2 T_1},  \\
&    \rho^{\pdagger}_{C_2,u}+\rho^{\pdagger}_{\sigma_x,\Bar{u}} +\rho^{\pdagger}_{C_2,u}+\rho^{\pdagger}_{\sigma_x,\Bar{u}}=\theta^{\pdagger}_2-\frac{\theta^{\pdagger}_{\sigma_x T_1}}{2}. \notag\\
\mbox{(ii) }&   w^{\pdagger}_{C_2}=1,w^{\pdagger}_{\sigma_x}=0 \;:\notag \\
&    \theta^{\pdagger}_{C_2 T_1} =\theta^{\pdagger}_{\sigma_x T_1},  \label{eq:U1_C2T1T2_2} \\
&    \rho^{\pdagger}_{C_2,u}-\rho^{\pdagger}_{\sigma_x,\Bar{u}} -\rho^{\pdagger}_{C_2,u}+\rho^{\pdagger}_{\sigma_x,\Bar{u}}=\theta^{\pdagger}_2- \theta^{\pdagger}_{\sigma_x T_2} -\theta^{\pdagger}_{C_2 T_2}. \notag\\
\mbox{(iii) }&   w^{\pdagger}_{C_2}=0,w^{\pdagger}_{\sigma_x}=1 \;:\notag \\
&    \theta^{\pdagger}_{C_2 T_1} =-\theta^{\pdagger}_{\sigma_x T_1},  \label{eq:U1_C2T1T2_3} \\
&    \rho^{\pdagger}_{C_2,u}+\rho^{\pdagger}_{\sigma_x,\Bar{u}} -\rho^{\pdagger}_{C_2,u}-\rho^{\pdagger}_{\sigma_x,\Bar{u}}=\theta^{\pdagger}_2- \theta^{\pdagger}_{\sigma_x T_2} -\theta^{\pdagger}_{C_2 T_2}. \notag\\
\mbox{(iv) }&   w^{\pdagger}_{C_2}=1,w^{\pdagger}_{\sigma_x}=1 \;:\notag \\
&    \theta^{\pdagger}_{\sigma_x T_1} +\theta^{\pdagger}_{C_2 T_1}=2( \theta^{\pdagger}_{\sigma_x T_2} +\theta^{\pdagger}_{C_2 T_2}) ,\label{eq:U1_C2T1T2_4} \\
&    \rho^{\pdagger}_{C_2,u}-\rho^{\pdagger}_{\sigma_x,\Bar{u}} +\rho^{\pdagger}_{C_2,u} - \rho^{\pdagger}_{\sigma_x,\Bar{u}}=\theta^{\pdagger}_2- \frac{\theta^{\pdagger}_{\sigma_x T_1}}{2}. \notag
\end{align}
Finally, we are left with the constraint \eqref{eq: G_C2_s2}, which yields the following conditions:
\begin{align}
\mbox{(i) }&   w^{\pdagger}_{C_2}=0,w^{\pdagger}_{\sigma_x}=0\;: \notag \\
&    2\theta^{\pdagger}_{\sigma_x T_2} +\theta^{\pdagger}_{C_2 T_1} =2 \theta^{\pdagger}_{C_2 T_2} +\theta^{\pdagger}_{{\sigma_x T_1}},\notag \\
&     \rho^{\pdagger}_{C_2,u}+\rho^{\pdagger}_{\sigma_x,\Bar{u}}+\rho^{\pdagger}_{C_2,u}+ \rho^{\pdagger}_{\sigma_x,\Bar{u}}=\theta^{\pdagger}_{C_2 \sigma_x}.\\
\mbox{(ii) }&    w^{\pdagger}_{C_2}=1,w^{\pdagger}_{\sigma_x}=0\;: \notag \\
&    \theta^{\pdagger}_{C_2 T_1} = \theta^{\pdagger}_{\sigma_x T_1},\notag \\
&    \rho^{\pdagger}_{C_2,u}-\rho^{\pdagger}_{\sigma_x,\Bar{u}}-\rho^{\pdagger}_{C_2,u}+ \rho^{\pdagger}_{\sigma_x,\Bar{u}}=\theta^{\pdagger}_{C_2 \sigma_x}.\\
\mbox{(iii) }&     w^{\pdagger}_{C_2}=0,w^{\pdagger}_{\sigma_x}=1\;: \notag \\
&     \theta^{\pdagger}_{C_2 T_1} = -\theta^{\pdagger}_{\sigma_x T_1}, \notag \\
&   \rho^{\pdagger}_{C_2,u}+\rho^{\pdagger}_{\sigma_x,\Bar{u}}-\rho^{\pdagger}_{C_2,u}- \rho^{\pdagger}_{\sigma_x,\Bar{u}}=\theta^{\pdagger}_{C_2 \sigma_x}.  \\
\mbox{(iv) }&      w^{\pdagger}_{C_2}=1,w^{\pdagger}_{\sigma_x}=1\;: \notag \\
&     2(\theta^{\pdagger}_{C_2 T_2}+ \theta^{\pdagger}_{\sigma_x T_2})= \theta^{\pdagger}_{\sigma_x T_1}+ \theta^{\pdagger}_{C_2 T_1},\notag \\
&    \rho^{\pdagger}_{C_2,u}-\rho^{\pdagger}_{\sigma_x,\Bar{u}}+\rho^{\pdagger}_{C_2,u}- \rho^{\pdagger}_{\sigma_x,\Bar{u}}=\theta^{\pdagger}_{C_2 \sigma_x}. 
\end{align}

It is worth emphasizing that all the abovementioned U(1) phases may not be gauge independent. Some of them can be fixed using a local gauge transformation of the form
\begin{equation}
    W(x,y,u)=\mathcal{F}^{\pdagger}_z(x\theta^{\pdagger}_x + y \theta^{\pdagger}_y).
\end{equation}
This transformation does not alter the structure of the translational gauges, except for introducing global phases. These global phases do not affect the {\it Ansatz} and are therefore irrelevant. The gauge transformations for $C^{\pdagger}_2$ and $\sigma^{\pdagger}_x$ are now given by 
\begin{widetext}
    \begin{align}
        \Tilde{\phi}^{\pdagger}_{C_2} & =x[-\theta^{\pdagger}_x - (-1)^{w^{\pdagger}_{C_2}} \theta^{\pdagger}_{{C_2 T_1}} - (-1)^{w^{\pdagger}_{C_2}} \theta^{\pdagger}_x] + y[-\theta^{\pdagger}_y - (-1)^{w^{\pdagger}_{C_2}} \theta^{\pdagger}_{{C_2 T_2}} - (-1)^{w^{\pdagger}_{C_2}} \theta^{\pdagger}_y]  +\rho^{\pdagger}_{C_2,u} \\
        \Tilde{\phi}^{\pdagger}_{\sigma_x} & =x[-\theta^{\pdagger}_x -(-1)^{w^{\pdagger}_{\sigma_x}} \theta^{\pdagger}_{{\sigma_x T_1}}+(-1)^{w^{\pdagger}_{\sigma_x}} \theta^{\pdagger}_{x}]+y[-\theta^{\pdagger}_y -(-1)^{w^{\pdagger}_{\sigma_x}} \theta^{\pdagger}_{\sigma_x T_2}+(-1)^{w^{\pdagger}_{\sigma_x}} \theta^{\pdagger}_x -(-1)^{w^{\pdagger}_{\sigma_x}} \theta^{\pdagger}_y \notag \\
&+\frac{1}{2}(-1)^{w^{\pdagger}_{\sigma_x}}(y-1) \theta] +\rho^{\pdagger}_{\sigma_x,u}\, .
\end{align}

In the following, we provide the details of the gauge fixing for the four distinct cases of $(w^{\pdagger}_{C_2},w^{\pdagger}_{\sigma_x})$ using the gauge-transformed solutions obtained above.

\begin{enumerate}[label=(\roman*)]
\item $w^{\pdagger}_{C_2}=0$, $w^{\pdagger}_{\sigma_x}=0$:\\  
Choosing $2 \theta^{\pdagger}_x=-\theta^{\pdagger}_{C_2 T_1}$ and $2 \theta^{\pdagger}_y=-\theta^{\pdagger}_{C_2 T_2}$ in Eq.~\eqref{eq:sigma_cyc_1} yields
\begin{align}
& \Tilde{\phi}^{\pdagger}_{C_2}(x,y,u) =\rho^{\pdagger}_{C_2,u}, \\
& \Tilde{\phi}^{\pdagger}_{\sigma_x}(x,y,u)=-x \theta^{\pdagger}_{\sigma_x T_1}-y\theta^{\pdagger}_{\sigma_x T_2} +\tfrac{1}{2}y(y-1)\theta +\rho^{\pdagger}_{\sigma_x,u}\, .
\end{align}
\item $w^{\pdagger}_{C_2}=1$, $w^{\pdagger}_{\sigma_{x}}=0$: \\ 
Here, selecting $\theta^{\pdagger}_x -2 \theta^{\pdagger}_y =\theta^{\pdagger}_{\sigma_x T_2}$ sets $\theta^{\pdagger}_{\sigma_xT_2}=0$.  
From Eqs.~\eqref{eq:C2_cyclic},~\eqref{eq:sigma_cyc_1}, and ~\eqref{eq:U1_C2T1T2_2} we obtain 
$\theta^{\pdagger}_{C_2 T_1}=\theta^{\pdagger}_{\sigma_x T_1}=\theta=n \pi$ and $\theta^{\pdagger}_{C_2 T_2}= n^{\pdagger}_{C_2 T_2}\pi$.  
Thus,
\begin{align}
& \Tilde{\phi}^{\pdagger}_{C_2}(x,y,u)=x \theta^{\pdagger}_{C_2 T_1}+y \theta^{\pdagger}_{C_2 T_2}+\rho^{\pdagger}_{C_2,u}, \\
& \Tilde{\phi}^{\pdagger}_{\sigma_x}(x,y,u)=-x \theta^{\pdagger}_{\sigma_x T_1}+\tfrac{1}{2}y(y-1)\theta+\rho^{\pdagger}_{\sigma_x,u}\, .
\end{align}
\item $w^{\pdagger}_{C_2}=0$, $w^{\pdagger}_{\sigma_x}=1$: \\ 
Upon choosing $2\theta^{\pdagger}_x=-\theta^{\pdagger}_{C_2 T_1}$ and $2\theta^{\pdagger}_y=-\theta^{\pdagger}_{C_2 T_2}$, and using ~\eqref{eq:sigma_cyc_2} and Eqs.~\eqref{eq:U1_C2T1T2_3} , we find
\begin{align}
& \Tilde{\phi}^{\pdagger}_{C_2}(x,y,u)=\rho^{\pdagger}_{C_2,u}, \\
& \Tilde{\phi}^{\pdagger}_{\sigma_x}(x,y,u)=y\theta^{\pdagger}_{\sigma_x T_2}-\tfrac{1}{2}y(y-1)\theta+\rho^{\pdagger}_{\sigma_x,u}\, .
\end{align}
\item $w^{\pdagger}_{C_2}=1$, $w^{\pdagger}_{\sigma_x}=1$:\\
From Eqs.~\eqref{eq:C2_cyclic},~\eqref{eq:sigma_cyc_2}, and~\eqref{eq:U1_C2T1T2_4}, we have  
$\theta^{\pdagger}_{C_2 T_i}=n^{\pdagger}_{C_2 T_i} \pi$,  
$\theta^{\pdagger}_{\sigma_x T_1}=2 \theta^{\pdagger}_{\sigma_x T_2}+\theta$, and  
$\theta^{\pdagger}_{\sigma_x T_1}+\theta^{\pdagger}_{C_2 T_1}=2(\theta^{\pdagger}_{\sigma_x T_2}+\theta^{\pdagger}_{C_2 T_2})$.  
The choice $\theta^{\pdagger}_x=\theta^{\pdagger}_{\sigma_x T_2}$ gives
\begin{align}
& \Tilde{\phi}^{\pdagger}_{C_2}(x,y,u)=\theta^{\pdagger}_{C_2 T_1}x+\theta^{\pdagger}_{C_2 T_2}y+\rho^{\pdagger}_{C_2,u}, \\
& \Tilde{\phi}^{\pdagger}_{\sigma_x}(x,y,u)=\theta^{\pdagger}_{C_2 T_1}x-\tfrac{1}{2}y(y-1)\theta+\rho^{\pdagger}_{\sigma_x,u}\, .
\end{align}
\end{enumerate}

\end{widetext}

To fix the  parameters $\rho_{\ldots}$ for all relevant symmetry operations, i.e., $\sigma_x$ and $C_2$, we apply a gauge transformation $W(r,u)=\mathcal{F}_z(\phi^{\pdagger}_{u})$ that depends only on the sublattice index $u$. This yields
\begin{align}
& \Tilde{\rho}^{\pdagger}_{C_2,u}=-\phi^{\pdagger}_u+\rho^{\pdagger}_{C_2,u}+(-1)^{w^{\pdagger}_{C_2}} \phi^{\pdagger}_{\Bar{u}}, \\
& \Tilde{\rho}^{\pdagger}_{\sigma_x,u}=-\phi^{\pdagger}_u+\rho^{\pdagger}_{\sigma_x,u}+(-1)^{w^{\pdagger}_{\sigma_x}} \phi^{\pdagger}_{\Bar{u}}\, .
\end{align}
From the cyclic relations~\eqref{eq:G_C22} and~\eqref{eq:G_sigma2}, we obtain
\begin{align}
& \rho^{\pdagger}_{C_2,u}+(-1)^{w^{\pdagger}_{C_2}} \rho^{\pdagger}_{C_2,\Bar{u}}=\theta^{\pdagger}_{C_2}, \label{eq:U1_C2_cyclic}\\
& \rho^{\pdagger}_{\sigma_x,u}+(-1)^{w^{\pdagger}_{\sigma_x}} \rho^{\pdagger}_{\sigma_x,\Bar{u}}=\theta^{\pdagger}_{\sigma_x}. \label{eq:U1_rho_sigma}
\end{align}
Using Eq.~\eqref{eq: G_C2_s2}, we also find
\begin{align}
 \label{eq:U1_rho_C2_s2}
 \rho^{\pdagger}_{C_2,u}&+(-1)^{w^{\pdagger}_{C_2}} \rho^{\pdagger}_{\sigma_x,\Bar{u}} \\ 
&+(-1)^{w^{\pdagger}_{C_2}+w^{\pdagger}_{\sigma_x}} \rho^{\pdagger}_{C_2,u}+(-1)^{w^{\pdagger}_{\sigma_x}} \rho^{\pdagger}_{\sigma_x,\Bar{u}}=\theta^{\pdagger}_{C_2\sigma_x}.\notag 
\end{align}
These conditions allow us to fix the $\rho$ parameters for the four cases of $(w^{\pdagger}_{C_2},w^{\pdagger}_{\sigma_x})$:

\begin{enumerate}[label=(\roman*)]
\item $w^{\pdagger}_{C_2}=0$, $w^{\pdagger}_{\sigma_x}=0$:\\  
Choosing $\phi^{\pdagger}_2=\phi^{\pdagger}_1-\rho^{\pdagger}_{C_2,1}+\theta^{\pdagger}_{C_2}/2$ and combining with Eq.~\eqref{eq:U1_C2_cyclic} yields $\Tilde{\rho}^{\pdagger}_{C_2,u}=\theta^{\pdagger}_{C_2}/2$.  
Using the IGG freedom to eliminate any global phase, we fix ${\rho}^{\pdagger}_{C_2,u}=0$.  
From Eq.~\eqref{eq:U1_rho_C2_s2}, we find $2\rho^{\pdagger}_{\sigma_x,u}=\theta^{\pdagger}_{C_2 \sigma_x}$, which after fixing a global phase gives $\rho^{\pdagger}_{\sigma_x,u}=\{0,n^{\pdagger}_{\sigma_x}\pi\}$.

\item $w^{\pdagger}_{C_2}=0$, $w^{\pdagger}_{\sigma_x}=1$:\\  
As in the previous case, we can set $\Tilde{\rho}^{\pdagger}_{C_2,u}=0$.  
Equation~\eqref{eq:U1_rho_sigma} then gives $\rho^{\pdagger}_{\sigma_x,2}=\rho^{\pdagger}_{\sigma_x,1}+n^{\pdagger}_{\sigma_x}\pi$.  
Using the IGG freedom to set $\rho^{\pdagger}_{\sigma_x,1}=0$, we obtain $\rho^{\pdagger}_{\sigma_x,u}=\{0,n^{\pdagger}_{\sigma_x}\pi\}$.

\item $w^{\pdagger}_{C_2}=1$, $w^{\pdagger}_{\sigma_x}=0$: \\ 
Choosing $\phi^{\pdagger}_1+\phi^{\pdagger}_2=\rho^{\pdagger}_{C_2,1}$ sets $\Tilde{\rho}^{\pdagger}_{C_2,1}=0$.  
From Eq.~\eqref{eq:U1_C2_cyclic}, we then obtain $\rho^{\pdagger}_{C_2,u}=\{0,n^{\pdagger}_{C_2}\pi\}$.  
Similarly, Eq.~\eqref{eq:U1_rho_sigma} gives $\rho^{\pdagger}_{\sigma_x,u}=\{0,\theta^{\pdagger}_{\sigma_x}\}$.

\item $w^{\pdagger}_{C_2}=1$, $w^{\pdagger}_{\sigma_x}=1$:\\  
As previously, we can set $\rho^{\pdagger}_{C_2,u}=\{0,n^{\pdagger}_{C_2}\pi\}$.  
Equation~\eqref{eq:U1_rho_sigma} then implies $\rho^{\pdagger}_{\sigma_x,u}=\{0,n^{\pdagger}_{\sigma_x}\pi\}$.
\end{enumerate}

\subsubsection{Time-reversal symmetry}
Having determined all solutions corresponding to the lattice space-group symmetries, we now turn to time-reversal symmetry. As a first step, we substitute the solution~\eqref{eq:g_U1_T1T2} into Eqs.~\eqref{eq:U1_g_T_T1} and~\eqref{eq:U1_g_T_T2}, yielding: 
\begin{align}
&  \Delta^{\pdagger}_{1}\phi^{\pdagger}_{\mathcal{T}} =\theta^{\pdagger}_{\mathcal{T} T_1} +y \theta[1-(-1)^{w^{\pdagger}_{\mathcal{T}}}] \label{eq:U1_TR_1}\\
& \Delta^{\pdagger}_2 \phi^{\pdagger}_{\mathcal{T}}(x,y,u)=\theta^{\pdagger}_{\mathcal{T}T_2} \, .\label{eq:U1_TR_2}
\end{align}
The consistency relations for time-reversal symmetry, analogous to Eqs.~\eqref{eq:c2_consis} and~\eqref{eq:sigma_consis}, are
\begin{align}\label{eq:time_consis}
& \Delta^{\pdagger}_1 \phi^{\pdagger}_{\mathcal{T}}(x,y,u)+ \Delta^{\pdagger}_2 \phi^{\pdagger}_{\mathcal{T}}(x-1,y,u) \notag \\
& =\Delta^{\pdagger}_2 \phi^{\pdagger}_{\mathcal{T}}(x,y,u) +\Delta^{\pdagger}_1 \phi^{\pdagger}_{\mathcal{T}}(x,y-1,u)
\end{align}
which impose the following constraint:
\begin{equation}
 \theta [1-(-1)^{w^{\pdagger}_{\mathcal{T}}}]=0\quad
 \begin{cases}\label{eq:w_time_constraint} 
\theta \in [0, 2\pi) & \text{if } w^{\pdagger}_{\mathcal{T}} = 0, \\
\theta = 0, \pi & \text{if } w^{\pdagger}_{\mathcal{T}} = 1\, . 
 \end{cases}
\end{equation}
Substituting these conditions into Eqs.~\eqref{eq:U1_TR_1} and~\eqref{eq:U1_TR_2}, we obtain the general solution
\begin{align}
& \phi^{\pdagger}_{\mathcal{T}}(x,y,u)= x \theta^{\pdagger}_{\mathcal{T} T_1} + y \theta^{\pdagger}_{\mathcal{T} T_2} + \rho^{\pdagger}_{\mathcal{T},u}\,,
\end{align}
where $\rho_{\mathcal{T},u}\equiv\phi_{\mathcal{T}}(0,0,u)$. We now consider the two cases $w^{\pdagger}_{\mathcal{T}}=0$ and $w^{\pdagger}_{\mathcal{T}}=1$ separately.

\smallskip
\textbf{$w^{}_{\mathcal{T}}$\,$=$\,$0$:} In this case, the consistency relation does not impose any constraint on $\theta$. Equation~\eqref{eq:U1_g_T2} further sets
\begin{equation}
    \theta^{\pdagger}_{\mathcal{T} T_1},\theta^{\pdagger}_{\mathcal{T},T_2} =0,\pi\,.
\end{equation}
The condition from Eq.~\eqref{eq:U1_g_T_C2} can be rewritten as
\begin{equation}
    \phi^{\pdagger}_{\mathcal{T}}(x,y,u) -(-1)^{w^{\pdagger}_{C_2}} \phi^{\pdagger}_{\mathcal{T}}(-x,-y,\Bar{u}) =\theta^{\pdagger}_{\mathcal{T} C_2}
\end{equation}  
which yields
\begin{align}
&    w^{\pdagger}_{C_2}=0 ,\,\theta^{\pdagger}_{\mathcal{T} T_i} =n^{\pdagger}_{\mathcal{T} T_i} \pi \\
&    w^{\pdagger}_{C_2}=1,\,  \theta^{\pdagger}_{\mathcal{T} T_i} \neq 0 \\
&     \rho^{\pdagger}_{\mathcal{T},u} -(-1)^{w^{\pdagger}_{C_2}} \rho^{\pdagger}_{\mathcal{T},\Bar{u}}=\theta^{\pdagger}_{ \mathcal{T}C_2}\, . \label{eq:U1_T_C2}
\end{align}
Similarly, Eq.~\eqref{eq:U1_g_T_sigma} becomes
\begin{equation}
    \phi^{\pdagger}_{\mathcal{T}}(x,y,u) -(-1)^{w^{\pdagger}_{\sigma_x}} \phi^{\pdagger}_{\mathcal{T}}(x+y,-y,\Bar{u}) =\theta^{\pdagger}_{\mathcal{T} \sigma_x}
\end{equation}  
leading to
\begin{align}
&    w^{\pdagger}_{\sigma_x}=0 ,\, 2 \theta^{\pdagger}_{\mathcal{T} T_2}=\theta^{\pdagger}_{\mathcal{T} T_1} \\
&    w^{\pdagger}_{\sigma_x}=1, \,\theta^{\pdagger}_{\mathcal{T} T_1}= 0\\
&    \rho^{\pdagger}_{\mathcal{T},u} -(-1)^{w^{\pdagger}_{\sigma_x}} \rho^{\pdagger}_{\mathcal{T},\Bar{u}}=\theta^{\pdagger}_{ \mathcal{T}\sigma_x} \,.\label{eq:U1_T_sigma}
\end{align}
The solutions for $\rho^{\pdagger}_{\mathcal{T},u}$ follow from Eqs.~\eqref{eq:U1_T_C2} and~\eqref{eq:U1_T_sigma}:
\begin{align}
& ( w^{\pdagger}_{C_2},w^{\pdagger}_{\sigma_x})=(0,0) : \hspace{5pt} \rho^{\pdagger}_{\mathcal{T},u}=\{0,n^{\pdagger}_{\mathcal{T}} \pi\}, \\   
&   ( w^{\pdagger}_{C_2},w^{\pdagger}_{\sigma_x})=(1,0) : \hspace{5pt} \rho^{\pdagger}_{\mathcal{T},u}=   \{0,n^{\pdagger}_{\mathcal{T}} \pi\}, \\   
&  ( w^{\pdagger}_{C_2},w^{\pdagger}_{\sigma_x})=(0,1) : \hspace{5pt} \rho^{\pdagger}_{\mathcal{T},u}=  \{0,n^{\pdagger}_{\mathcal{T}} \pi\}, \\    
&   ( w^{\pdagger}_{C_2},w^{\pdagger}_{\sigma_x})=(1,1) :\hspace{5pt} \rho^{\pdagger}_{\mathcal{T},u}= \{0,\theta^{\pdagger}_{\mathcal{T} C_2} \}\,. \   
\end{align}

\smallskip
\textbf{$w^{}_{\mathcal{T}}$\,$=$\,$1$:} In this case, the consistency relation Eq.~\eqref{eq:w_time_constraint} enforces $2\theta=0$, leaving only $\theta=0,\pi$. Here, Eq.~\eqref{eq:U1_g_T2} does not impose further constraints, while Eqs.~\eqref{eq:U1_g_T_C2} and~\eqref{eq:U1_g_T_sigma} become
\begin{align}
    \phi^{\pdagger}_{\mathcal{T}}(x,y,u)-(-1)^{w^{\pdagger}_{C_2}}&\phi^{\pdagger}_{\mathcal{T}}(-x,-y,\Bar{u})\notag \\
&-2  \phi^{\pdagger}_{C_2}(x,y,u)=\theta^{\pdagger}_{\mathcal{T}C_2} \label{eq:U1_T_C2_phi}\\
    \phi^{\pdagger}_{\mathcal{T}}(x,y,u)-(-1)^{w^{\pdagger}_{\sigma_x}}&\phi^{\pdagger}_{\mathcal{T}}(x+y,-y,\Bar{u}) \notag\\
& -2 \phi^{\pdagger}_{\sigma_x}(x,y,u)=\theta^{\pdagger}_{\mathcal{T}\sigma_x}\,.\label{eq:U1_T_sigma_phi} 
\end{align}
Then, solving Eqs.~\eqref{eq:U1_T_C2_phi} and~\eqref{eq:U1_T_sigma_phi} for the four possible $( w^{\pdagger}_{C_2},w^{\pdagger}_{\sigma_x})$ cases yields
\begin{align}
( w^{\pdagger}_{C_2},w^{\pdagger}_{\sigma_x})&=(0,0)\text{: }\notag\\
&      \theta=n\pi,\theta^{\pdagger}_{C_2 T_i}=0,\,\rho^{\pdagger}_{\mathcal{T},u} =\{0,n^{\pdagger}_{\mathcal{T}} \pi\},  \\
&      \theta^{\pdagger}_{\sigma_x T_1}=n\pi,\theta^{\pdagger}_{\sigma_x T_2}=\frac{\theta}{2}+n^{\pdagger}_{\sigma_x T_2}\pi, \\     
&     \theta^{\pdagger}_{\mathcal{T}T_1}=n\pi,\theta^{\pdagger}_{\mathcal{T}T_2}=n^{\pdagger}_{T T_2} \pi. 
\end{align}
\begin{align}
( w^{\pdagger}_{C_2},w^{\pdagger}_{\sigma_x})&=(1,0)\text{: } \notag \\
&     \theta=n\pi,\,\rho^{\pdagger}_{\mathcal{T},u} =\{0,n^{\pdagger}_{\mathcal{T}} \pi\},\\
&      \theta^{\pdagger}_{C_2 T_1}=n\pi,\theta^{\pdagger}_{C_2 T_2}=n^{\pdagger}_{C_2 T_2} \pi,\\
&     \theta^{\pdagger}_{\sigma_x T_1}=n\pi,\theta^{\pdagger}_{\sigma_x T_2}=0,\,2\theta^{\pdagger}_{\mathcal{T}T_2}=\theta^{\pdagger}_{\mathcal{T}T_1}.
\end{align}
\begin{align}
( w^{\pdagger}_{C_2},w^{\pdagger}_{\sigma_x})&=(0,1)\text{: }  \notag \\
&     \theta=n\pi,\,\rho^{\pdagger}_{\mathcal{T},u} =\{0,n^{\pdagger}_{\mathcal{T}} \pi\}, \\
&      \theta^{\pdagger}_{C_2 T_1}=0,\theta^{\pdagger}_{C_2 T_2}=0\\
&     \theta^{\pdagger}_{\sigma_x T_1}=0,\theta^{\pdagger}_{\sigma_x T_2}=n^{\pdagger}_{\sigma_x T_2}\pi-\frac{\theta}{2}, \\ 
&     \theta^{\pdagger}_{\mathcal{T}T_1}=-n\pi,\,\theta^{\pdagger}_{\mathcal{T}T_2}=n^{\pdagger}_{ \mathcal{T} T_2} \pi. 
\end{align}
\begin{align}
( w^{\pdagger}_{C_2},w^{\pdagger}_{\sigma_x})&=(1,1)\text{: } \notag \\
&     \theta=n\pi,\,\theta^{\pdagger}_{\mathcal{T}T_1}=0,\,\rho^{\pdagger}_{\mathcal{T},u} =\{0,\theta^{\pdagger}_{\mathcal{T}}\}, \\
&       \theta^{\pdagger}_{C_2 T_1}=n\pi,\theta^{\pdagger}_{C_2 T_2}=n^{\pdagger}_{C_2 T_2} \pi,\\
&   \theta^{\pdagger}_{\sigma_x T_1}=n\pi,\theta^{\pdagger}_{\sigma_x T_2}=0.
\end{align}

It is possible, and often convenient, to choose a gauge such that $\phi^{\pdagger}_\mathcal{T}(x,y,u)=0$ for $w^{\pdagger}_{\mathcal{T}}=1$, which ensures the resulting \textit{Ans\"atze} contain only real hopping parameters. The required gauge transformation is
\begin{equation}
    W(x,y,u)= \mathcal{F}^{\pdagger}_z\left(\frac{x \theta^{\pdagger}_{\mathcal{T}T_1}}{2}+ \frac{y \theta^{\pdagger}_{\mathcal{T}T_2}}{2}+\frac{\rho^{\pdagger}_{\mathcal{T},u}}{2}\right).
\end{equation}
Applying this transformation for $\mathcal{O} \in \{\mathcal{T},C_2,\sigma_x\}$ gives
\begin{align}
    \Tilde{\phi}^{\pdagger}_{\mathcal{T}}(x,y,u)&=0, \\ 
   \Tilde{ \phi}^{\pdagger}_{C_2}(x,y,u)&=-\phi^{\pdagger}_{T}(x,y,u) + \phi^{\pdagger}_{C_2}(x,y,u)\notag \\
&       + (-1)^{w^{\pdagger}_{C_2}} \phi^{\pdagger}_{T}(-x,-y,\Bar{u}), \label{eq:U1_C2_new}\\
    \Tilde{\phi}^{\pdagger}_{\sigma^{\pdagger}_x}(x,y,u) &=-\phi^{\pdagger}_{T}(x,y,u) +\phi^{\pdagger}_{\sigma_x}(x,y,u) \notag \\
     &+ (-1)^{w^{\pdagger}_{\sigma_x}} \phi^{\pdagger}_{T}(x+y,-y,\Bar{u})\,.\label{eq:U1_sigma_new}
\end{align}

The algebraic PSG solutions for $(w^{\pdagger}_{T_1},w^{\pdagger}_{T_2})=(0,0)$ are summarized in Eqs.~\eqref{eq:U1_wT1=0,U1_wT2=0}--\eqref{eq:U1_wT1=0,U1_wT2=02} and in Table~\ref{tab:U1_PSG_combined} (rows 1---8).

\subsection[]{U(1) class with $w^{\pdagger}_{T_1}=0,\,w^{\pdagger}_{T_2}=1$}
\subsubsection{Lattice symmetries}

In this class of U(1) PSGs, the translation gauge transformations take the general form
\begin{align}
    W^{\pdagger}_{T_1}(x,y,u)
    &=\mathcal{F}^{\pdagger}_z(\phi^{\pdagger}_{T_1}(x,y,u)),\\
    W^{\pdagger}_{T_2}(x,y,u)
    &=\mathcal{F}^{\pdagger}_z(\phi^{\pdagger}_{T_2}(x,y,u))(\dot \iota \tau^{1})\,.
\end{align}
Analogous to the case with $w^{\pdagger}_{T_1}=w^{\pdagger}_{T_2}=0$, one can perform a local gauge transformation such that
\begin{equation}
 \phi^{\pdagger}_{T_2}(x,y,u)=\phi^{\pdagger}_{T_1}(x,0,u)=0\,.
\end{equation}
To determine $\phi^{\pdagger}_{T_1}(x,y,u)$, we use Eq.~\eqref{eq:g_translations} together with the above gauge choice, which yields
\begin{equation}
    \phi^{\pdagger}_{T_1}(x,y,u) = \tfrac{1}{2}\bigl(1-(-1)^{y+1}\bigr)\theta\,.
\end{equation}
After a global phase shift and redefining $\theta\rightarrow2\theta$, the resulting projective translations are
\begin{align}
    W^{\pdagger}_{T_1}(x,y,u)
    &=\mathcal{F}^{\pdagger}_z[(-1)^y \theta], \label{eq:w_T1_b}\\
    W^{\pdagger}_{T_2}(x,y,u)
    &=\dot \iota \tau^x\,. \label{eq:w_t2_b}
\end{align}
Substituting Eqs.~\eqref{eq:w_T1_b} and \eqref{eq:w_t2_b} into Eqs.~\eqref{eq:g_C2_T1} and \eqref{eq:g_C2_T2}, respectively, we obtain
\begin{align}
    \Delta^{\pdagger}_{1} \phi^{\pdagger}_{C_2}(x,y,u)
    &= -(-1)^{w^{\pdagger}_{C_2}}\theta^{\pdagger}_{C_2 T_1}
       + (-1)^{y} \theta \bigl[1+(-1)^{w^{\pdagger}_{C_2}}\bigr], \notag\\
    \Delta^{\pdagger}_{2} \phi^{\pdagger}_{C_2}(x,y,u)
    &= (-1)^{w^{\pdagger}_{C_2}-y}\theta^{\pdagger}_{C_2 T_2}\,, \label{eq:delta_12}
\end{align}
which must satisfy the consistency relation~\eqref{eq:c2_consis}. This leads to the following constraint on $\theta$:
\begin{align}\label{eq:w_sigma_constraint_0} 
    \bigl[1+(-1)^{w^{\pdagger}_{C_2}}\bigr]\theta
    &=0,  \quad 
  \begin{cases}
\theta =0,\pi & \text{if }  w^{\pdagger}_{C_2}=0,\\
\theta \in [0,2\pi) & \text{if }  w^{\pdagger}_{C_2}=1. 
\end{cases}
\end{align}
Using this constraint, the solution for $\phi^{\pdagger}_{C_2}$ follows from Eq.~\eqref{eq:delta_12}:
\begin{align}
     \phi^{\pdagger}_{C_2}(x,y,u)
    =-(-1)^{w^{\pdagger}_{C_2}}\Bigl(x \theta^{\pdagger}_{C_2 T_1}- \zeta^{\pdagger}_y\theta^{\pdagger}_{C_2T_2}\Bigr)
    +\rho^{\pdagger}_{C_2,u}\,,  \label{eq:C2_even}
\end{align}
where $\zeta^{\pdagger}_y=\tfrac{1}{2}\bigl(1+(-1)^y\bigr)$.  

A similar analysis can be applied to $\sigma^{\pdagger}_x$ using Eqs.~\eqref{eq:g_sigma_T1} and \eqref{eq:g_sigma_T2}, together with the consistency condition~\eqref{eq:sigma_consis}, which imposes
\begin{equation}\label{eq:w_sigma_constraint_1} 
\bigl(1-(-1)^{w^{\pdagger}_{\sigma_x}}\bigr)\theta
=0,\quad
\begin{cases}
 \theta\in[0,2\pi)  & \text{if }  w^{\pdagger}_{\sigma_x}=0\,, \\
 \theta=0,\pi  & \text{if } w^{\pdagger}_{\sigma_x}=1\,.
\end{cases}    
\end{equation}
This yields
\begin{align}
    \phi^{\pdagger}_{\sigma_x}(x,y,u)
    =&-(-1)^{w^{\pdagger}_{\sigma_x}}
      \bigl[x \theta^{\pdagger}_{\sigma_xT_1}- \zeta^{\pdagger}_y\theta^{\pdagger}_{\sigma_{x T_2}}\bigr] \notag \\
    &-(-1)^{w^{\pdagger}_{\sigma_x}}y \theta 
      +\rho^{\pdagger}_{\sigma_x,u}\,. 
 \label{eq:rho_b}
\end{align}

The combined constraints from Eqs.~\eqref{eq:w_sigma_constraint_0} and \eqref{eq:w_sigma_constraint_1} are summarized in Table~\ref{tab:U1_theta_2a}.

\begin{table}[h!]
    \centering
    \begin{ruledtabular}
    \begin{tabular}{c ccc}
        Case & $w^{\pdagger}_{C_2}$ (rotations) & $w^{\pdagger}_{\sigma_x}$ (reflections) & Constraints on $\theta$\\
        \hline
        (i) & 0 & 0 & $0,\pi$\\
        (ii) & 1 & 0 &$[0,2\pi)$\\
        (iii) & 0 & 1 & $0,\pi$\\
        (iv) & 1 & 1 & $0,\pi$\\
    \end{tabular}
    \end{ruledtabular}
    \caption{Allowed values of $\theta$ for different combinations of $(w^{\pdagger}_{C_2},w^{\pdagger}_{\sigma_x})$.}
    \label{tab:U1_theta_2a}
\end{table}

We now substitute the solutions~\eqref{eq:C2_even} and~\eqref{eq:rho_b} into the
cyclic relations~\eqref{eq:G_C22} and~\eqref{eq:G_sigma2}, which yield the following constraints:
\begin{align}
   \text{$C^{\pdagger}_2$:}\quad &w^{\pdagger}_{C_2}=0 : \theta^{\pdagger}_{C_2 T_1}\neq 0,\;
   \theta^{\pdagger}_{C_2 T_2}=n^{\pdagger}_{C_2 T_2}\pi, \notag\\
   &w^{\pdagger}_{C_2}=1: \theta^{\pdagger}_{C_2 T_1}=n^{\pdagger}_{C_2 T_1}\pi, \notag\\
   &\rho^{\pdagger}_{C_2,u}+(-1)^{w^{\pdagger}_{C_2}}\rho^{\pdagger}_{C_2,\bar{u}}
   =\theta^{\pdagger}_{C_2}. \\[6pt]
   \text{$\sigma^{\pdagger}_x$:}\quad &w^{\pdagger}_{\sigma_x}=0:\theta^{\pdagger}_{\sigma_x T_1}=0,\;
   \theta^{\pdagger}_{\sigma_x T_2}=n^{\pdagger}_{\sigma_x T_2}\pi, \notag\\
   &w^{\pdagger}_{\sigma_x}=1:\theta^{\pdagger}_{\sigma_x T_1}=0, \notag\\
   &\rho^{\pdagger}_{\sigma_x,u}+(-1)^{w^{\pdagger}_{\sigma_x}}\rho^{\pdagger}_{\sigma_x,\bar{u}}
   =\theta^{\pdagger}_{\sigma_x}.
\end{align}

Next, we perform the gauge transformation
\begin{equation}
    W(x,y,u)=\mathcal{F}^{\pdagger}_z\!\left[(-1)^y \phi^{\pdagger}_u\right],
\end{equation}
and choose the phases $\phi^{\pdagger}_u$ appropriately to simplify the solutions in each case:
\begin{enumerate}[label=(\roman*)]
\item $w^{\pdagger}_{C_2}=0,\;w^{\pdagger}_{\sigma_x}=0$: \\ 
No particular choice of $\phi^{\pdagger}_u$ further simplifies the solutions.

\item $w^{\pdagger}_{C_2}=0,\;w^{\pdagger}_{\sigma_x}=1$:\\  
Choosing $\phi^{\pdagger}_{1}+\phi^{\pdagger}_{2}=-\theta^{\pdagger}_{\sigma_x T_2}/2$ sets $\theta^{\pdagger}_{\sigma_x T_2}=0$, yielding
\begin{align}
    \phi^{\pdagger}_{C_2}(x,y,u)&=-x \theta^{\pdagger}_{C_2 T_1}+\zeta^{\pdagger}_y \theta^{\pdagger}_{C_2 T_2}+\rho^{\pdagger}_{C_2,u}, \\
    \phi^{\pdagger}_{\sigma_x}(x,y,u)&=y\theta+\rho^{\pdagger}_{\sigma_x,u}.
\end{align}

\item $w^{\pdagger}_{C_2}=1,\;w^{\pdagger}_{\sigma_x}=0$: \\ 
Choosing $\phi^{\pdagger}_{u}+\phi^{\pdagger}_{\bar{u}}=-\theta^{\pdagger}_{C_2 T_2}/2$ sets $\theta^{\pdagger}_{C_2 T_2}=0$, leading to
\begin{align}
    \phi^{\pdagger}_{C_2}(x,y,u)&=x \theta^{\pdagger}_{C_2 T_1}+\rho^{\pdagger}_{C_2,u}, \\
    \phi^{\pdagger}_{\sigma_x}(x,y,u)&=\zeta^{\pdagger}_y \theta^{\pdagger}_{\sigma_x T_2}-y\theta+\rho^{\pdagger}_{\sigma_x,u}.
\end{align}

\item $w^{\pdagger}_{C_2}=1,\;w^{\pdagger}_{\sigma_x}=1$:  \\
Using the same choice and setting $\theta^{\pdagger}_{C_2 T_2}=0$ gives
\begin{align}
    \phi^{\pdagger}_{C_2}(x,y,u)&=x \theta^{\pdagger}_{C_2 T_1}+\rho^{\pdagger}_{C_2,u}, \\
    \phi^{\pdagger}_{\sigma_x}(x,y,u)&=-\zeta^{\pdagger}_y \theta^{\pdagger}_{\sigma_x T_2}+y\theta+\rho^{\pdagger}_{\sigma_x,u}.
\end{align}
\end{enumerate}
Furthermore, Eq.~\eqref{eq: G_C2_s2} enforces $\theta^{\pdagger}_{C_2 T_1}=0$ for all four combinations of $(w^{\pdagger}_{C_2},w^{\pdagger}_{\sigma_x})$, together with the following additional constraints:
\begin{align}
&w^{\pdagger}_{C_2}=w^{\pdagger}_{\sigma_x}=0:\quad 
   \theta^{\pdagger}_{C_2 T_2}=m\pi-\theta^{\pdagger}_{\sigma_x T_2}, \\
&w^{\pdagger}_{C_2}=w^{\pdagger}_{\sigma_x}=1:\quad 
   \theta^{\pdagger}_{\sigma_x T_2}=n^{\pdagger}_{\sigma_x T_2}\pi.
\end{align}
In addition, Eq.~\eqref{eq:G_C2_T1_sigma_x} imposes:
\begin{align}
&w^{\pdagger}_{C_2}=0,\;w^{\pdagger}_{\sigma_x}=0:\quad 
   \theta^{\pdagger}_{C_2 T_2}=m\pi-\theta^{\pdagger}_{\sigma_x T_2}, \\
&w^{\pdagger}_{C_2}=1,\;w^{\pdagger}_{\sigma_x}=0:\quad 
   \theta=n\pi, \\
&w^{\pdagger}_{C_2}=0,\;w^{\pdagger}_{\sigma_x}=1:\quad 
   \theta=n\pi, \\
&w^{\pdagger}_{C_2}=1,\;w^{\pdagger}_{\sigma_x}=1:\quad 
   \theta^{\pdagger}_{\sigma_x T_2}=n^{\pdagger}_{\sigma_x T_2}\pi.
\end{align}

Last, let us discuss how to fix all the $\rho$ parameters.
From the constraints imposed by \eqref{eq: G_C2_s2}, Eqs.~\eqref{eq:G_C22}, \eqref{eq:G_sigma2}, and~\eqref{eq:G_C2_T1_sigma_x}, we obtain
\begin{align*}
    \rho^{\pdagger}_{C_2,u}+(-1)^{w^{\pdagger}_{C_2}}\rho^{\pdagger}_{C_2,\bar{u}}&=\theta^{\pdagger}_{C_2}, \\
    \rho^{\pdagger}_{\sigma_x,u}+(-1)^{w^{\pdagger}_{\sigma_x}}\rho^{\pdagger}_{\sigma_x,\bar{u}}&=\theta^{\pdagger}_{\sigma_x}, \\
    \rho^{\pdagger}_{C_2,u}+(-1)^{w^{\pdagger}_{C_2}}\rho^{\pdagger}_{\sigma_x,\bar{u}}+(-1)^{w^{\pdagger}_{C_2}+w^{\pdagger}_{\sigma_x}}
    \rho^{\pdagger}_{C_2,u} \notag\\
    +(-1)^{w^{\pdagger}_{\sigma_x}}\rho^{\pdagger}_{\sigma_x,\bar{u}}&=\theta^{\pdagger}_{C_2 \sigma_x}, \label{eq:rho_C2s2}\\
    \rho^{\pdagger}_{C_2,u}-(-1)^{w^{\pdagger}_{C_2}}\rho^{\pdagger}_{\sigma_x,\bar{u}}-(-1)^{w^{\pdagger}_{C_2}+w^{\pdagger}_{\sigma_x}}
    \rho^{\pdagger}_{C_2,u} \notag\\
    +(-1)^{w^{\pdagger}_{\sigma_x}}\rho^{\pdagger}_{\sigma_x,\bar{u}}&=\theta^{\pdagger}_{C_2 \sigma_x T_2}.
\end{align*}
These relations yield the solutions
\begin{align}
w^{\pdagger}_{C_2}=0,\;w^{\pdagger}_{\sigma_x}=0:\quad & 
\begin{cases}
    \rho^{\pdagger}_{C_2,u} \in \{0,\;\theta^{\pdagger}_{C_2}\}, \\[3pt]
    \rho^{\pdagger}_{\sigma_x,u} \in \{0,\;\theta^{\pdagger}_{\sigma_x}\},
\end{cases} \\[6pt]
\text{All other cases}:\quad &
\begin{cases}
    \rho^{\pdagger}_{C_2,u} \in \{0,\;n^{\pdagger}_{C_2}\pi\}, \\[3pt]
    \rho^{\pdagger}_{\sigma_x,u} \in \{0,\;n^{\pdagger}_{\sigma_x}\pi\}.
\end{cases}
\end{align}

\subsubsection{Time-reversal symmetry}
We now analyze the PSG solutions associated with time-reversal symmetry. From Eqs.~\eqref{eq:U1_g_T_T1} and \eqref{eq:U1_g_T_T2}, we arrive at
\begin{align}
    \Delta^{\pdagger}_1 \phi^{\pdagger}_{\mathcal{T}}&=\theta^{\pdagger}_{\mathcal{T}T_1}+(-1)^y \theta[1-(-1)^{w^{\pdagger}_\mathcal{T}}],\label{eq:time_trans_1_1} \\
    \Delta^{\pdagger}_2 \phi^{\pdagger}_{\mathcal{T}}&=(-1)^y \theta^{\pdagger}_{\mathcal{T}T_2}\label{eq:time_trans_1_2}\,,
\end{align}
which, when substituted into the consistency condition~\eqref{eq:time_consis}, result in
\begin{equation}\label{eq:w_time_constraint_1} 
\left(1-(-1)^{w^{\pdagger}_{\sigma_x}}\right)
=0,\quad
\begin{cases}
 \theta\in[0,2\pi)  & \text{if }  w^{\pdagger}_{\mathcal{T}}=0, \\
 \theta=0,\pi  & \text{if } w^{\pdagger}_{\mathcal{T}}=1.
\end{cases}    
\end{equation}
A general solution for $\phi^{\pdagger}_{\mathcal{T}}(x,y,u)$ can then be written by combining these constraints on $\theta$ with Eqs.~\eqref{eq:time_trans_1_1} and \eqref{eq:time_trans_1_2}:
\begin{equation}
    \theta^{\pdagger}_{\mathcal{T}}(x,y,u)=x \theta^{\pdagger}_{\mathcal{T}T_1}+\zeta^{\pdagger}_y\theta^{\pdagger}_{\mathcal{T}T_2}+\rho^{\pdagger}_{\mathcal{T},u}\,.
\end{equation}
We now consider separately the two cases $w^{\pdagger}_{\mathcal{T}}=0$ and $w^{\pdagger}_{\mathcal{T}}=1$.

\smallskip
\textbf{$w^{}_{\mathcal{T}}$\,$=$\,$0$:}
Equation~\eqref{eq:U1_g_T2} requires $\theta^{\pdagger}_{\mathcal{T}T_i}=n^{\pdagger}_{\mathcal{T}T_i}\pi$. Furthermore, solving Eqs.~\eqref{eq:U1_g_T_C2} and \eqref{eq:U1_g_T_sigma} gives $\theta^{\pdagger}_{\mathcal{T}T_1}=0$, with
\begin{align}
 w^{\pdagger}_{C_2}=1,\;w^{\pdagger}_{\sigma_x}=1:&\quad\rho^{\pdagger}_{\mathcal{T},u}=\{0,\theta^{\pdagger}_{\mathcal{T}}\},  \\
 \text{All other cases}:&\quad\rho^{\pdagger}_{\mathcal{T},u}=\{0,n^{\pdagger}_{\mathcal{T}}\pi\}\,.
\end{align}

\smallskip
\textbf{$w^{}_{\mathcal{T}}$\,$=$\,$1$:}
Solving Eqs.~\eqref{eq:U1_g_T2}, \eqref{eq:U1_g_T_C2}, and \eqref{eq:U1_g_T_sigma} yields $\theta^{\pdagger}_{\mathcal{T}T_1}=0$, with the following possibilities:
\begin{align*}
w^{\pdagger}_{C_2}=0,\;w^{\pdagger}_{\sigma_x}=0:\,&\theta^{\pdagger}_{\mathcal{T}T_2}\in[0,2\pi),\quad
\rho^{\pdagger}_{\mathcal{T},u}=\{0,\theta^{\pdagger}_{\mathcal{T}}\},  \\
w^{\pdagger}_{C_2}=0,\;w^{\pdagger}_{\sigma_x}=1:\,&\theta^{\pdagger}_{\mathcal{T}T_2}=n^{\pdagger}_{\mathcal{T}T_2}\pi,\quad
\rho^{\pdagger}_{\mathcal{T},u}=\{0,n^{\pdagger}_{\mathcal{T}}\pi\}, \\
w^{\pdagger}_{C_2}=1,\;w^{\pdagger}_{\sigma_x}=0:\,&\theta^{\pdagger}_{\mathcal{T}T_2}=n^{\pdagger}_{\mathcal{T}T_2}\pi,\quad
\rho^{\pdagger}_{\mathcal{T},u}=\{0,n^{\pdagger}_{\mathcal{T}}\pi\}, \\
w^{\pdagger}_{C_2}=1,\;w^{\pdagger}_{\sigma_x}=1:\,&\theta^{\pdagger}_{\mathcal{T}T_2}=n^{\pdagger}_{\mathcal{T}T_2}\pi,\quad
\rho^{\pdagger}_{\mathcal{T},u}=\{0,\theta^{\pdagger}_{\mathcal{T}}\}\,.
\end{align*}

Finally, the complete set of algebraic PSG solutions for the $(w^{\pdagger}_{T_1},w^{\pdagger}_{T_2})$\,$=$\,$(0,1)$ class is summarized in Eqs.~\eqref{eq:U1_wT1=0,U1_wT2=0}--\eqref{eq:U1_wT1=0,U1_wT2=02} and in Table~\ref{tab:U1_PSG_combined} (rows 9--16).

\begin{table*}
\caption{PSG parameter values $\theta$ and $\rho$ for the U(1) states on the trellis lattice, corresponding to the \textit{Ans\"atze} illustrated in Fig.~\ref{fig:u1_ansatz}.}
\begin{ruledtabular}
\begin{tabular}{ccccccccccc}
Label & PSG No. & $\theta$ & $\theta^{}_{C_2 T_1}$ & $\theta^{}_{C_2 T_2}$ & $\theta^{}_{\sigma_x T_1}$ & $\theta^{}_{\sigma_x T_2}$ & $\theta^{}_{\mathcal{T} T_2}$ & $\rho^{}_{C_2,u}$ & $\rho^{}_{\sigma_x,u}$ & $\rho^{}_{\mathcal{T},u}$ \\
\hline
U1 & 5 & 0 & 0 & 0 & 0 & 0 & 0 & 0 & 0 & 0 \\
U2 & 8 & 0 & 0 & 0 & 0 & 0 & 0 & $\pi$ & $\pi$ & 0 \\
U3 & 5 & $\pi$ & $\pi$ & 0 & 0 & 0 & 0 & 0 & 0 & 0 \\
U4 & 8 & $\pi$ & $\pi$ & 0 & 0 & 0 & 0 & $\pi$ & $\pi$ & 0 \\
U5 & 13 & $0$ & - & $\pi$ & - & $\pi$ & 0 & 0 & 0 & 0 \\
U6 & 16 & $0$ & - & $0$ & - & 0 & $\pi$ & $\pi$ & $\pi$ & 0 \\
\hline
U7 & 2 & $\theta$ & $0$ & $0$ & 0 & 0 & $\pi$ & 0 & $\pi$ & 0 \\
\end{tabular}
\end{ruledtabular}
\label{tab:theta_rho_PSG}
\end{table*}

\section{$\mathbb{Z}_2$ projective symmetry groups}
\label{app:z2_PSG_derivation}

In this section, we derive the algebraic solutions for projective symmetry groups assuming a $\mathbb{Z}_2$ invariant gauge group. The \textit{Ansatz} now incorporates both hopping and pairing terms, leading to the general form
\begin{equation}
    u^{\pdagger}_{ij}= \dot \iota \chi^{0}_{ij} \tau^{0} +\chi_{ij}^{3}\tau^{3}+\Delta_{ij}^{1}\tau^{1}+\Delta_{ij}^{2}\tau^{2}.
\end{equation}
Here, the coefficients of $\tau^{3}$ and $\tau^{0}$ correspond to real and imaginary hopping amplitudes, respectively, while the coefficients of $\tau^{1}$ and $\tau^{2}$ correspond to real and imaginary pairing amplitudes. The elements of the $\mathbb{Z}_2$ IGG are parameterized by $\eta_{\ldots} \in \{ \pm 1 \}$. We derive the projective realizations of the lattice and time-reversal symmetries separately in the following subsections.

\subsection{Lattice symmetries}

Analogous to the U(1) case, an appropriate choice of local gauge allows one to set $ W^{\pdagger}_{T_1}(x,0,u)= W^{\pdagger}_{T_2}(x,y,u)=\tau^0$. Substituting this into the gauge condition~\eqref{eq:g_translations} yields the projective representations of the translations:
\begin{equation}
     W^{\pdagger}_{T_1}(x,y,u)=\eta_{T_1}^y \tau^0,\quad W^{\pdagger}_{T_2}(x,y,u) = \tau^0 \,.\label{eq: gaugeT2}
\end{equation}

To determine the PSG for $C_2$, we insert Eq.~\eqref{eq: gaugeT2} into Eqs.~\eqref{eq:g_C2_T1} and~\eqref{eq:g_C2_T2}, leading to the recursion relations
\begin{align}
     W^{\pdagger}_{C_2}(x,y,u)&= \eta^{\pdagger}_{C_2 T_1} W^{\pdagger}_{C_2}(x-1,y,u), \label{eq: gaugeC2}\\
     W^{\pdagger}_{C_2}(x,y,u)&=\eta^{\pdagger}_{C_2 T_2} W^{\pdagger}_{C_2}(x,y-1,u),  \label{eq:C22}\,
\end{align}
solving which gives
\begin{equation}
    W^{\pdagger}_{C_2}(x,y,u)= \eta_{C_2 T_2}^y \eta_{C_2 T_1}^x W^{\pdagger}_{C_2}(0,0,u) \,.\label{eq:G_C2}
\end{equation}

Similarly, using Eqs.~\eqref{eq:g_sigma_T1},~\eqref{eq:g_sigma_T2} and~\eqref{eq: gaugeT2}, we obtain
\begin{align}
    W^{\pdagger}_{\sigma_x}(x,y,u)&=\eta^{\pdagger}_{\sigma_x T_1} W^{\pdagger}_{\sigma_x}(x-1,y,u), \label{eq:gauge_sigma1}\\
    W^{\pdagger}_{\sigma_x}(x,y,u) &=\eta_{T_1}^{-y+1}\eta^{\pdagger}_{\sigma_x T_2} W^{\pdagger}_{\sigma_x}(x,y-1,u), \label{eq:gauge_sigma2}      
\end{align}
which are solved by
\begin{align}
    W^{\pdagger}_{\sigma_x}(x,y,u) = \eta_{T_1}^{-\frac{y(y-1)}{2}}\eta_{\sigma_x T_2}^y \eta_{\sigma_x T_1}^x W^{\pdagger}_{\sigma_x}(0,0,u)\, . \label{eq:G_sigma}
\end{align}
From Eq.~\eqref{eq: G_C2_s2}, we obtain  
\begin{align}
    W^{\pdagger}_{C_2,u}W^{\pdagger}_{\sigma_x,\Bar{u}}W^{\pdagger}_{C_2,u} W^{\pdagger}_{\sigma_x,\Bar{u}} &=\eta^{\pdagger}_{C_2\sigma_x}\tau^0, \\
   \eta^{\pdagger}_{C_2 T_1} &=\eta^{\pdagger}_{\sigma_x T_1}.
\end{align}
Similarly, Eq.~\eqref{eq:G_C22} yields
\begin{align}
    W^{\pdagger}_{C_2,u}W^{\pdagger}_{C_2,\Bar{u}}=\eta^{\pdagger}_{C_2} \tau^0,
\end{align}
while Eq.~\eqref{eq:G_sigma2} gives
\begin{align}
    W^{\pdagger}_{\sigma_x,u}W^{\pdagger}_{\sigma_x,\bar{u}}&=\eta^{\pdagger}_{\sigma_x}\tau^{0}, \\
    \eta^{\pdagger}_{\sigma_x T_1}&=\eta^{\pdagger}_{T_1}.
\end{align}

Further gauge fixing can be implemented through a transformation of the form 
\begin{align}
    W(x,y,u)=\eta_x^x \eta_y^y \tau^0,
\end{align}
which leaves the translational gauge structure unchanged, up to global sign factors. These global signs can be absorbed into the IGG freedom, allowing us to set one of them to $+1$ without loss of generality. Under this transformation, the operators ${W}_{C_2}(x,y,u)$ and ${W}_{\sigma_x}(x,y,u)$ transform as
\begin{align}
    &\Tilde{W}^{\pdagger}_{C_2}(x,y,u)=\eta_{C_2 T_2}^{y} \eta_{C_2 T_1}^{x} W^{\pdagger}_{C_2,u}, \\
    &\Tilde{W}^{\pdagger}_{\sigma_x}(x,y,u) =\eta_{x}^y \eta_{T_1}^{-\frac{y(y-1)}{2}} \eta_{\sigma_x T_2}^{y} \eta_{\sigma_x T_1}^{x} W^{\pdagger}_{\sigma_x ,u}.
\end{align}
We observe that $W_{C_2}(x,y,u)$ remains unaffected, while $W_{\sigma_x}(x,y,u)$ acquires an additional sign structure. By choosing $\eta_{x}=\eta_{\sigma_x T_2}$, one can eliminate the parameter $\eta_{\sigma_x T_2}$ from $W_{\sigma_x}$. In this gauge, the solutions take the form
\begin{align}
    &W^{\pdagger}_{T_1}(x,y,u)=\eta_{T_1}^y \tau^0,\quad 
      W^{\pdagger}_{T_2}(x,y,u) =\tau^0, \\
    &W^{\pdagger}_{C_2}(x,y,u)=\eta_{C_2 T_2}^{y} \eta_{T_1}^{x} W^{\pdagger}_{C_2,u}, \\
    &W^{\pdagger}_{\sigma_x}(x,y,u) =\eta_{T_1}^{-\frac{y(y-1)}{2}+x} W^{\pdagger}_{\sigma_x ,u}.
\end{align}
For notational convenience, we omit the tilde in what follows.  

Next, combining  \eqref{eq: G_C2_s2}--\eqref{eq:G_sigma2}, we obtain the constraints
\begin{align}
   & W^{\pdagger}_{C_2,u}W^{\pdagger}_{C_2,\Bar{u}}=\eta^{\pdagger}_{C_2}\tau^0, \label{eq:W_C2}\\
   & W^{\pdagger}_{\sigma_x,u}W^{\pdagger}_{\sigma_x,\Bar{u}}=\eta^{\pdagger}_{\sigma_x} \tau^0, \label{eq:z_cyclic_sigma} \\
   & W^{\pdagger}_{C_2,u}W^{\pdagger}_{\sigma_x,\Bar{u}}W^{\pdagger}_{C_2,u}W^{\pdagger}_{\sigma_x,\Bar{u}}=\eta^{\pdagger}_{C_2 \sigma_x} \tau^0.
\end{align}

At this stage, one may still perform a sublattice-dependent gauge transformation of the form $W(x,y,u)=W^{\pdagger}_u$, under which the sublattice-dependent components transform as
\begin{align}
    &\Tilde{W}^{\pdagger}_{C_2,u} = W^{\dagger}_u W^{\pdagger}_{C_2,u} W^{\pdagger}_{\Bar{u}}, \label{eq:C2_tilde} \\
    &\Tilde{W}^{\pdagger}_{\sigma_x,u} = W^{\dagger}_u W^{\pdagger}_{\sigma_x,u} W^{\pdagger}_{\Bar{u}}. \label{eq:sigma_tilde}
\end{align}
By choosing $W_1=W_{\sigma_x,u} W_{2}$, we can fix ${W}_{\sigma_x,1} \to\tau^0$. Substituting into Eq.~\eqref{eq:z_cyclic_sigma}, we obtain
\begin{equation}
    W^{\pdagger}_{\sigma_x}=\{\tau^0,\eta^{\pdagger}_{\sigma_x }\tau^0\}. \label{eq:W_sigmax}
\end{equation}
Combining Eq.~\eqref{eq:W_C2} with Eq.~\eqref{eq:W_sigmax}, we find
\begin{align}
   W_{C_2,u}^2=\eta^{\pdagger}_{C_2 \sigma_x} \tau^0 
   \;\;\Rightarrow\;\; W_{C_2,1}^2=W_{C_2,2}^2=\eta^{\pdagger}_{C_2 \sigma_x} \tau^0,
\end{align}
leading to two distinct cases:
\begin{align}
    \eta^{\pdagger}_{C_2 \sigma_x}=+1 &: W^{\pdagger}_{C_2,u} =\eta^{\pdagger}_{C,u} \tau^0, \label{eq:eta_C2_1}\\
    \eta^{\pdagger}_{C_2 \sigma_x}=-1&: W^{\pdagger}_{C_2,u}=i\Vec{\alpha}_{C_2,u} \cdot \Vec{\tau}. \label{eq:eta_C2_-1}
\end{align}
Plugging Eq.~\eqref{eq:eta_C2_1} back into Eq.~\eqref{eq:W_C2}, we obtain the condition 
\begin{align}
    \eta^{\pdagger}_{C,1} \eta^{\pdagger}_{C,2} =\eta^{\pdagger}_{C_2},
\end{align}
which implies 
\begin{align}
    W^{\pdagger}_{C_2,u}=\{\tau^0,\eta^{\pdagger}_{C_2} \tau^0\}.
\end{align}
On the other hand, substituting Eq.~\eqref{eq:eta_C2_-1} into Eq.~\eqref{eq:W_C2} gives
\begin{align}
    W^{\pdagger}_{C_2,u}=\{1,\eta^{\pdagger}_{C_2}\}\, \dot \iota \,\tau^z.
\end{align}

\subsection{Time-reversal symmetry}
Akin to the U(1) case, one can employ Eqs.~\eqref{eq:U1_g_T2}--\eqref{eq:U1_g_T_sigma} to determine the $\mathbb{Z}_2$ PSG associated with time-reversal symmetry. On substituting Eq.~\eqref{eq: gaugeT2} into Eqs. \eqref{eq:U1_g_T_T1} and~\eqref{eq:U1_g_T_T2}, we obtain
\begin{align}
    W^{\pdagger}_{\mathcal{T}}(x,y,u)=\eta_{\mathcal{T}x}^x \eta_{\mathcal{T}y}^y W^{\pdagger}_{\mathcal{T},u}. \label{eq:G_T}
\end{align}
Next, inserting the solutions given by Eqs.~\eqref{eq:G_C2},~\eqref{eq:G_sigma}, and~\eqref{eq:G_T} into Eqs.~\eqref{eq:U1_g_T2},~\eqref{eq:U1_g_T_C2}, and~\eqref{eq:U1_g_T_sigma}, we arrive at the following constraints:
\begin{align}
   W_{\mathcal{T},u}^2 &=\eta^{\pdagger}_{\mathcal{T}} \tau^0, \label{eq:W_T2} \\
   W^{\pdagger}_{\mathcal{T},u} W^{\pdagger}_{C_2,u} W_{\mathcal{T},\Bar{u}}^{-1} W_{C_2,u}^{-1}&= \eta^{\pdagger}_{\mathcal{T} C_2} \tau^0\,,\label{eq:W_TC2}\\
    \eta^{\pdagger}_{\mathcal{T}x}&=1,\\
   W^{\pdagger}_{\mathcal{T},u} W^{\pdagger}_{\sigma_x,u} W_{\mathcal{T},\Bar{u}}^{-1} W_{\sigma_x,u}^{-1}&=  \eta^{\pdagger}_{\mathcal{T} \sigma_x} \tau^0\,.\label{eq:W_T_sigma}
\end{align}
From Eqs.~\eqref{eq:W_sigmax} and~\eqref{eq:W_T_sigma}, one deduces that 
\begin{align}
    W^{\pdagger}_{\mathcal{T},2}=\eta^{\pdagger}_{\mathcal{T} \sigma_x}W^{\pdagger}_{\mathcal{T},1}.
\end{align}
Combining this with Eq.~\eqref{eq:W_T2} leads to
\begin{align} 
    & \eta^{\pdagger}_{\mathcal{T}}=+1 : \quad W^{\pdagger}_{\mathcal{T},u}=\{1,\eta^{\pdagger}_{\mathcal{T} \sigma_x}\}\tau^0, \label{eq:eta_+1} \\
    & \eta^{\pdagger}_{\mathcal{T}}=-1 : \quad W^{\pdagger}_{\mathcal{T},u}=\{1,\eta^{\pdagger}_{\mathcal{T} \sigma_x}\}\, \dot \iota \,\boldsymbol{\hat{v}}\cdot\mathbf{\hat{\tau}},
\end{align}
where $\boldsymbol{\hat{v}} \in \mathbb{R}^3$ is a unit vector. Importantly, the case $\eta_{\mathcal{T} \sigma_x} = +1$ with $\eta_{\mathcal{T}} = +1$ does not yield a PSG compatible with nonvanishing mean-field amplitudes, and is therefore excluded from the classification. The remaining ambiguity associated with $\boldsymbol{\hat{v}}$ is resolved through the constraint in Eq.~\eqref{eq:W_TC2}. All gauge-inequivalent solutions, including those derived here, are summarized in Table~\ref{table:z2_psg1}.

\begin{table}[h!]
    \centering
    \caption{Representative PSG phase parameters $\theta$ for $\mathbb{Z}_2$ states on the trellis lattice. Each row corresponds to a gauge-inequivalent {\it Ansatz} labeled Z1--Z25. The labeling scheme follows the U(1) case: each $\eta_i$ is represented by its phase angle $\theta_i = \text{Arg}(\eta_i)$. For example, $\eta_{T_1} = 1$ corresponds to $\theta_{T_1} = 0$, while $\eta_{T_1} = -1$ corresponds to $\theta_{T_1} = \pi$. This convention is applied to all $\eta_i$ with $i =T_1, C_2T_2, \sigma_x, \mathcal{T}_y, C_2, \mathcal{T}$.}
    \begin{ruledtabular}
    \begin{tabular}{cccccccc}
    Label & PSG No. & $\theta^{\pdagger}_{T_1}$ & $\theta^{\pdagger}_{C_2 T_2}$ & $\theta^{\pdagger}_{\sigma_x}$ & $\theta^{\pdagger}_{\mathcal{T}_y}$ & $\theta^{\pdagger}_{C_2}$ & $\theta^{\pdagger}_\mathcal{T}$ \\
    \hline
    Z1  & 2 & 0 & 0 & 0 & 0 & 0 & 0 \\
    Z2  & 2 & 0 & 0 & 0 & $\pi$ & 0 & 0 \\
    Z3  & 2 & $\pi$ & 0 & 0 & 0 & 0 & 0 \\
    Z4  & 2 & $\pi$ & 0 & 0 & $\pi$ & 0 & 0 \\
    Z5  & 2 & 0 & $\pi$ & 0 & $\pi$ & 0 & 0 \\
    Z6  & 2 & $\pi$ & $\pi$ & 0 & $\pi$ & 0 & 0 \\
    Z7  & 2 & 0 & 0 & 0 & 0 & 0 & $\pi$ \\
    Z8  & 2 & $\pi$ & 0 & 0 & 0 & 0 & $\pi$ \\
    Z9  & 2 & 0 & 0 & $\pi$ & 0 & $\pi$ & $\pi$ \\
    Z10 & 2 & $\pi$ & 0 & $\pi$ & 0 & $\pi$ & $\pi$ \\
    Z11 & 2 & 0 & $\pi$ & 0 & 0 & 0 & $\pi$ \\
    Z12 & 2 & 0 & $\pi$ & $\pi$ & 0 & $\pi$ & $\pi$ \\
    Z13 & 2 & $\pi$ & $\pi$ & $\pi$ & 0 & $\pi$ & $\pi$\\
    Z14 & 2 & 0 & $\pi$ & $\pi$ & $\pi$ & $\pi$ & $\pi$ \\
    Z15 & 2 & $\pi$ & 0 & 0 & $\pi$& 0 & $\pi$ \\
    Z16 & 2 & 0 & 0 & 0 & $\pi$ & 0 & $\pi$ \\
    Z17 & 2 & $\pi$ & $\pi$ & $\pi$ & $\pi$ & $\pi$ & $\pi$ \\
    \hline
    Z18 & 5 & 0 & $\pi$ & 0 & $\pi$ & 0 & 0 \\
    Z19 & 5 & $\pi$ & $\pi$ & 0 & $\pi$ & 0 & 0 \\
    Z20 & 5 & 0 & 0 & 0 & $\pi$ & $\pi$ & 0 \\
    Z21 & 5 & $\pi$ & 0 & 0 & $\pi$ & $\pi$ & 0 \\
    Z22 & 5 & 0 & 0 & 0 & 0 & $\pi$ & $\pi$ \\
    Z23 & 5 & $\pi$ & 0 & 0 & 0 & $\pi$ & $\pi$ \\
    Z24 & 5 & 0 & 0 & $\pi$ & 0 & $\pi$ & $\pi$ \\
    Z25 & 5 & $\pi$ & 0 & $\pi$ & 0 & $\pi$ & $\pi$ \\
    \end{tabular}
    \end{ruledtabular}
    \label{tab:z2_theta}
\end{table}

\section{Construction of mean-field \textit{Ans\"atze}}
\label{app:u1_ansatz}

Having obtained all projective realizations of the space-group symmetry elements, one can construct lattice-symmetric mean-field \textit{Ans\"atze} using the condition
\begin{equation}
    W^{\dagger}_{\mathcal{O}} (\mathcal{O}(i))\, u^{\pdagger}_{\mathcal{O}(i) \mathcal{O}(j)}\, W^{\pdagger}_{\mathcal{O}} (\mathcal{O}(j)) = u^{\pdagger}_{ij}.
\end{equation}
For a composite symmetry operator $\mathcal{O}_3 = \mathcal{O}_2 \mathcal{O}_1$, this condition becomes
\begin{align}
      \label{gauge_uij_1}
    W^{\dagger}_{\mathcal{O}_1}(\mathcal{O}_1(i))\, &W^{\dagger}_{\mathcal{O}_2} (\mathcal{O}_3(i))\, &u^{\pdagger}_{\mathcal{O}_3(i) \mathcal{O}_3(j)} \notag\\
    \times &W^{\pdagger}_{\mathcal{O}_2} (\mathcal{O}_3(j))\, W^{\pdagger}_{\mathcal{O}_1} (\mathcal{O}_1(j)) = u^{\pdagger}_{ij}.
\end{align}
Analogous conditions can be derived for any \textit{Ansatz} involving longer strings of symmetry operations.

We define symmetry-allowed mean-field parameters on representative bonds within a unit cell. The corresponding parameters on all other bonds are generated via translations:
\begin{align}
\text{On $J_v$ bonds: } & u^{\pdagger}_{(0,0,1),(0,0,2)} = u^{\pdagger}_{v},\\  
\text{On $J_h$ bonds: } & u^{\pdagger}_{(0,0,1),(1,0,1)} = u^{\pdagger}_{h},\\
 & u^{\pdagger}_{(0,0,2),(1,0,2)} = u^\prime_{h},\\
\text{On $J_z$ bonds: } & u^{\pdagger}_{(0,0,1),(0,1,2)} = u^{\pdagger}_{z}, \\
 & u^{\pdagger}_{(0,0,1),(-1,1,2)} = u_{z}^{\prime}.
\end{align}

A symmetry operation $\mathcal{O}$ can affect a link field $u^{\pdagger}_{ij}$ in one of three ways:
\begin{enumerate}
    \item Leave it invariant: $\mathcal{O}: u^{\pdagger}_{ij} \rightarrow u^{\pdagger}_{ij}$.
    \item Reverse its direction: $\mathcal{O}: u^{\pdagger}_{ij} \rightarrow u^{\pdagger}_{ji} = u^\dagger_{ij}$.
    \item Map it to a different, symmetry-equivalent link: $\mathcal{O}: u^{\pdagger}_{ij} \rightarrow u^{\pdagger}_{i'j'}$, with $\langle i'j' \rangle \neq \langle ij \rangle$.
\end{enumerate}
The first two scenarios determine symmetry-allowed mean-field parameters, while the third relates distinct but symmetry-equivalent bonds across the lattice.

The conditions relevant  for our classification are
\begin{align}
\sigma^{\pdagger}_x: & \quad u^{\pdagger}_{v} \rightarrow u_{v}^{\dagger},\\
C^{\pdagger}_2: & \quad u^{\pdagger}_{v} \rightarrow u_{v}^{\dagger},\\
C^{\pdagger}_2 \sigma_x: & \quad u^{\pdagger}_v \rightarrow u^{\pdagger}_v,\\
T^{\pdagger}_1 C^{\pdagger}_2 \sigma_x: & \quad u^{\pdagger}_h \rightarrow u_h^{\dagger},\\
\sigma^{\pdagger}_x: & \quad u^{\pdagger}_h \rightarrow u^\prime_h,\\
T^{\pdagger}_2 C^{\pdagger}_2: & \quad u^{\pdagger}_z \rightarrow u_z^{\dagger},\\
\sigma^{\pdagger}_x C^{\pdagger}_2: & \quad u^{\pdagger}_z \rightarrow u^\prime_z.
\end{align}
The corresponding projective symmetry conditions read:
 \begin{align}
W_{\sigma_x}^{\dagger}(0,0,2)u_v^{\dagger} W^{\pdagger}_{\sigma_x}(0,0,1)&=u^{\pdagger}_v,\\ 
W_{C_2}^{\dagger}(0,0,2)u_v^{\dagger} W^{\pdagger}_{C_2}(0,0,1)&=u^{\pdagger}_v ,\\
W_{\sigma_x}^{\dagger}(0,0,2) W_{C_2}^{\dagger}(0,0,1) u^{\pdagger}_vW^{\pdagger}_{C_2}(0,0,2)&\notag \\ W^{\pdagger}_{\sigma_x}(0,0,1)&=u^{\pdagger}_v ,\\
W_{\sigma_x}^{\dagger}(0,0,2) W_{C_2}^{\dagger}(0,0,1)W_{T_1}^{\dagger}(1,0,1) u_{h}^{\dagger} &\notag\\W^{\pdagger}_{T_1}(0,0,1)W^{\pdagger}_{C_2}(-1,0,1)W^{\pdagger}_{\sigma_x}(1,0,2)&=u^{\pdagger}_h, \\
W_{\sigma_x}^{\dagger}(0,0,2)u^\prime_h W^{\pdagger}_{\sigma_x}(1,0,2)&=u^{\pdagger}_h,\\
 W_{C_2}^{\dagger}(0,0,2)W_{T_2}^{\dagger}(0,1,2) u_{z}^{\dagger} W^{\pdagger}_{T_2}(0,0,1)&\notag\\ W^{\pdagger}_{C_2}(0,-1,1)&= u^{\pdagger}_z, \\
   W_{C_2}^{\dagger}(0,0,2)W_{\sigma_x}^{\dagger}(0,0,1)u^\prime_z W^{\pdagger}_{\sigma_x}(-1,1,2)&\notag\\
   W^{\pdagger}_{C_2}(0,-1,1) &=u^{\pdagger}_z\,.
 \end{align}
Additionally, time-reversal invariance dictates that
\begin{equation}
    W_{\mathcal{T}}^{\dagger}(i)\, u^{\pdagger}_{ij}\, W^{\pdagger}_{\mathcal{T}}(j) = -u^{\pdagger}_{ij}.
\end{equation}

%\bibliography{trellis}


\end{document}